\documentclass[%
 reprint,
 amsmath,amssymb,
 aps,
 prb,
 floatfix,
]{revtex4-2}

\usepackage{xcolor}
\usepackage{graphicx}
\usepackage{dcolumn}
\usepackage{bm}

\begin{document}


\title{
Spin and quadrupole correlations by 
three-spin interaction in the frustrated pyrochlore magnet Tb$_{2+x}$Ti$_{2-x}$O$_{7+y}$
}

\author{Hiroaki Kadowaki}
\affiliation{Department of Physics, Tokyo Metropolitan University, Hachioji, Tokyo 192-0397, Japan}

\author{Mika Wakita}
\affiliation{Department of Physics, Tokyo Metropolitan University, Hachioji, Tokyo 192-0397, Japan}

\author{Bj\"{o}rn F{\aa}k}
\affiliation{Institut Laue-Langevin, CS 20156, 38042 Grenoble Cedex 9, France}

\author{Jacques Ollivier}
\affiliation{Institut Laue-Langevin, CS 20156, 38042 Grenoble Cedex 9, France}

\author{Seiko Ohira-Kawamura}
\affiliation{Neutron Science Section, MLF, J-PARC Center, Shirakata, Tokai, Ibaraki 319-1195, Japan}

\date{\today}

\begin{abstract}
We have investigated the origin of the magnetic dipole correlations $\langle \sigma_{\bm{Q}}^z \sigma_{\bm{-Q}}^z \rangle$ 
characterized by the modulation wave vector $\bm{k} \sim (\tfrac{1}{2},\tfrac{1}{2},\tfrac{1}{2})$ 
observed in the frustrated pyrochlore magnet Tb$_{2+x}$Ti$_{2-x}$O$_{7+y}$. 
This magnetic short-range order cannot be accounted for by adding further-neighbor 
exchange interactions to the nearest-neighbor pseudospin-$\tfrac{1}{2}$ Hamiltonian for quantum 
pyrochlore magnets. 
Using classical Monte Carlo simulation and 
quantum simulation based on thermally pure quantum (TPQ) states 
we have shown that the spin correlations with $\bm{k} \sim (\tfrac{1}{2},\tfrac{1}{2},\tfrac{1}{2})$ 
are induced at low temperatures by a three-spin interaction 
of a form $\sigma_{\bm{r}}^{\pm} \sigma_{\bm{r}^{\prime}}^z \sigma_{\bm{r}^{\prime \prime}}^z $, 
which is a correction to the Hamiltonian 
due to the low crystal-field excitation. 
Simulations using TPQ states have shown that the spin correlations coexist with electric quadrupole correlations 
$\langle \sigma_{\bm{Q}}^{\alpha} \sigma_{\bm{-Q}}^{\beta} \rangle$ ($\alpha, \beta = x, y$) with $\bm{k} \sim \bm{0}$. 
These results suggest that the putative quantum spin liquid state of Tb$_{2+x}$Ti$_{2-x}$O$_{7+y}$ 
is located close to phase boundaries of 
the spin-ice, quadrupole-ordered, and magnetic-ordered states in the classical approximation, 
and that the three-spin interaction brings about a quantum disordered ground state 
with both spin and quadrupole correlations. 
\end{abstract}


\maketitle

\section{\label{Introduction_section} Introduction}
Frustrated magnetic systems have been actively studied in decades \cite{Lacroix11}. 
Archetypal frustrated systems consist of spins or pseudospins residing on lattices 
built from triangular and tetrahedral units. 
For example, antiferromagnetically coupled Ising spins  
on a tetrahedron are prohibited from possessing a simple ground state configuration, 
being referred to as geometrical frustration. 
Geometrically frustrated classical and quantum magnets on 
two-dimensional (2D) triangle \cite{Wannier50,Mekata1977,Hirakawa1985} 
and kagome \cite{Shyozi51,Shores2005,Han2012} lattices, 
and three-dimensional (3D) pyrochlore-lattice systems \cite{Anderson1956,Gardner99,Bramwell01,Gardner10} 
have been investigated. 
Among frustrated classical magnets, 
the spin ice on a pyrochlore lattice is of crucial importance 
because of its macroscopically degenerate ground state \cite{Bramwell01} 
and fractionalized magnetic monopole excitations \cite{Castelnovo08,Kadowaki09,Fennell2009,Morris2009,Bramwell2009}. 
Possibilities of quantum spin liquid (QSL) states in frustrated magnets 
have been actively studied in a number of years \cite{Anderson73,Balents2010}. 
By introducing transverse interactions in a frustrated Ising system, 
a QSL ground state without conventional magnetic long-range order (LRO) 
can occur, which provides challenging theoretical problems \cite{Hermele04,Savary2017}. 
Investigations of real (or candidate) QSL magnets 
are fascinating experimental explorations \cite{Hirakawa1985,Han2012,Sibille2017NC,Fak2017,Dai2021}. 

A non-Kramers pyrochlore magnet Tb$_{2}$Ti$_{2}$O$_{7}$ 
has attracted much attention for decades as a QSL candidate \cite{Gardner99,Gardner10}.
For this system any conventional magnetic LRO has never been reported. 
However, our careful studies using off-stoichiometry controlled samples 
Tb$_{2+x}$Ti$_{2-x}$O$_{7+y}$ (TTO) \cite{Taniguchi13,Wakita2016,Kadowaki2018} 
showed that TTO samples in the range $x > x_{\text{c}} \simeq -0.0025$ 
have a ground state with a conventional LRO with a hidden order parameter. 
We proposed that this LRO is an electric quadrupole (or multipole) 
order \cite{Takatsu2016prl,Kadowaki2015,Kadowaki2018prb}, 
which was predicted for general non-Kramers pyrochlore $f$-electron magnets \cite{Onoda10,Onoda11,Lee12}. 
Recently, an ultrasound experiment proved more firmly that 
a phase transition from the paramagnetic state to a quadrupole ordered (QO) state 
actually occurs \cite{Gritsenko2020}. 
On the other hand, for TTO samples in the range $x < x_{\text{c}}$ 
we showed that they have a disordered ground state without any conventional LRO \cite{Taniguchi13,Wakita2016,Kadowaki2018,Kadowaki2019}, 
being the putative QSL ground state of TTO debated in many years \cite{Gardner10,Rau_Gingras2019}. 
In spite of these experimental advances, 
theoretical challenges of clarifying the nature of this disordered ground state remain 
very difficult to date \cite{Rau_Gingras2019}. 

One can naturally expect that the QSL state of TTO can be understood 
within a framework of the pseudospin-$\tfrac{1}{2}$ nearest-neighbor (NN) exchange Hamiltonian [Eq.~(\ref{H_0})] 
for non-Kramers pyrochlore magnets \cite{Onoda10,Onoda11}. 
In this understanding it is referred to as a U(1) QSL state \cite{Hermele04,Lee12} 
or the quantum spin ice (QSI) state \cite{Molavian07,Gingras14}. 
However, it is not obvious 
whether the state in question is really the QSI state (or a state adiabatically connected to QSI) 
or another disordered ground state. 
From an experimental viewpoint there are at least two observed facts which do not conform to the QSI state, 
posing two problems to be solved. 

The first problem is why specific heat of QSL samples of TTO ($x < x_{\text{c}}$) behaves almost temperature independent, 
$C(T) \sim \text{const}$ in a range $T < 2$ K \cite{Taniguchi13}, 
while for the QSI model $C(T)$ shows the single-peak structure (anticipated at $T \sim 1$ K for TTO), 
which is a characteristic of the classical spin ice model 
and appears also in QSI \cite{Kato2015}. 
To resolve this problem the effective Hamiltonian of TTO has to be determined 
more precisely than that proposed in our previous study \cite{Takatsu2016prl},  
and a theoretical model calculation has to be performed. 
It seems that an important term is absent in our proposed Hamiltonian \cite{Takatsu2016prl}. 

The second problem is why spin correlations observed by neutron scattering experiments 
show pronounced magnetic short-range order (SRO) close to 
the wave vector $\bm{k} \sim (\tfrac{1}{2},\tfrac{1}{2},\tfrac{1}{2})$ \cite{Kadowaki2019}, 
while for the QSI model spin correlations show the pinch-point like structure 
at $\bm{k} \sim \bm{0}$ \cite{Kato2015}, that is commonly seen in spin ice models. 
In order to solve this problem, 
we studied a simple hypothesis that magnetic further-neighbor exchange interactions would 
modify the spin correlations by lifting the spin ice degeneracy. 
But we had to reject this naive hypothesis 
because unrealistically further-neighbor interactions were required to 
reproduce the observed spin correlations \cite{Kadowaki2019}. 

In this study, to solve the second problem we attempt to make use of another hint from 
an experimental fact that QO samples of TTO show a very small 
magnetic LRO with $\bm{k} = (\tfrac{1}{2},\tfrac{1}{2},\tfrac{1}{2})$, 
where the magnitude of the ordered moment is as small as $\sim 0.1 \mu_{\text{B}}$, 
which is much smaller than the moment $\simeq 5 \mu_{\text{B}}$ of the crystal-field (CF) ground state doublet 
\cite{Taniguchi13,Takatsu2016prl,Kadowaki2019,Guitteny2015}. 
If this is an intrinsic effect, 
one can come up with an idea 
that there is a weak interaction term in the Hamiltonian which couples magnetic dipole 
and electric quadrupole degrees of freedom. 
This coupling term may be at work, 
thereby spin and quadrupole correlations (and LRO) develop and affect each other at low temperatures.  

This kind of weak interaction was theoretically pointed out to exist as a three-spin interaction term 
in TTO and generally in non-Kramers pyrochlore magnets with low CF excited states \cite{Molavian2009,Rau_Gingras2019}. 
This interaction is derived from a perturbation expansion 
via virtual CF excitations \cite{Molavian2009}. 
To date, however, few investigations focusing on the three-spin interaction have been carried out. 
In the present study, we explore a possibility that the observed spin correlations 
with $\bm{k} \sim (\tfrac{1}{2},\tfrac{1}{2},\tfrac{1}{2})$ 
are accounted for by adding the three-spin interaction term to 
the NN bilinear Hamiltonian [Eq.~(\ref{H_0})]. 
More specifically, we compare the structure factor 
$S(\bm{Q}) = \int S(\bm{Q},E) dE $, where $S(\bm{Q},E)$ is the dynamic structure factor 
obtained from our previous inelastic neutron scattering data \cite{Kadowaki2018,Kadowaki2019}, 
with theoretical model calculations 
to find appropriate parameters of the three-spin interaction term. 

Among theoretical tools of model calculations for frustrated systems 
we chose two methods. 
One is the classical Monte Carlo (MC) simulation technique for Heisenberg models \cite{LandauBinder15}, 
which has been expected to be valuable for a phase transition 
with a finite critical temperature $T_{\text{c}}$ \cite{Zhitomirsky14,Yan2017,Kadowaki2018prb}.
The other is a quantum simulation technique based on 
the typicality of quantum statistical mechanics \cite{Popescu2006,Goldstein2006,Raedt2021}
and the thermally pure quantum (TPQ) states \cite{HamsRaedt2000,Sugiura2012,Sugiura2013}. 
Two methods 
using a microcanonical TPQ (mTPQ) state \cite{Sugiura2012,Kawamura2017}
and a canonical TPQ (cTPQ) state \cite{HamsRaedt2000,Sugiura2013} were employed. 
These simulation methods using the TPQ states are useful for frustrated quantum magnets, 
and have been applied for those on 
kagome \cite{Sugiura2013,Shimokawa2016}, 
honeycomb \cite{Yamaji2016,Sala2021}, 
square \cite{Misawa2018}, 
and pyrochlore \cite{Uematsu2019,Schafer2020} lattices. 
They enable approximation-free quantum simulation 
down to relatively low temperatures 
for systems as large as those of exact diagonalization. 
Using these two simulation methods 
we have found that the spin correlations with $\bm{k} \sim (\tfrac{1}{2},\tfrac{1}{2},\tfrac{1}{2})$ 
can be induced by the three-spin interaction. 

In the following sections, 
we summarize effective Hamiltonians, our previous work \cite{Takatsu2016prl}, 
and a technical target of this study in Sec.~\ref{Hamiltonian_section}. 
We present methods of neutron scattering experiments and 
the simulations in Sec.~\ref{method_section}, 
and experimental and simulation results in Sec.~\ref{Results_section}, 
which are discussed in Sec.~\ref{Discussion_section}. 

Considering that the simulation technique using the TPQ states 
is applied to analysis of $S(\bm{Q})$ observed by neutron scattering for the first time 
and that this technique itself has several limitations, 
we decide to show a number of figures of calculated $S(\bm{Q})$ 
for careful readers 
especially who will use this technique for other quantum pyrochlore magnets 
and who will examine the present results for further theoretical investigations. 
When these figures are inspected, we recommend using two (or more) displays 
to maximize the reader's image-recognition-processing ability. 
For readers who are interested in mainly results of the TTO analysis (and for first-time readers), 
to spare them the technical details 
we suggest that they read Sec.~\ref{Hamiltonian_section} first, 
and then observe 
Figs.~\ref{SQobs}(a0,c0), 
Fig.~\ref{SQcalMCp2}(k0) (with Fig.~\ref{phase_diagram_Dnn0p478}), 
and 
Figs.~\ref{SQTPQL2J3b30p1J1qP}(c1,d1) and \ref{CTST_TPQ_J3b30p1}(a) (with Fig.~\ref{phase_diagram_Dnn0p0}), 
before proceeding to the conclusion section. 

\section{\label{Hamiltonian_section} pseudospin-$\frac{1}{2}$ Hamiltonian}
\begin{figure}[ht]
\centering
\includegraphics[width=8cm,clip]{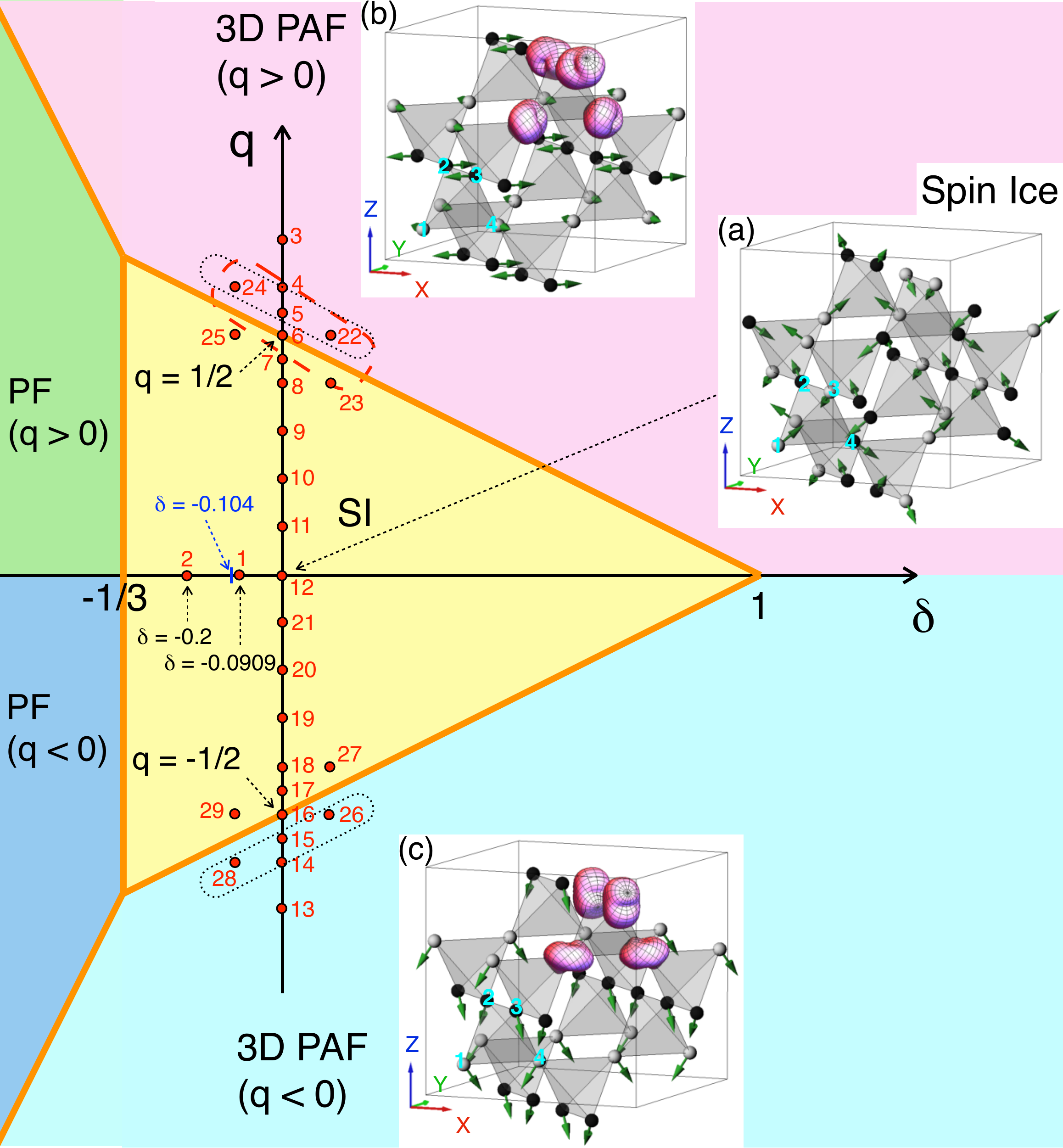}
\caption{
Classical phase diagram of the effective Hamiltonian $\mathcal{H}_{0}$ [Eq.~(\ref{H_0})] for $J_{ \text{nn} }>0$ at $T=0$ \cite{Onoda11}. 
Red circles, points 1--29, denote parameter sets of $(\delta,q)$ where simulations using the TPQ states were performed. 
The two regions enclosed by the black dotted lines represent 
the acceptable parameters for the QO sample of TTO proposed in our previous analyses \cite{Takatsu2016prl}. 
The region enclosed by the red dashed line represents 
the suggested parameters for QO and QSL samples of TTO ($-0.007<x<0.003$) 
by the present study. 
(a) A spin configuration of the classical spin ice state for $(\delta,q) = (0,0)$, 
where $\bm{\sigma}_{\bm{t}_n+\bm{d}_{\nu}} = \pm \bm{z}_{\nu}$. 
(b) A pseudospin configuration of the quadrupole order 3D PAF ($q>0$) \cite{Kadowaki2018prb}, 
where $\langle \bm{\sigma}_{\bm{t}_n+\bm{d}_{\nu}} \rangle = \bm{y}_{\nu}$ ($\nu=1,4$) 
and $\langle \bm{\sigma}_{\bm{t}_n+\bm{d}_{\nu}} \rangle = - \bm{y}_{\nu}$ ($\nu=2,3$), 
and schematic view of the deformation of the $f$-electron charge density. 
(c) A pseudospin configuration of the quadrupole order 3D PAF ($q<0$), 
where $\langle \bm{\sigma}_{\bm{t}_n+\bm{d}_{\nu}} \rangle = \bm{x}_{\nu}$ ($\nu=1,4$) 
and $\langle \bm{\sigma}_{\bm{t}_n+\bm{d}_{\nu}} \rangle = - \bm{x}_{\nu}$ ($\nu=2,3$), 
and schematic view of the deformation of the $f$-electron charge density. 
}
\label{phase_diagram_Dnn0p0}
\end{figure}

A minimal theoretical model 
for general non-Kramers $f$-electron magnets on a pyrochlore lattice \cite{Onoda10,Onoda11,Lee12} 
is the effective pseudospin-$\tfrac{1}{2}$ Hamiltonian due to electronic superexchange interactions. 
It is expressed as 
\begin{align}
\mathcal{H}_0 = 
J_{ \text{nn} } \sum_{ \langle \bm{r} , \bm{r}^{\prime} \rangle } 
\sigma_{\bm{r}}^{z} \sigma_{\bm{r}^{\prime}}^{z} 
& + J_{\text{nn}} \sum_{ \langle \bm{r} , \bm{r}^{\prime} \rangle } 
 [ 2 \delta ( \sigma_{\bm{r}}^+ \sigma_{\bm{r}^{\prime}}^- + \sigma_{\bm{r}}^- \sigma_{\bm{r}^{\prime}}^+ ) \nonumber \\
& + 2 q ( e^{2 i \phi_{\bm{r},\bm{r}^{\prime}} } \sigma_{\bm{r}}^+ \sigma_{\bm{r}^{\prime}}^+ + \text{H.c.} ) ] \; , 
\label{H_0}
\end{align}
where magnetic dipole and electric quadrupole moments at each site $\bm{r}$ are represented by 
Pauli matrices $\sigma_{\bm{r}}^z$ and $\sigma_{\bm{r}}^{\pm} = (\sigma_{\bm{r}}^{x} \pm i \sigma_{\bm{r}}^{y})/2 $, 
respectively, which are defined within the CF ground state doublet. 
The summation of Eq.~(\ref{H_0}) runs over NN site pairs $\langle \bm{r} , \bm{r}^{\prime} \rangle$. 
Detailed definitions of the Hamiltonian for TTO, 
the CF ground state doublet, lattice sites, phases $\phi_{\bm{r},\bm{r}^{\prime}}$ etc. 
are described in Appendix~\ref{appendix_def}. 

The classical phase diagram of the effective Hamiltonian [Eq.~(\ref{H_0})] for $J_{ \text{nn} }>0$ 
at $T=0$ \cite{Onoda11,Rau_Gingras2019} is reproduced in Fig.~\ref{phase_diagram_Dnn0p0} 
to briefly explain the results of our previous work of applying Eq.~(\ref{H_0}) to TTO \cite{Takatsu2016prl} 
and a technical target of this investigation. 
At the origin of Fig.~\ref{phase_diagram_Dnn0p0}, $(\delta,q) = (0,0)$, 
the Hamiltonian consists of the first term of Eq.~(\ref{H_0}) representing the classical spin ice (SI) model. 
A spin configuration of the macroscopically degenerate SI state is illustrated in Fig.~\ref{phase_diagram_Dnn0p0}(a). 
In a region close to the origin ($|\delta|, |q| \ll 1$), 
the second transverse term of Eq.~(\ref{H_0}) lifts the macroscopic degeneracy 
and the system has the U(1) QSL (QSI) ground state \cite{Hermele04,Onoda11,Lee12}. 
On the other hand, in regions far from origin ($|\delta| \gg 1$ or $|q| \gg 1 $), 
there are four classical LRO ground states: 
3D PAF ($q>0$), 3D PAF ($q<0$), PF ($q>0$), and PF ($q<0$) 
using the notations of Refs.~\cite{Onoda11,Kadowaki2018prb}, 
which correspond to PC, SFM, $\psi_2$, and $\psi_3$ of Ref.~\cite{Rau_Gingras2019}, respectively. 
Pseudospin configurations of the 3D-PAF ($q>0$) and 3D-PAF ($q<0$) states 
are shown in Figs.~\ref{phase_diagram_Dnn0p0}(b) and \ref{phase_diagram_Dnn0p0}(c), respectively, 
where electric quadrupole (multipole) orders of these states are also 
illustrated by deformation of the $f$-electron charge density from the SI state \cite{Takatsu2016prl,Kadowaki2018prb,Kadowaki2015}. 
Intermediate states between the QSI and classical LRO states 
have not been fully studied \cite{Lee12,Benton2018,Hagymasi2021}. 
On the negative $\delta$-axis 
large-scale quantum Monte-Carlo simulation was performed \cite{Kato2015}, 
which showed that the classical critical point $(\delta,q)_{\text{c}} = (-1/3,0)$ 
moves to $(\delta,q)_{\text{c}} = (-0.104,0)$ for the quantum system. 

In the previous study \cite{Takatsu2016prl}, we made arguments based mostly on classical approximations 
that the QO sample of TTO with $x=0.005$ is located close to the phase boundary 
between the SI and 3D-PAF phases. 
The acceptable $(\delta,q)$ ranges for the QO sample correspond to 
the two regions enclosed by black dotted lines in Fig.~\ref{phase_diagram_Dnn0p0} \cite{Takatsu2016prl}. 
We note that the pseudospin configuration of the LRO with $q<0$ 
is related to that with $q>0$ by the transformation 
of rotating $\bm{\sigma}_{\bm{r}}$ about the local $\bm{z}_{\bm{r}}$ axis by $\pi/2$ \cite{Onoda11}. 

In classical MC (CMC) simulation, 
we include the magnetic dipolar interaction \cite{Takatsu2016prl,Kadowaki2015}
described by 
\begin{equation}
\mathcal{H}_{\text{d}} = D r_{\rm nn}^3 \sum 
\left\{ 
\frac{ \bm{z}_{\bm{r}} \cdot \bm{z}_{\bm{r}'} }{ | \Delta \bm{r} |^3 } 
- \frac{ 3[ \bm{z}_{\bm{r}} \cdot \Delta \bm{r} ] [ \bm{z}_{\bm{r}'} \cdot \Delta \bm{r} ] }{ |\Delta \bm{r}|^5 } 
\right\}
\sigma_{\bm{r}}^{z} \sigma_{\bm{r}'}^{z} , 
\label{H_d}
\end{equation}
where the summation runs over all pairs of sites, 
$r_{\rm nn}$ is the NN distance, and $\Delta \bm{r} = \bm{r} - \bm{r}'$. 
This interaction can be approximated by 
$D_{\text{nn}} \sum_{ \langle \bm{r} , \bm{r}^{\prime} \rangle } 
\sigma_{\bm{r}}^{z} \sigma_{\bm{r}^{\prime}}^{z} $ 
with $D_{\text{nn}} = \tfrac{5}{3} D$ \cite{Hertog2000,Isakov2005}. 
Thus the effective Hamiltonian of $\mathcal{H}_{0}+\mathcal{H}_{\text{d}}$ 
can be approximated by $\mathcal{H}_{0}$ with replacements 
$J_{\text{nn}} \rightarrow J_{\text{nn}}+D_{\text{nn}}$, 
$\delta \rightarrow J_{\text{nn}}\delta/(J_{\text{nn}}+D_{\text{nn}})$, and 
$q \rightarrow J_{\text{nn}}q/(J_{\text{nn}}+D_{\text{nn}})$. 

\begin{figure}[ht]
\centering
\includegraphics[width=8.0cm,clip]{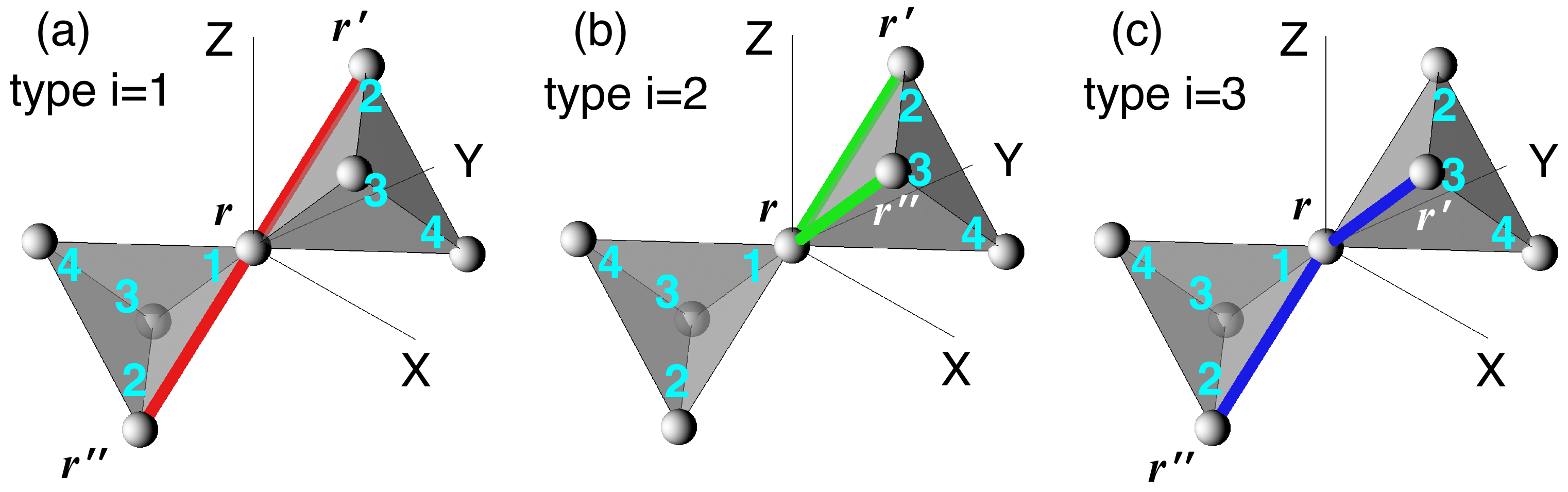}
\caption{ 
Three geometrically distinct triplets 
$\langle \bm{r}, \bm{r}^{\prime}, \bm{r}^{\prime \prime} \rangle$ of Eq.~(\ref{H_3s})
are illustrated in (a), (b), and (c), which 
correspond to the type $i=1$, $2$, and $3$ three-spin interaction 
with the coupling constant $J_{3\text{s},i}$, respectively. 
}
\label{J3s_type123}
\end{figure}

In addition to $\mathcal{H}_{0}$ and $\mathcal{H}_{\text{d}}$, 
we consider a three-spin interaction term 
expressed as 
\begin{equation}
\mathcal{H}_{3\text{s}} = 
\sum_{i=1}^3 J_{3\text{s},i} 
\sum_{\langle \bm{r}, \bm{r}^{\prime}, \bm{r}^{\prime \prime} \rangle} 
[ e^{i \phi^{(i)}_{\bm{r}, \bm{r}^{\prime}, \bm{r}^{\prime \prime} } } 
\sigma_{\bm{r}}^+ \sigma_{\bm{r}^{\prime}}^z \sigma_{\bm{r}^{\prime \prime}}^z + \text{H.c.} ] \; , 
\label{H_3s}
\end{equation}
where $\langle \bm{r}, \bm{r}^{\prime} \rangle$ and $\langle \bm{r}, \bm{r}^{\prime \prime} \rangle$ 
are NN pairs and $\bm{r}^{\prime} \ne \bm{r}^{\prime \prime}$, 
which are satisfied by three distinct types of triplet 
$\langle \bm{r}, \bm{r}^{\prime}, \bm{r}^{\prime \prime} \rangle$ 
shown in Fig.~\ref{J3s_type123} \cite{Rau_Gingras2019,Molavian2009}. 
By imposing the condition of invariance under the space group symmetry 
to $\mathcal{H}_{3\text{s}}$, 
one can show that the three-spin interaction term has the form of Eq.~(\ref{H_3s}) 
with three real coupling constants $J_{3\text{s},i}$ ($i=1,2,3$) 
and phases $\phi^{(i)}_{\bm{r}, \bm{r}^{\prime}, \bm{r}^{\prime \prime} }$ 
listed in Tables~\ref{phase_type1}, \ref{phase_type2}, and \ref{phase_type3} 
in Appendix~\ref{appendix_three_spin_interaction}. 
We note that the phases $\phi^{(i)}_{\bm{r}, \bm{r}^{\prime}, \bm{r}^{\prime \prime} }$ ($i=1,2,3$) 
with the site triplet $\langle \bm{r}, \bm{r}^{\prime}, \bm{r}^{\prime \prime} \rangle$ 
illustrated in Figs.~\ref{J3s_type123}(a), \ref{J3s_type123}(b), and \ref{J3s_type123}(c) are listed 
in the first lines of Table~\ref{phase_type1}, \ref{phase_type2}, and \ref{phase_type3}, 
respectively. 
Since the phases $\phi^{(i)}_{\bm{r}, \bm{r}^{\prime}, \bm{r}^{\prime \prime} }$ are fixed by the symmetry, 
the adjustable parameters of $\mathcal{H}_{3\text{s}}$ are the three coupling constants $J_{3\text{s},i}$. 

We used total effective Hamiltonians 
$\mathcal{H} = \mathcal{H}_{0} + \mathcal{H}_{d} + \mathcal{H}_{3\text{s}}$ 
and 
$\mathcal{H} = \mathcal{H}_{0} + \mathcal{H}_{3\text{s}}$ 
for the classical and quantum simulations, respectively. 
The magnitude of the coupling constants of $\mathcal{H}_{0} + \mathcal{H}_{d}$, 
scaled by $J_{\text{nn}}$ and $D_{\text{nn}}$, 
should be close to that of our previous study \cite{Takatsu2016prl}, in which 
$J_{\text{nn}}=1.0$ K and $D_{\text{nn}} = 0.48$ K. 
As for the parameters $(\delta,q)$, 
they should be close to one of the two regions 
enclosed by the black dotted lines in Fig.~\ref{phase_diagram_Dnn0p0} \cite{Takatsu2016prl}. 
Thus the technical target of the present study is to find 
parameter sets $(J_{3\text{s},1},J_{3\text{s},2},J_{3\text{s},3})$ of $\mathcal{H}_{3\text{s}}$ 
which can explain the spin correlations of TTO. 
The magnitude of $J_{3\text{s},i}$ is the order of $J_{\text{nn}}^2/\Delta \simeq 0.1 J_{\text{nn}} \simeq 0.1$ K, 
where $\Delta$ is the energy of the first CF excited state \cite{Rau_Gingras2019,Molavian2009}. 
It should be noted that 
since the theoretical tools we applied are far from perfect for many-body quantum states, 
what we can do best at present would be to qualitatively reproduce the spin correlations of TTO using the simulations. 

\section{\label{method_section} Methods}
\subsection{\label{method_exp} Experimental Methods}
Single crystalline samples of Tb$_{2+x}$Ti$_{2-x}$O$_{7+y}$ 
with $x=-0.007, 0.000$ and $0.003$ 
used in this study are those of Refs.~\cite{Kadowaki2018,Kadowaki2019}, 
where methods of the sample preparation and the estimation of $x$ values 
are described. 
The QSL sample with $x=-0.007$ remains in the paramagnetic 
state down to 0.1 K.
The QO samples with $x=0.000$ and $x=0.003$ very likely have small and 
large electric quadrupole orders, respectively, 
in $T \ll T_c \sim 0.4 $ K \cite{Taniguchi13,Wakita2016}. 

Neutron scattering experiments were carried out on the time-of-flight (TOF) spectrometer 
IN5 operated with $\lambda = 8$ {\AA} at ILL for the $x = -0.007$ and 0.000 crystal samples 
\cite{Fak2015,Fak2016,Kadowaki2018,Kadowaki2019}. 
The energy resolution of this condition was $\Delta E = 0.021$ meV (FWHM) 
at the elastic position. 
Neutron scattering experiments for the $x = 0.003$ crystal sample were performed on the TOF 
spectrometer AMATERAS operated with $\lambda = 7$ {\AA} at J-PARC 
\cite{Kadowaki2018,Kadowaki2019}. 
The energy resolution of this condition was $\Delta E = 0.024$ meV (FWHM) at the elastic position. 
Each crystal sample was mounted in a dilution refrigerator 
so as to coincide its $(h,h,l)$ plane with 
the horizontal scattering plane of the spectrometer. 
The observed intensity data were 
corrected for background and absorption using a home-made program \cite{Kadowaki2018github}. 
Construction of a four dimensional $S(\bm{Q},E)$ data object 
from a set of the TOF data taken by rotating each crystal sample 
was performed using {H}{\footnotesize ORACE} \cite{Horace2016}. 

\subsection{\label{method_CMC} classical MC simulation}
\begin{figure}[ht]
\centering
\includegraphics[width=6cm,clip]{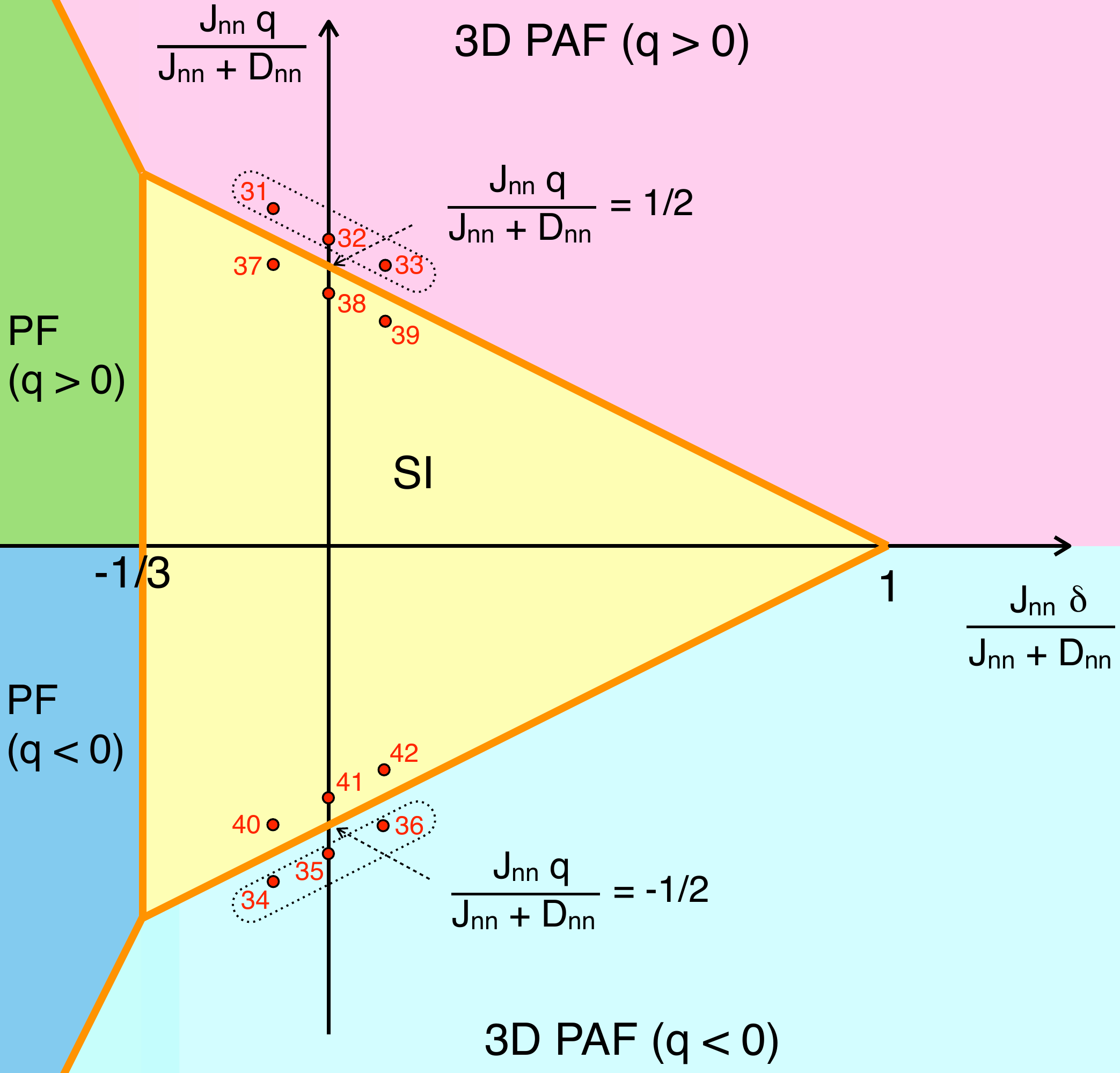}
\caption{ 
Classical phase diagram of the effective Hamiltonian 
$\mathcal{H}_{0}+\mathcal{H}_{\text{d}}$ [Eqs.~(\ref{H_0}) and (\ref{H_d})] 
where the magnetic dipolar interaction is approximated by the NN coupling 
$D_{\text{nn}} \sum_{ \langle \bm{r} , \bm{r}^{\prime} \rangle } \sigma_{\bm{r}}^{z} \sigma_{\bm{r}^{\prime}}^{z} $. 
Red circles, points 31--42, denote parameter sets of $(\delta,q)$ where CMC simulations were performed. 
The two regions enclosed by the black dotted lines represent 
the acceptable parameters for the QO sample of TTO proposed in our previous analyses \cite{Takatsu2016prl}. 
}
\label{phase_diagram_Dnn0p478}
\end{figure}
Classical MC simulations of the model described by 
$\mathcal{H} = \mathcal{H}_0 + \mathcal{H}_{\text{d}} + \mathcal{H}_{3\text{s}}$ [Eqs.~(\ref{H_0}), (\ref{H_d}), and (\ref{H_3s})]
were carried out 
by treating the pseudospin $\bm{\sigma}_{\bm{r}}$ as a classical unit vector \cite{LandauBinder15}. 
The NN exchange constant and the dipole interaction parameter were fixed to 
$J_{\text{nn}} = 1.0$ K and $D_{\text{nn}} = 0.48 $ K \cite{Takatsu2016prl}. 
The parameter sets of $(\delta,q)$ were 
$J_{\text{nn}} \delta / (J_{\text{nn}}+D_{\text{nn}}) = -0.1, 0, 0.1$ 
and 
$J_{\text{nn}}|q| / (J_{\text{nn}}+D_{\text{nn}}) + J_{\text{nn}} \delta / [2(J_{\text{nn}}+D_{\text{nn}})] = 1.1/2, 0.9/2 $, 
encompassing the 3D-PAF and classical SI states. 
These are shown by red circles, the points 31--42, in Fig.~\ref{phase_diagram_Dnn0p478}. 
The CMC simulations were performed 
with typically $\sim 4 \times 10^5$ MC steps per spin 
and on periodic clusters with $N= 16 L^3$ spins (sites), 
where $L$ ($=4,10$) stands for a linear dimension parallel to the [100] direction. 
We used the Metropolis single spin-flip update \cite{LandauBinder15} 
and the exchange Monte-Carlo method \cite{Hukushima96}. 

\subsection{\label{method_TPQ} quantum simulation using TPQ states}
We have adopted methods of the quantum simulation based on the mTPQ and cTPQ states 
which are described in Refs.~\cite{Sugiura2012,Sugiura2013,Kawamura2017}. 
These methods enable us to calculate expectation values of observables 
and thermodynamic quantities at finite temperatures 
by applying a computational technique which is commonly used for the Lanczos method \cite{Nishimori1991,titpack2url,Kawamura2017}. 
The algorithm of the simulation using the mTPQ state can be performed by 
storing only two vectors in the $2^N$ dimensional Hilbert space, 
where $N$ is the number of pseudospins (sites). 
It is not difficult to carry out this simulation for a system with $N=32$ sites 
without special techniques on a PC with ca. 160 GB memory. 
We calculated expectation values of pseudospin correlations 
and thermodynamic quantities 
using the methods of the mTPQ and cTPQ states, respectively.  

A series of mTPQ states are generated by iteratively operating the Hamiltonian $\mathcal{H}$ to 
a random normalized vector $| \psi_0 \rangle$ in the Hilbert space \cite{Sugiura2012}. 
More specifically, the mTPQ states are calculated by 
\begin{equation}
| \psi_k \rangle = \frac{1}{\sqrt{Q_k}} \left( \ell - \hat{h} \right)^k | \psi_0 \rangle , 
\label{phi_k}
\end{equation}
where $k=0,1,2, \cdots$,  $\hat{h} = \mathcal{H}/N $, 
$\ell$ is a constant larger than the maximum eigenvalue of $\hat{h} $, 
and $Q_k = \left| \left( \ell - \hat{h} \right)^k | \psi_0 \rangle \right|^{2}$ is 
a normalization constant. 
The temperature corresponding to $| \psi_k \rangle$ is 
\begin{equation}
T_k = \frac{N}{2k k_{\text{B}}} \left( \ell - \langle \psi_k | \hat{h} | \psi_k \rangle \right) .
\label{phi_k}
\end{equation}

An equilibrium expectation value of an observable 
represented by an operator $\hat{A}$ for the mTPQ state $| \psi_k \rangle$ is 
\begin{equation}
\langle \hat{A} \rangle_k = \langle \psi_k | \hat{A} | \psi_k \rangle .
\label{A_k}
\end{equation}
By applying this equation to pseudospin correlations 
$ \hat{A} = \sigma_{\bm{r}}^{\alpha} \sigma_{\bm{r}^{\prime}}^{\alpha} $ ($\alpha = x,z$), 
the Fourier transform of their expectation values is calculated by 
\begin{equation}
\langle \sigma_{\bm{Q}}^{\alpha} \sigma_{-\bm{Q}}^{\alpha} \rangle 
\propto \sum_{\bm{r},\bm{r}^{\prime}} 
\langle \sigma_{\bm{r}}^{\alpha} \sigma_{\bm{r}^{\prime}}^{\alpha} \rangle_k 
\exp[-i \bm{Q} \cdot (\bm{r} - \bm{r}^{\prime}) ]  .
\label{sigma_xz_Q}
\end{equation}
Similarly the structure factor 
$S(\bm{Q})$ at $T_k$ is calculated by 
\begin{align}
S(\bm{Q}) \propto f(Q)^2 \sum_{\bm{r},\bm{r}^{\prime}} 
& [ \bm{z}_{\bm{r}} \cdot \bm{z}_{\bm{r}^{\prime}} 
- (\hat{\bm{Q}} \cdot \bm{z}_{\bm{r}})(\hat{\bm{Q}} \cdot \bm{z}_{\bm{r}^{\prime}}) ] 
\nonumber \\
& \langle \sigma_{\bm{r}}^z \sigma_{\bm{r}^{\prime}}^z \rangle_k 
\exp[-i \bm{Q} \cdot (\bm{r} - \bm{r}^{\prime}) ] ,
\label{SQ}
\end{align}
where $f(Q)$ is the magnetic form factor and $\hat{\bm{Q}}=\bm{Q}/|\bm{Q}|$. 
These expectation values [Eqs.~(\ref{sigma_xz_Q}) and (\ref{SQ})] 
are averaged over different (typically four) realizations of 
the series of the mTPQ states.

Once all expectation values of $\langle \hat{h}^n \rangle_k $ ($n=1,2,3$) 
for the mTPQ states are obtained, 
it is straightforward to evaluate specific heat and entropy  
at a temperature $T = 1/(k_{\text{B}} \beta)$ 
using the cTPQ state $| \beta, N \rangle$ which is defined by 
\begin{equation}
| \beta, N \rangle = \exp [ - \beta \mathcal{H}/2 ] | \psi_0 \rangle .
\label{phi_beta}
\end{equation}
In the cTPQ method 
an equilibrium value of $\hat{A}$ at $T$ is 
\begin{equation}
\langle \hat{A} \rangle_T = 
\frac{ \left[ \langle \beta, N | \hat{A} | \beta, N \rangle \right]_{\text{av}} }
     { \left[ \langle \beta, N |           \beta, N \rangle \right]_{\text{av}} }  , 
\label{A_beta}
\end{equation}
where $\left[ \bullet \right]_{\text{av}} $ stands for 
the arithmetic mean over the initial states $| \psi_0 \rangle $ \cite{Sugiura2013}. 
Specific heat and entropy per pseudospin at $T$ are expressed as 
\begin{equation}
C(T) = \frac{N}{T^2} \left[ \langle \hat{h}^2 \rangle_T -  \left( \langle \hat{h} \rangle_T  \right)^2 \right] 
\label{C_beta}
\end{equation}
and 
\begin{equation}
S(T) = \frac{\langle \hat{h} \rangle_T}{T} 
+ \frac{1}{N} \ln  \left[ \langle \beta, N | \beta, N \rangle \right]_{\text{av}} + \ln 2 ,
\label{S_beta}
\end{equation}
respectively. 
These $C(T)$ and $S(T)$ can be calculated using 
\begin{align}
\langle \beta, N | \hat{h}^n | \beta, N \rangle 
= 
e^{-N \beta \ell} 
\sum_{k=0}^{\infty} \frac{(N \beta)^{2 k}}{(2 k)!} Q_k & \nonumber \\
 \left[ \langle \hat{h}^n \rangle_k 
+ \frac{N \beta}{2k+1} ( \ell \langle \hat{h}^n \rangle_k - \langle \hat{h}^{n+1} \rangle_k ) \right] & ,
\label{Hn_beta}
\end{align}
where $n=0,1,$ and $2$. 

\begin{figure}[ht]
\centering
\includegraphics[width=8.0cm,clip]{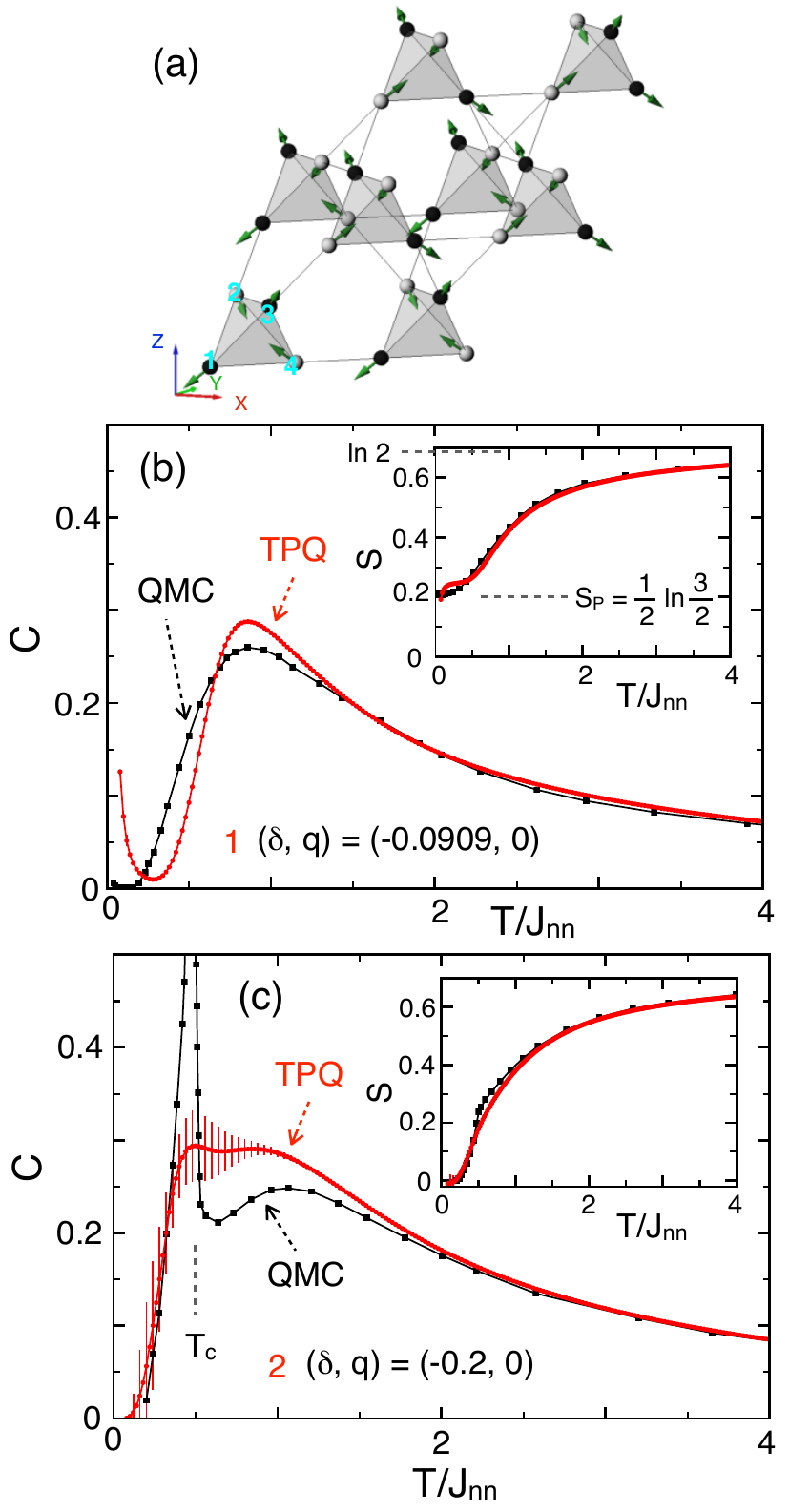}
\caption{
(a) Periodic 32-site cluster for the simulation using the TPQ states. 
(b,c) Specific heat and entropy as a function of temperature 
obtained by 6912-site QMC (black curve) \cite{Kato2015} and the 32-site simulations using the TPQ states 
without three-spin interaction (red curve) 
are shown in 
(b) for the parameters $(\delta,q)=(-0.0909,0)$ (the point 1 in Fig.~\ref{phase_diagram_Dnn0p0}) 
and in (c) for $(\delta,q)=(-0.2,0)$ (the point 2 in Fig.~\ref{phase_diagram_Dnn0p0}), 
which correspond to the QSI and quadrupole LRO ground states, respectively. 
}
\label{QMC_TPQ}
\end{figure}

Simulations using the TPQ states were carried out 
using the simplified Hamiltonian $\mathcal{H} = \mathcal{H}_0 + \mathcal{H}_{3\text{s}}$ [Eqs.~(\ref{H_0}) and (\ref{H_3s})]. 
They were performed on a periodic cluster with $N= 4 L^{\prime 3}=32$ sites ($L^{\prime}=2$), 
which is illustrated in Fig.~\ref{QMC_TPQ}(a), 
where $L^{\prime}$ stands for a linear dimension 
parallel to the FCC translation vector $(\tfrac{1}{2},\tfrac{1}{2},0)$. 
We note that this $2 \times 2 \times 2$ lattice is the minimal cluster size, 
by which one can study 
whether a peak in pseudospin correlations 
is $\bm{k} \sim (\tfrac{1}{2},\tfrac{1}{2},\tfrac{1}{2})$ or $(0,0,0)$. 

To examine limitations of the TPQ methods especially due to finite size effects 
we compare the 32-site simulation using the TPQ states 
with the large-scale quantum MC (QMC) simulation 
on a cluster of $N= 4 L^{\prime 3}=6912$ sites ($L^{\prime}=12$) \cite{Kato2015}, 
which were performed for the Hamiltonian $\mathcal{H}_0$ 
in the negative $\delta$ direction $(\delta < 0, q=0)$ (Fig.~\ref{phase_diagram_Dnn0p0}). 
We performed 32-site simulations with two parameter sets 
corresponding to the points 1 and 2 in Fig.~\ref{phase_diagram_Dnn0p0}, 
where QMC data are available \cite{Kato2015}. 

At the point 1 in Fig.~\ref{phase_diagram_Dnn0p0}, $(\delta,q)=(-0.0909,0)$, 
$\delta$ is larger than the critical value $\delta_{\text{c}} = -0.104$, 
and the system is in the QSI state at $T=0$. 
Specific heat and entropy as a function of temperature 
are shown in Fig.~\ref{QMC_TPQ}(b). 
In a high temperature range of $T/J_{\text{nn}} > 0.2$, specific heat and entropy 
show similar behavior of the classical SI for the both simulations. 
On the other hand, in a lower $T$-range of $T/J_{\text{nn}} < 0.2$ 
the TPQ result of $C(T)$ shows considerable upturn, which is very different from the QMC result. 
This is probably a small-size artifact, 
which is commonly seen in TPQ results at low temperatures \cite{Sugiura2013,Shimokawa2016,Schnack2018}. 

At the point 2 in Fig.~\ref{phase_diagram_Dnn0p0}, $(\delta,q)=(-0.2,0)$,
$\delta$ is smaller than the critical value, 
and the system is in a quadrupole LRO state at $T=0$. 
Specific heat and entropy as a function of temperature 
are plotted in Fig.~\ref{QMC_TPQ}(c). 
One can see from this figure that the specific heat peak at the phase transition temperature 
$T_{\text{c}}/J_{\text{nn}} \simeq 0.5$ 
is only slightly seen for the TPQ result, which is the well-known finite-size effect, 
and that statistical errors of the TPQ result become very large at low temperatures ($T/J_{\text{nn}}< 1$). 
The large errors at low temperatures are commonly seen in other TPQ results \cite{Yamaji2016,Misawa2018,Raedt2021}. 

From the two comparisons shown in Fig.~\ref{QMC_TPQ}, 
we can infer 
that despite the small system size TPQ results can provide useful information on 
low-$T$ states which have high entropy down to about $T/J_{\text{nn}}= 0.2$ (for the present case). 
This is in agreement with the studies of 
the frustrated Heisenberg antiferromagnet on the kagome lattice \cite{Sugiura2013,Shimokawa2016}. 
While we have to cautiously interpret TPQ results, 
when ground states have classical LROs 
and finite-temperature phase transitions occur. 

\section{\label{Results_section} Results}
\subsection{\label{results_SQ_obs} $S(\bm{Q})$ observed by neutron scattering experiments}
\begin{figure*}[ht]
\centering
\includegraphics[width=18cm,clip]{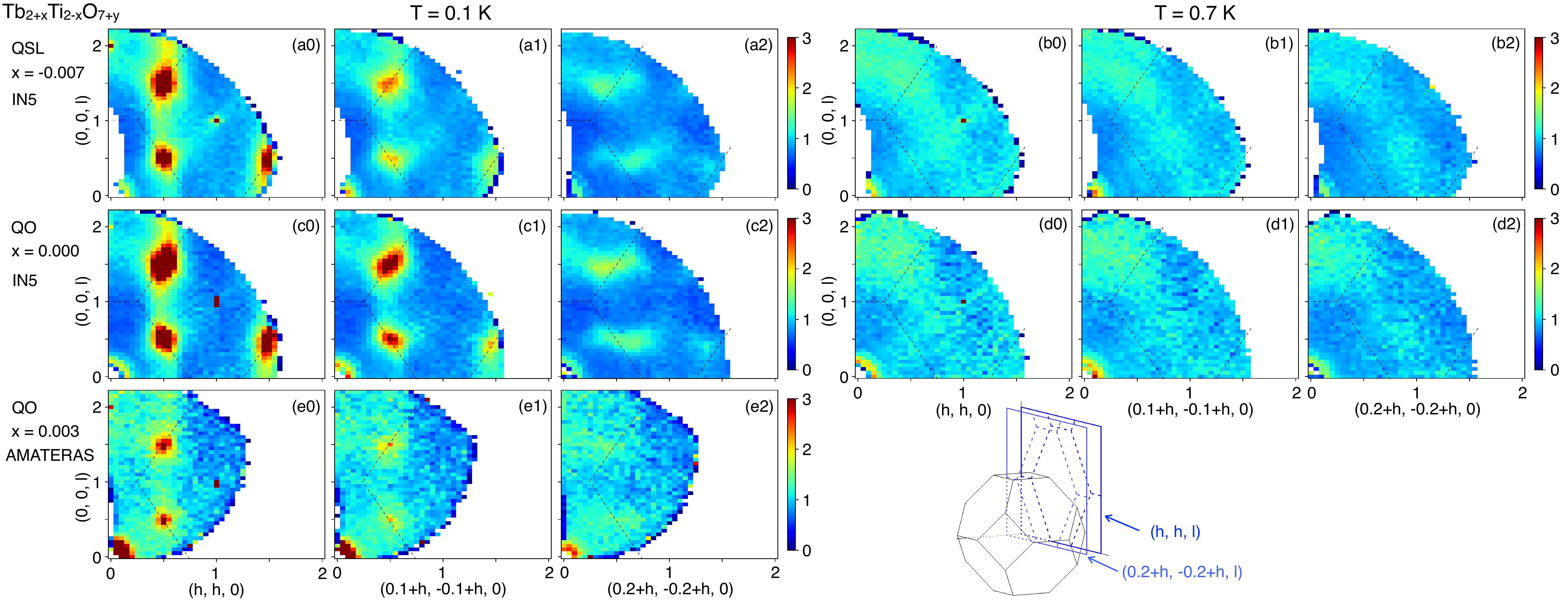}
\caption{
(a,b) Intensity maps of 3D data $S(\bm{Q})$ of the QSL sample with $x = -0.007$, 
which were measured on IN5 at 0.1 K and 0.7 K, 
are shown in (a) and (b), respectively. 
(c,d) Intensity maps of 3D data $S(\bm{Q})$ of the QO sample with $x = 0.000$, 
which were measured on IN5 at 0.1 K and 0.7 K, 
are shown in (c) and (d), respectively. 
(e) Intensity maps of 3D data $S(\bm{Q})$ of the QO sample with $x = 0.003$, 
which were measured on AMATERAS at 0.1 K. 
The 3D data are viewed by 2D slices, which are parallel cross-sections 
of $\bm{Q}=(k+h,-k+h,l)$ with fixed $k=0,0.1$, and $0.2$, 
are shown in (w0), (w1), and (w2) (w=a--e), respectively. 
Dashed lines in the 2D slices (a)--(e) are boundaries of Brillouin zones. 
The bottom right corner shows the first Brillouin zone of 
the FCC lattice (thin black lines),  
and two 2D slice planes with $k=0$ and $0.2$ (blue lines). 
}
\label{SQobs}
\end{figure*}
In the simulations using CMC and the TPQ states equal-time spin correlations are calculated. 
These correspond to the structure factor 
$S(\bm{Q}) = \int S(\bm{Q},E) dE$. 
To compare results of the simulations with 
the previous neutron scattering data \cite{Kadowaki2018,Kadowaki2019}, 
we integrated $S(\bm{Q},E)$ in an energy range $-0.3 < E < 0.5$ meV, 
which covers most of the energy spread around $E=0$ and excludes CF excitations. 
We constructed 3D data sets of 
$S(\bm{Q})=\int_{-0.3 \text{meV}}^{0.5 \text{meV}} S(\bm{Q},E) dE$ 
which are normalized using the same methods as those described in Refs.~\cite{Kadowaki2018,Kadowaki2019}. 
Consequently intensities of $S(\bm{Q})$ can be compared mutually among the three samples of TTO. 

In Fig.~\ref{SQobs} we show intensity maps of the observed $S(\bm{Q})$ 
of the QSL sample with $x=-0.007$ and of the two QO samples with $x=0.000$ and $0.003$. 
It is obvious that the pronounced peaks in $S(\bm{Q})$ 
at $\bm{k} \sim (\tfrac{1}{2},\tfrac{1}{2},\tfrac{1}{2})$ 
appear only at 0.1 K. 
An interesting point of these data, 
which is not seen in the $[S(\bm{Q})]_{\text{el}}$ (nominally elastic scattering) data of Ref.~\cite{Kadowaki2019}, 
is that there are pinch-point like structures in the $S(\bm{Q})$ data at 0.1 K 
for the QSL sample around $\bm{Q}=(0,0,2)$ and $(1,1,1)$, 
and that they become weak for the QO samples. 
This fact is consistent with the interpretation that the QSL sample 
is located closer to the SI phase (Fig.~\ref{phase_diagram_Dnn0p0}) 
than the QO samples \cite{Takatsu2016prl}. 
It should be noted that the pinch-point like structures in $S(\bm{Q})$ are inelastic scattering. 

\subsection{\label{results_CMC} classical MC simulations }
Classical simulations based on the MC method using the Hamiltonian 
$\mathcal{H}_0 + \mathcal{H}_{\text{d}} + \mathcal{H}_{3\text{s}}$ were carried out. 
By these CMC simulations we can search for candidate parameter sets for TTO 
in a wider parameter space than the TPQ methods. 
A guideline of this search is that the spin correlations of TTO are most enhanced 
in the QO sample with $x=0.000$ [Fig.~\ref{SQobs}(c0)], where the quadrupole order is probably small. 
Assuming small quadrupole LRO $\langle \sigma_{\bm{r}}^+ \rangle$, 
it is expected that an effective bilinear magnetic coupling term 
\begin{equation}
\sum_{i=1}^3 J_{3\text{s},i} 
\sum_{\langle \bm{r}, \bm{r}^{\prime}, \bm{r}^{\prime \prime} \rangle} 
[ e^{i \phi^{(i)}_{\bm{r}, \bm{r}^{\prime}, \bm{r}^{\prime \prime} } } 
\langle \sigma_{\bm{r}}^+ \rangle \sigma_{\bm{r}^{\prime}}^z \sigma_{\bm{r}^{\prime \prime}}^z + \text{H.c.} ] 
\label{H_3s_LRO}
\end{equation}
becomes at work to lift the SI degeneracy due to $\mathcal{H}_0 + \mathcal{H}_{\text{d}}$, 
and consequently spin correlations with different wave-vector dependence 
appear at low temperatures. 
Therefore, there is a chance to find candidate coupling constants $J_{3\text{s},i}$, 
if the parameters $(\delta,q)$ are 
close to the boundaries of the SI and 3D-PAF phases (Fig.~\ref{phase_diagram_Dnn0p478}), 
in particular, on the 3D-PAF phase sides. 

\subsubsection{\label{results_CMC_C_3DPAF} specific heat in 3D-PAF phases}
\begin{figure}[hbt]
\centering
\includegraphics[width=8cm,clip]{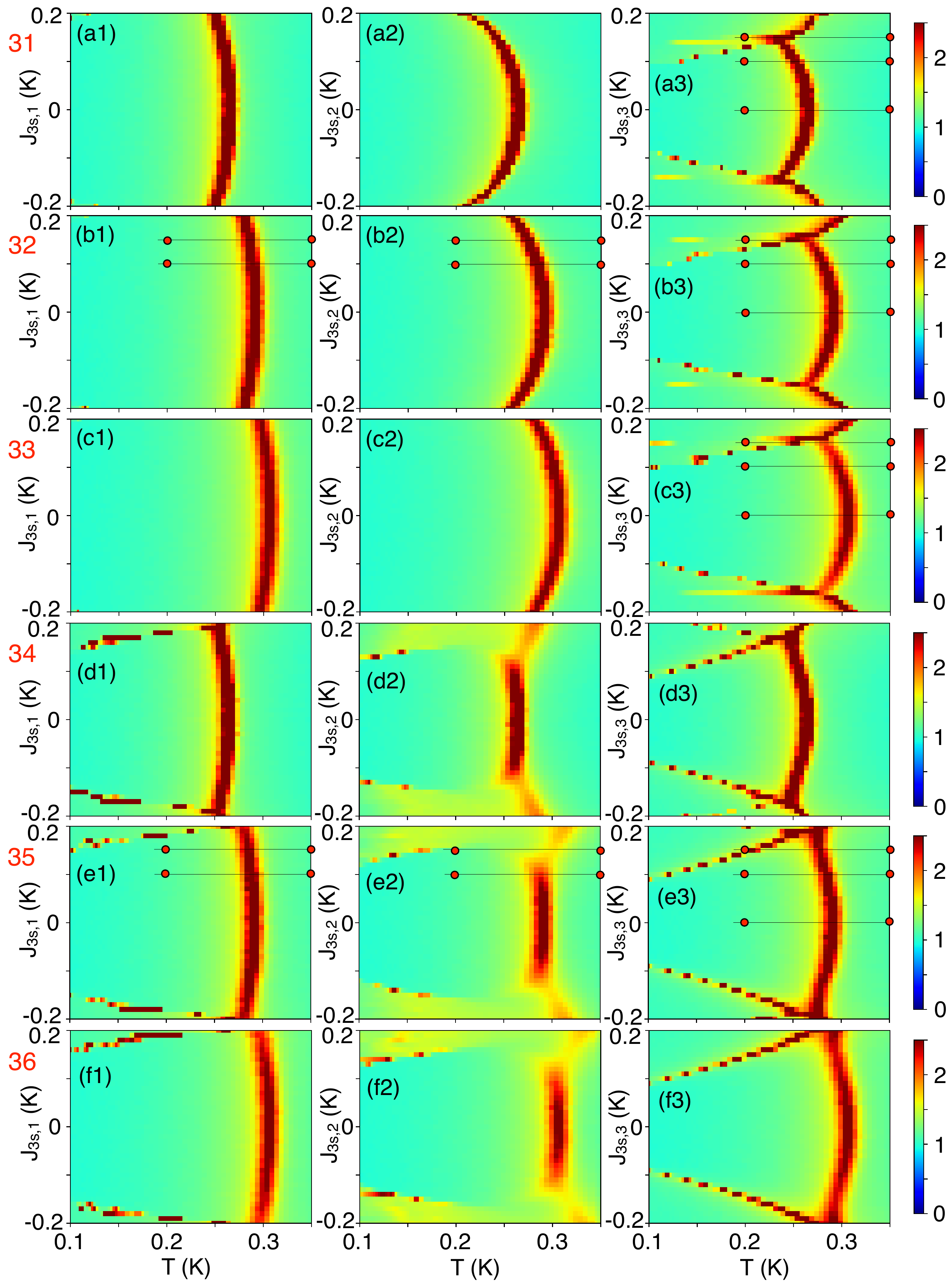}
\caption{
(a--f) Specific heat $C(T,J_{3\text{s},i})$ ($i=1,2,3$; $J_{3\text{s},j \neq i}=0$) calculated by the 1024-site CMC simulations are shown 
as color maps: (w1) $C(T,J_{3\text{s},1})$, (w2) $C(T,J_{3\text{s},2})$, and (w3) $C(T,J_{3\text{s},3})$, 
where w=a--f. 
The parameters $(\delta,q)$ of (a), (b), (c), (d), (e), and (f) 
are $\frac{J_{\text{nn}}}{J_{\text{nn}} + D_{\text{nn}}} (\delta,q)=(-0.1,0.6)$, 
$(0.0,0.55)$, $(0.1,0.5)$, $(-0.1,-0.6)$, $(0.0,-0.55)$, and $(0.1,-0.5)$, respectively, 
which correspond to the points 31--36 in Fig.~\ref{phase_diagram_Dnn0p478}.
}
\label{CT_MC123456}
\end{figure}

A number of CMC simulations with a system size $L=4$ (1024 sites) were performed 
to study effects of each three-spin interaction 
on the 3D-PAF phase sides of neighborhoods of the SI and 3D-PAF phase boundaries. 
The parameters $(\delta,q)$ were fixed to the six selected sets: 
$\frac{J_{\text{nn}}}{J_{\text{nn}} + D_{\text{nn}}} (\delta,q)=(-0.1,0.6)$, 
$(0.0,0.55)$, $(0.1,0.5)$, $(-0.1,-0.6)$, $(0.0,-0.55)$, and $(0.1,-0.5)$.  
These correspond to the points 31--36 in Fig.~\ref{phase_diagram_Dnn0p478}, 
which are in the proposed parameter ranges for the QO sample \cite{Takatsu2016prl}. 
The CMC simulations were carried out with many 
three-spin coupling constants $J_{3\text{s},i}$ in a range $|J_{3\text{s},i}|<0.2$ K, 
where one $J_{3\text{s},i} \neq 0$ is finite and the other two $J_{3\text{s},j \neq i}=0$ are zero. 
Resulting specific heat data are plotted in Fig.~\ref{CT_MC123456} 
as color maps of $C(T,J_{3\text{s},i})$. 

\begin{figure}[hbt]
\centering
\includegraphics[width=8cm,clip]{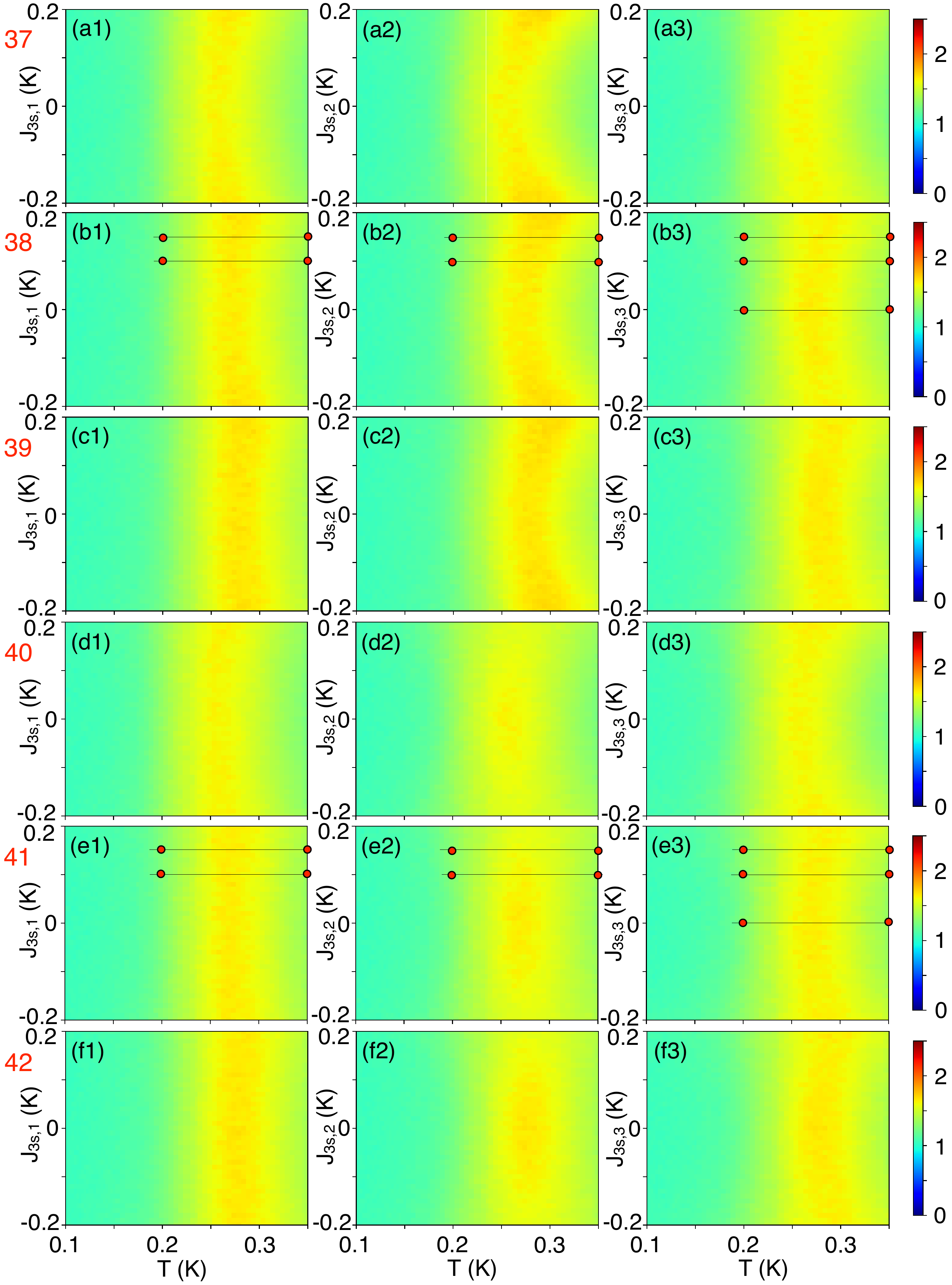}
\caption{ 
(a--f) Specific heat $C(T,J_{3\text{s},i})$ ($i=1,2,3$; $J_{3\text{s},j \neq i}=0$) calculated by the 1024-site CMC simulation are shown 
as color maps: (w1) $C(T,J_{3\text{s},1})$, (w2) $C(T,J_{3\text{s},2})$, and (w3) $C(T,J_{3\text{s},3})$, 
where w=a--f. 
The parameters $(\delta,q)$ of (a), (b), (c), (d), (e), and (f) 
are $\frac{J_{\text{nn}}}{J_{\text{nn}} + D_{\text{nn}}} (\delta,q)=(-0.1,0.5)$, 
$(0.0,0.45)$, $(0.1,0.4)$, $(-0.1,-0.5)$, $(0.0,-0.45)$, and $(0.1,-0.4)$, respectively, 
which correspond to the points 37--42 in Fig.~\ref{phase_diagram_Dnn0p478}. 
}
\label{CT_MC111213141516}
\end{figure}
The temperature dependence of specific heat with $J_{3\text{s},i}=0$, 
$C(T,J_{3\text{s},i}=0)$ [Figs.~\ref{CT_MC123456}(a--f)], 
is consistent with our previous CMC simulation study \cite{Kadowaki2018prb}, 
where a single first-order phase transition occurs 
from the paramagnetic to 3D-PAF states. 
For finite $J_{3\text{s},i} \ne 0$ another phase transition at a lower temperature occurs, 
which is noticeably seen in Figs.~\ref{CT_MC123456}(a3--f3).
The lower critical temperature is a phase transition 
to a state with both quadrupole and magnetic LROs, as will be discussed later. 
We note that the invariance of the Hamiltonian due to the transformation, 
$\sigma_{\bm{r}}^+ \rightarrow -\sigma_{\bm{r}}^+$ and 
$J_{3\text{s},i} \rightarrow  -J_{3\text{s},i}$, 
are seen in Figs.~\ref{CT_MC123456}(a--f) 
as $C(T,J_{3\text{s},i}) \simeq C(T,-J_{3\text{s},i})$. 
We also note that the symmetry between positive-$q$ and negative-$q$ states for $J_{3\text{s},i} = 0$ 
does not hold for $J_{3\text{s},i} \ne 0$, 
which is seen in Figs.~\ref{CT_MC123456}(a--f) as, e.g., 
a fact that Fig.~\ref{CT_MC123456}(a1) ($q=0.6$) is different from Fig.~\ref{CT_MC123456}(d1) ($q=-0.6$). 
Intriguingly, this implies that it is possible 
to distinguish the 3D-PAF ($q>0$) order from the 3D-PAF ($q<0$) order 
even though the quadrupole order is experimentally invisible, 
if the three-spin interaction term is finite. 

\subsubsection{\label{results_CMC_C_SI} specific heat in SI phase}
To study effects of each three-spin interaction 
on the SI phase sides of neighborhoods of the SI and 3D-PAF phase boundaries, 
a number of 1024-site CMC simulations were performed with 
the six selected parameter sets of $(\delta,q)$: 
$\frac{J_{\text{nn}}}{J_{\text{nn}} + D_{\text{nn}}} (\delta,q)=(-0.1,0.5)$, 
$(0.0,0.45)$, $(0.1,0.4)$, $(-0.1,-0.5)$, $(0.0,-0.45)$, and $(0.1,-0.4)$.  
These correspond to the points 37--42 in Fig.~\ref{phase_diagram_Dnn0p478}. 
The CMC simulations were carried out 
with many three-spin coupling constants $J_{3\text{s},i}$ in a range $|J_{3\text{s},i}|<0.2$ K,
where one $J_{3\text{s},i} \neq 0$ is finite and the other two $J_{3\text{s},j \neq i}=0$ are zero. 
Resulting specific heat data are plotted in Fig.~\ref{CT_MC111213141516} 
as color maps of $C(T,J_{3\text{s},i})$. 

Figures~\ref{CT_MC111213141516}(a--f) show that in each simulation 
there is a single broad peak in the temperature dependence of 
$C(T,J_{3\text{s},i})$ at $T \sim$ 0.3 K, which is the characteristic of the SI model, 
and that no phase transition appears. 
These imply that the mechanism expressed by Eq.~(\ref{H_3s_LRO}) 
is much less clear, if the quadrupole moments remain SRO. 
It seems that long-lived fluctuations of quadrupole moments $ \sigma_{\bm{r}}^+ $ 
do not well function in the mechanism compared to 
the average $\langle \sigma_{\bm{r}}^+ \rangle$ 
within the CMC simulations. 

\subsubsection{\label{results_CMC_SQ_3DPAF} $S(\bm{Q})$ in 3D-PAF phases ($\delta = 0$)}
\begin{figure*}[ht]
\centering
\includegraphics[width=18cm,clip]{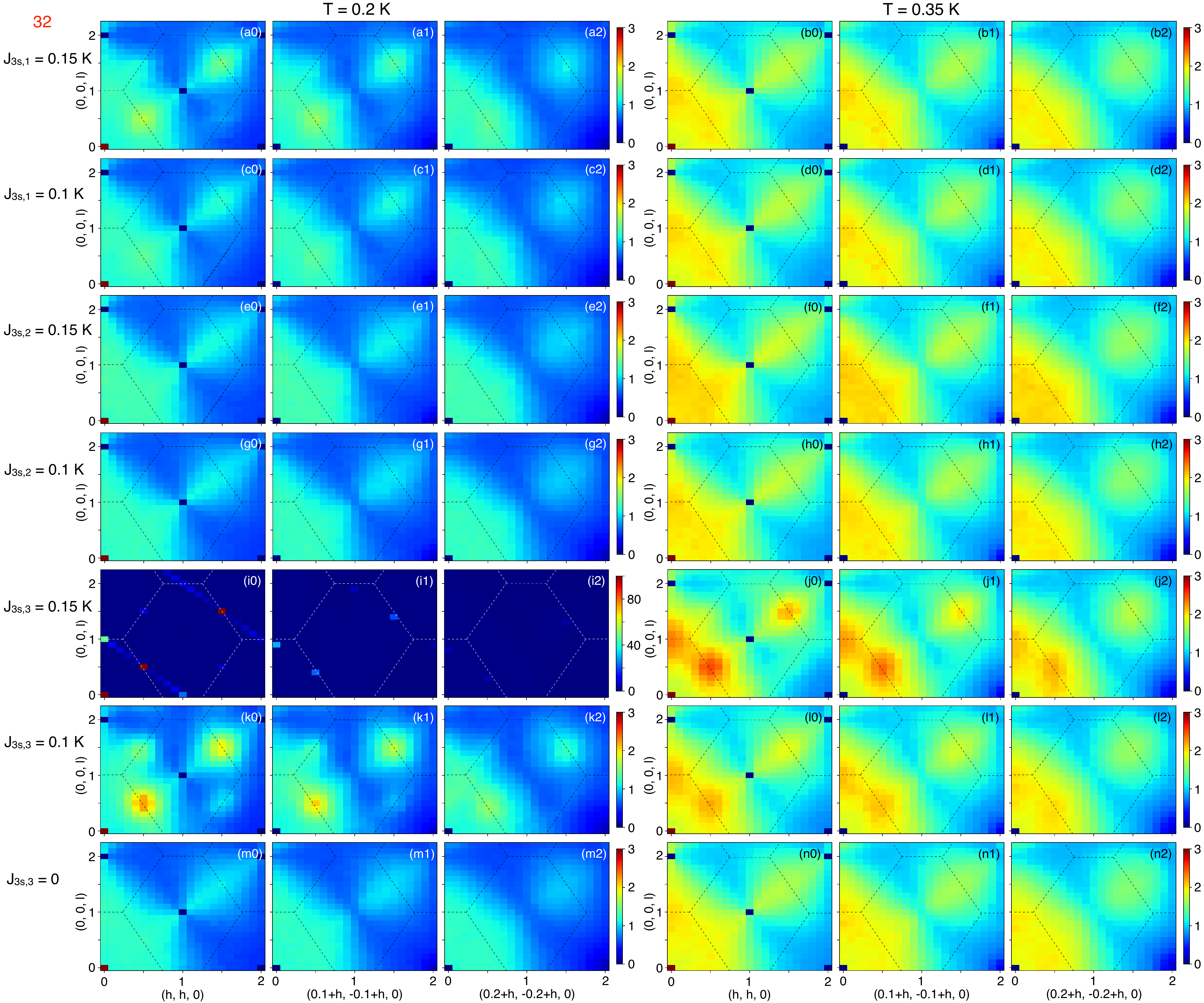}
\caption{ 
Intensity maps of $S(\bm{Q})$ calculated by the 16000-site CMC simulations 
using parameters indicated by red circles shown in Figs.~\ref{CT_MC123456}(b1--b3) ($J_{3\text{s},i}$; $i=1,2,3$; $J_{3\text{s},j \neq i}=0$) and 
by the point 32 in Fig.~\ref{phase_diagram_Dnn0p478} [$\frac{J_{\text{nn}}}{J_{\text{nn}} + D_{\text{nn}}}(\delta,q)=(0.0,0.55)$]. 
They are viewed by 2D slices of $\bm{Q}=(k+h,-k+h,l)$ with fixed $k=0,0.1$, and $0.2$, 
which are shown in (w0), (w1), and (w2) (w=a--n), respectively. 
They are calculated at two temperatures 0.2 K (a,c,e,g,i,k,m) and 0.35 K (b,d,f,h,j,l,n), 
below and above the phase transition temperature of the 3D-PAF ($q>0$) LRO.
Intensity maps for $J_{3\text{s},1} = 0.15$ and $0.1$ K [Fig.~\ref{CT_MC123456}(b1)] are shown in (a,b) and (c,d), respectively. 
Intensity maps for $J_{3\text{s},2} = 0.15$ and $0.1$ K [Fig.~\ref{CT_MC123456}(b2)] are shown in (e,f) and (g,h), respectively. 
Intensity maps for $J_{3\text{s},3} = 0.15$, $0.1$, and 0 K [Fig.~\ref{CT_MC123456}(b3)] are shown in (i,j), (k,l), and (m,n), respectively. 
}
\label{SQcalMCp2}
\end{figure*}
\begin{figure*}[ht]
\centering
\includegraphics[width=18cm,clip]{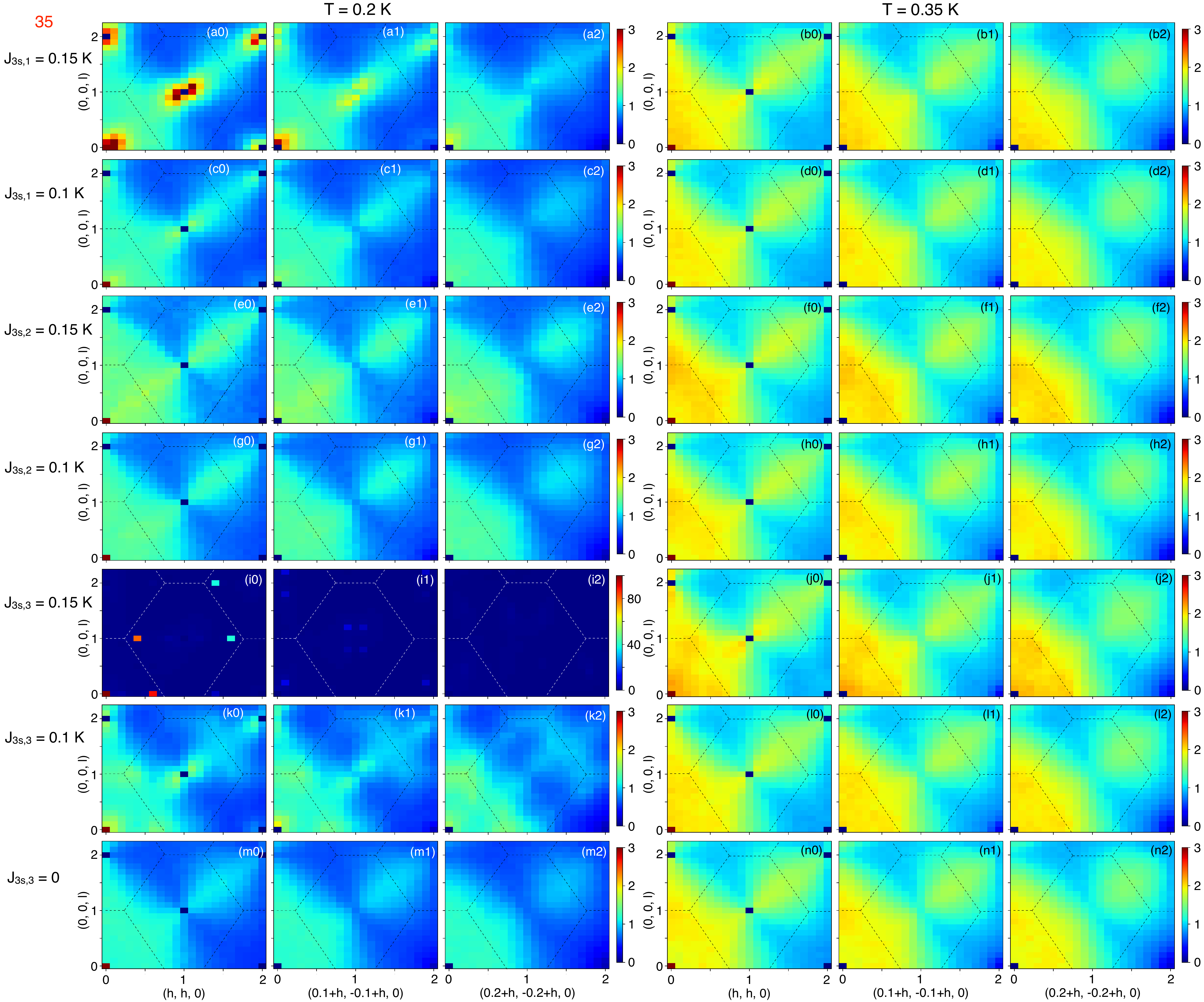}
\caption{ 
Intensity maps of $S(\bm{Q})$ calculated by the 16000-site CMC simulations 
using parameters indicated by red circles shown in Figs.~\ref{CT_MC123456}(e1--e3) ($J_{3\text{s},i}$; $i=1,2,3$; $J_{3\text{s},j \neq i}=0$) and 
by the point 35 in Fig.~\ref{phase_diagram_Dnn0p478} [$\frac{J_{\text{nn}}}{J_{\text{nn}} + D_{\text{nn}}}(\delta,q)=(0.0,-0.55)$]. 
They are viewed by 2D slices of $\bm{Q}=(k+h,-k+h,l)$ with fixed $k=0,0.1$, and $0.2$, 
which are shown in (w0), (w1), and (w2) (w=a--n), respectively. 
They are calculated at two temperatures 0.2 K (a,c,e,g,i,k,m) and 0.35 K (b,d,f,h,j,l,n), 
below and above the phase transition temperature of the 3D-PAF ($q<0$) LRO. 
Intensity maps for $J_{3\text{s},1} = 0.15$ and $0.1$ K [Fig.~\ref{CT_MC123456}(e1)] are shown in (a,b) and (c,d), respectively. 
Intensity maps for $J_{3\text{s},2} = 0.15$ and $0.1$ K [Fig.~\ref{CT_MC123456}(e2)] are shown in (e,f) and (g,h), respectively. 
Intensity maps for $J_{3\text{s},3} = 0.15$, $0.1$, and 0 K [Fig.~\ref{CT_MC123456}(e3)] are shown in (i,j), (k,l), and (m,n), respectively. 
}
\label{SQcalMCp5}
\end{figure*}
Classical MC simulations were carried out with a larger system size $L=10$ (16000 sites) 
to study effects of each three-spin interaction on the structure factor $S(\bm{Q})$. 
We calculated $S(\bm{Q})$ in the 3D-PAF phases. 
Considering the results of Sec.~\ref{results_CMC_C_3DPAF}, 
the parameters $(\delta,q)$ were fixed to the two sets: 
$\frac{J_{\text{nn}}}{J_{\text{nn}} + D_{\text{nn}}} (\delta,q)=(0.0,0.55)$ and $(0.0,-0.55)$, 
corresponding to the points 32 and 35 in Fig.~\ref{phase_diagram_Dnn0p478}. 
The three-spin interaction constant was fixed to two typical values: 
$J_{3\text{s},i}=0.1$ and $0.15$ K ($J_{3\text{s},j \neq i}=0$). 
Figure~\ref{SQcalMCp2} shows the resulting intensity maps of $S(\bm{Q})$ 
which are calculated with the parameters corresponding to the red circles in Figs.~\ref{CT_MC123456}(b1--b3), 
and at two temperatures $0.2$ and $0.35$ K,
below and above the phase transition temperature of the 3D-PAF ($q>0$) LRO. 
Figure~\ref{SQcalMCp5} shows the resulting intensity maps of $S(\bm{Q})$ 
which are calculated with the parameters corresponding to the red circles in Figs.~\ref{CT_MC123456}(e1--e3), 
and at $0.2$ and $0.35$ K, 
below and above the phase transition temperature of the 3D-PAF ($q<0$) LRO. 

When the three-spin interactions are set to zero, 
the calculated $S(\bm{Q})$ with $\frac{J_{\text{nn}}}{J_{\text{nn}} + D_{\text{nn}}} q=0.55$ and $-0.55$, 
which are Figs.~\ref{SQcalMCp2}(m,n) and Figs.~\ref{SQcalMCp5}(m,n), respectively, 
show almost the same characteristics: 
the pinch-point structure of the SI is seen around the $\Gamma$ points $(0,0,2)$ and $(1,1,1)$, 
the intensity pattern becomes weakened owing to the quadrupole order as temperature is lowered below $T_{\text{c}}$, 
the intensity pattern is scarcely affected by the quadrupole structures ($q>0$ or $q<0$). 

When the three-spin interactions are switched on, 
$S(\bm{Q})$ at 0.35 K [Figs.~\ref{SQcalMCp2}(b,d,f,h,j,l) and \ref{SQcalMCp5}(b,d,f,h,j,l)] show little dependence on $J_{3\text{s},i}$, 
while $S(\bm{Q})$ at 0.2 K [Figs.~\ref{SQcalMCp2}(a,c,e,g,i,k) and \ref{SQcalMCp5}(a,c,e,g,i,k)] show large changes 
depending on the value of $J_{3\text{s},i}$. 
In particular, for $J_{3\text{s},3}=0.15$ K 
magnetic Bragg peaks appear at 0.2 K [Figs.~\ref{SQcalMCp2}(i0) and \ref{SQcalMCp5}(i0)]. 
This is consistent with the interpretation that 
the second phase transitions [Figs.~\ref{CT_MC123456}(b3,e3)] are ascribed to magnetic ordering in addition to the 3D-PAF LRO. 
We will not delve into these magnetic phase transitions, order parameters, etc. in this study. 
The magnetic LRO of the pyrochlore magnet Tb$_2$Sn$_2$O$_7$ \cite{Mirebeau05} 
may possibly be accounted for by the three-spin interaction term. 

The most interesting results of the calculated $S(\bm{Q})$ are 
those for $J_{3\text{s},3}=0.1$ K and $q>0$ [Figs.~\ref{SQcalMCp2}(k,l)]. 
It is obvious that the calculated $S(\bm{Q})$ at 0.2 K [Fig.~\ref{SQcalMCp2}(k0)] 
bears a resemblance to the observed $S(\bm{Q})$ of TTO at 0.1 K [Figs.~\ref{SQobs}(a0,c0,e0)], 
in a sense that they commonly show peaks at $\bm{Q} = (\tfrac{1}{2},\tfrac{1}{2},\tfrac{1}{2})$ 
and $(\tfrac{1}{2},\tfrac{1}{2},\tfrac{3}{2})$. 
In addition, the temperature dependence of the calculated $S(\bm{Q})$ shows that 
the intensity pattern changes 
from the peaked structure around $\bm{Q} = (\tfrac{1}{2},\tfrac{1}{2},\tfrac{1}{2})$ at 0.2 K 
to a pinch-point like pattern of the classical SI at 0.35 K [Figs.~\ref{SQcalMCp2}(k,l)]. 
This roughly agrees with the temperature variation of the observed $S(\bm{Q})$ of TTO [Figs.~\ref{SQobs}(a--d)]. 
In contrast, for $J_{3\text{s},3}=0.1$ K and $q<0$ 
the calculated $S(\bm{Q})$ at 0.2 K [Fig.~\ref{SQcalMCp5}(k0)] shows 
a very different intensity pattern from $S(\bm{Q})$ shown in Fig.~\ref{SQcalMCp2}(k0). 
This can be understood by the difference in the quadrupole orders [Figs.~\ref{phase_diagram_Dnn0p0}(b,c)]
and in the effective bilinear interactions [Eq.~(\ref{H_3s_LRO})]. 
Thus we can conclude that the CMC simulation results 
suggest that a parameter set that should be further investigated, especially 
using techniques for many-body quantum states, 
is $\frac{J_{\text{nn}}}{J_{\text{nn}} + D_{\text{nn}}} (\delta,q) \sim (0.0,0.55)$, 
$J_{3\text{s},1}=J_{3\text{s},2}=0$, and 
$\frac{J_{3\text{s},3}}{J_{\text{nn}} + D_{\text{nn}}} \sim 0.1$ (or $-0.1$). 
We also conclude 
that the $q<0$ sides of the phase diagrams (Figs.~\ref{phase_diagram_Dnn0p0} 
and \ref{phase_diagram_Dnn0p478}) can be excluded from studies of TTO. 

It should be noted that 
the interesting results [Figs.~\ref{SQcalMCp2}(k,l)] are obtained 
for the parameters of the Hamiltonian, 
which are close to the classical phase boundaries [Fig.~\ref{CT_MC123456}(b3)]. 
Theoretically effects of proximity to phase boundaries separating two or three LROs 
in the pyrochlore magnets were studied in Refs.~\cite{Benton2016,Yan2017} 
based on the generic bilinear NN Hamiltonian. 
It was shown that a disordered ground state can be induced 
by non-trivial degeneracy of the order parameters around the phase boundary. 
This mechanism may be related to that of the disordered ground state of TTO. 

\subsubsection{\label{results_CMC_SQ_SI} other results of $S(\bm{Q})$}
To complement $S(\bm{Q})$ shown in Section~\ref{results_CMC_SQ_3DPAF} (Figs.~\ref{SQcalMCp2} and \ref{SQcalMCp5}) 
we performed several 16000-site CMC simulations with parameters: 
$\frac{J_{\text{nn}}}{J_{\text{nn}} + D_{\text{nn}}} (\delta,q)=(-0.1,0.6)$, $(0.1,0.5)$, $(0.0,0.45)$, and $(0.0,-0.45)$
corresponding to the points 31, 33, 38, and 41 in Fig.~\ref{phase_diagram_Dnn0p478}, respectively. 
These results are presented in Appendixes~\ref{appendix_CMC_3DPAF_delta} and \ref{appendix_CMC_SI}. 

\subsection{\label{results_TPQ} quantum simulations using TPQ states}
Quantum simulations using the TPQ states were carried out to confirm 
the classical MC results shown in Figs.~\ref{SQcalMCp2}(k0), \ref{SQcalMCp1}(c0), and \ref{SQcalMCp3}(c0). 
There were two practical problems. 
First, since the computation was very time-consuming, 
the number of the simulations was limited to far less than that of the CMC simulations. 
Second, since it was memory-intensive task, 
the system size was limited to only 32 sites, 
which precluded us from studying systematic size dependence. 
Thus we had to carefully interpret results of the simulations 
by paying particular attention to changes of results with varying interaction parameters. 
For example, by comparing results with $J_{3\text{s},3}=0$ and $J_{3\text{s},3} \ne 0$ 
it was not difficult to discern effects of $J_{3\text{s},3}$ from those due to the small size. 
Based on this idea many 32-site simulations using the TPQ states 
were carried out using the Hamiltonian 
$\mathcal{H}_0 + \mathcal{H}_{3\text{s}}$ 
with $J_{3\text{s},3}/J_{\text{nn}} = 0.1$ and $0$, 
where $J_{3\text{s},1}=J_{3\text{s},2}=0$ were fixed to zero. 
The parameters $(\delta,q)$ were systematically changed 
mainly on the $q$-axis ($\delta=0$) in Fig.~\ref{phase_diagram_Dnn0p0}, 
where $(\delta,q)$ values we selected are indicated by the red circles, the points 3--29. 

\subsubsection{\label{results_TPQ_C} specific heat and entropy on $q$-axis}
\begin{figure}[hbt]
\centering
\includegraphics[width=8cm,clip]{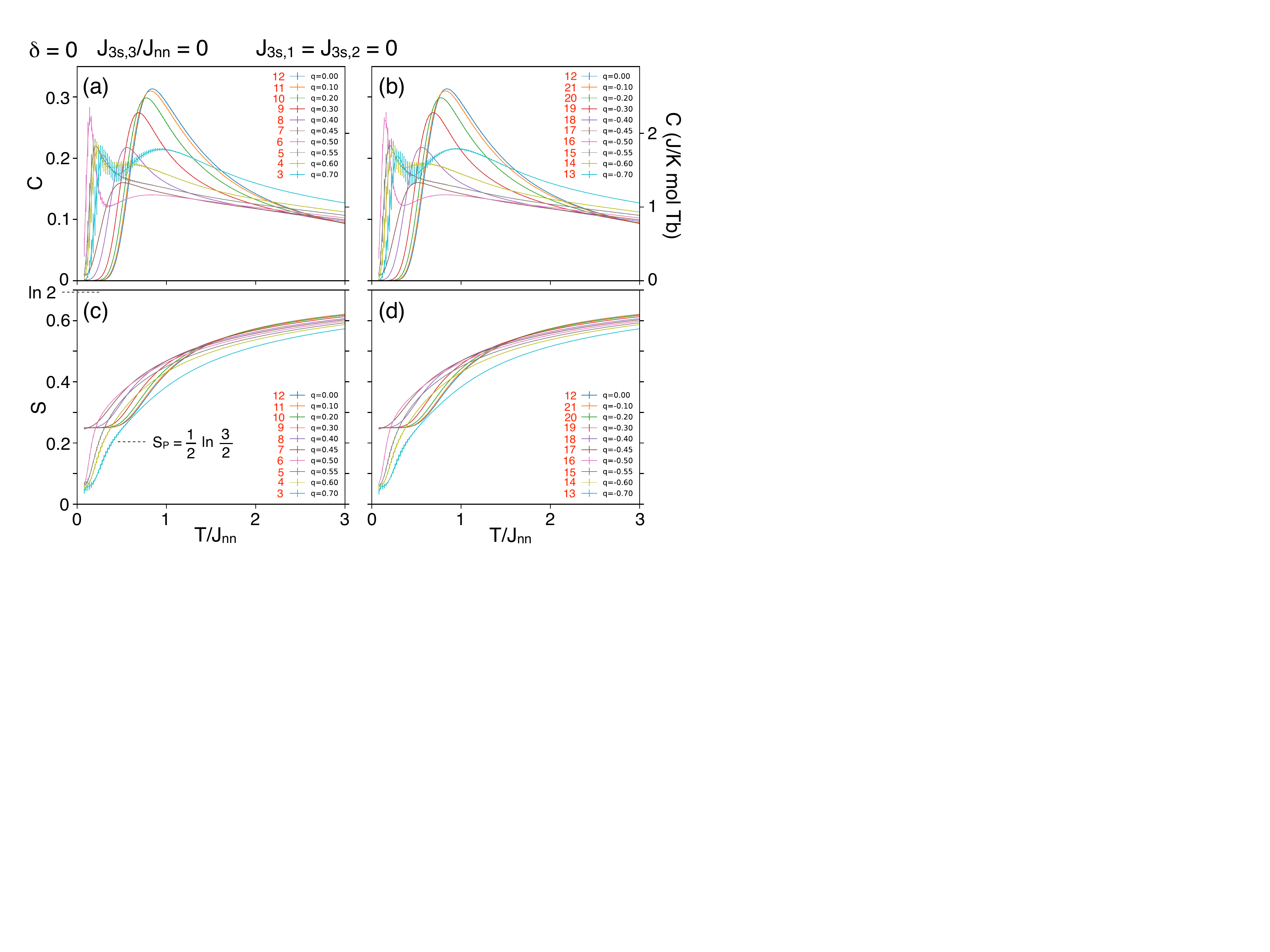}
\caption{
Temperature dependence of specific heat $C(T)$ and entropy $S(T)$ 
obtained by the 32-site simulations using the cTPQ state 
for $J_{3\text{s},3}/J_{\text{nn}}=0$ ($J_{3\text{s},1}=J_{3\text{s},2}=0$) 
with parameters $(\delta=0,q)$, the points 3--21 in Fig.~\ref{phase_diagram_Dnn0p0}. 
In (a) and (b) $C(T)$ for $q \geq 0$ and $q \leq 0$ are shown, respectively.
In (c) and (d) $S(T)$ for $q \geq 0$ and $q \leq 0$ are shown, respectively.
}
\label{CTST_TPQ_J3b30p0}
\end{figure}
\begin{figure}[hbt]
\centering
\includegraphics[width=8cm,clip]{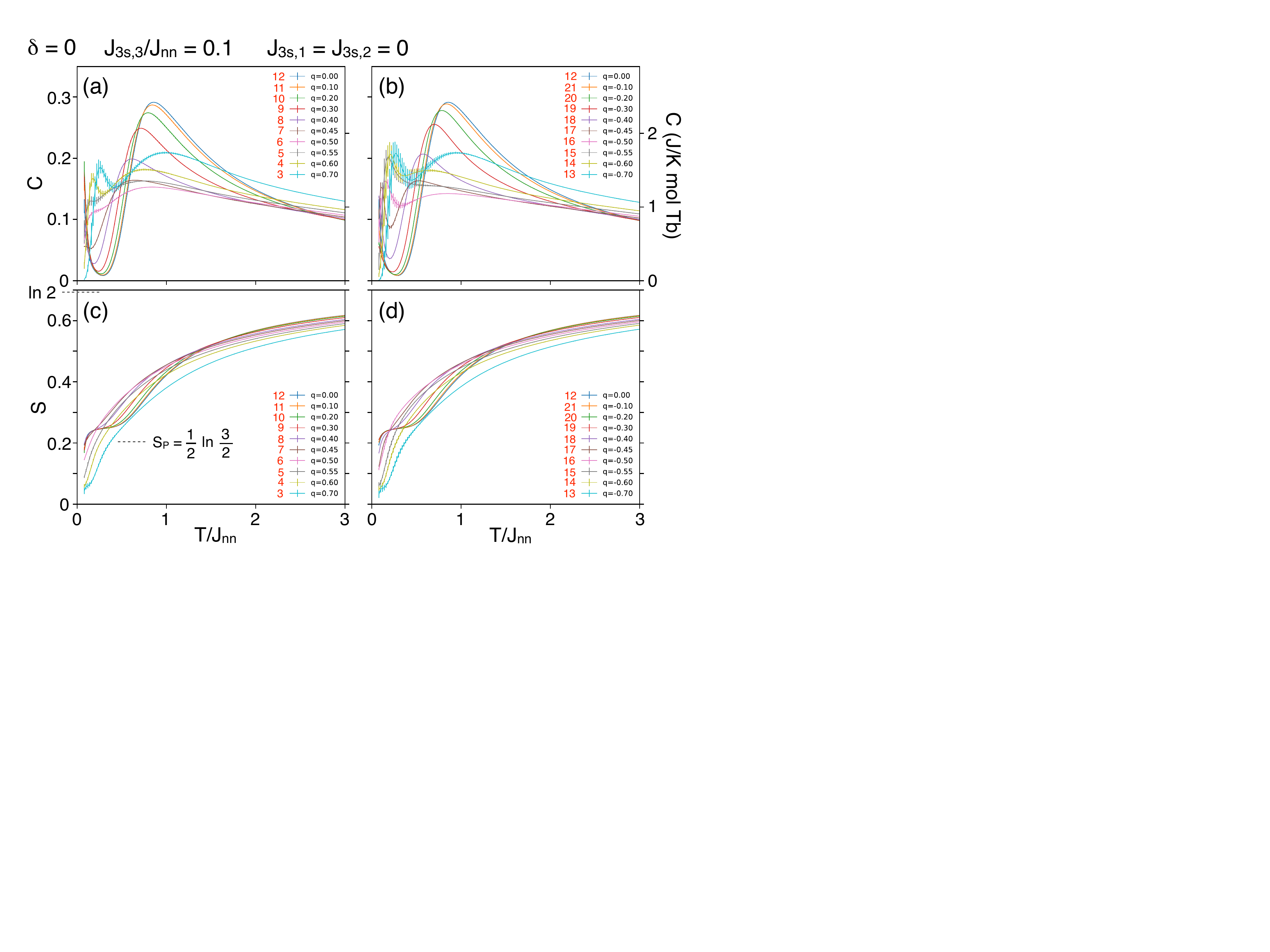}
\caption{
Temperature dependence of specific heat $C(T)$ and entropy $S(T)$ 
obtained by the 32-site simulations using the cTPQ state 
for $J_{3\text{s},3}/J_{\text{nn}}=0.1$  ($J_{3\text{s},1}=J_{3\text{s},2}=0$) 
with parameters $(\delta=0,q)$, the points 3--21 in Fig.~\ref{phase_diagram_Dnn0p0}. 
In (a) and (b) $C(T)$ for $q \geq 0$ and $q \leq 0$ are shown, respectively.
In (c) and (d) $S(T)$ for $q \geq 0$ and $q \leq 0$ are shown, respectively.
}
\label{CTST_TPQ_J3b30p1}
\end{figure}
Several 32-site simulations using the cTPQ state with the parameters $(\delta=0,q)$ on the $q$-axis 
were carried out for $J_{3\text{s},3}/J_{\text{nn}}=0$ and $0.1$ ($J_{3\text{s},1}=J_{3\text{s},2}=0$). 
Resulting temperature dependence of specific heat $C(T)$ and entropy $S(T)$ 
are plotted in Figs.~\ref{CTST_TPQ_J3b30p0} and \ref{CTST_TPQ_J3b30p1}. 
For $J_{3\text{s},3}=0$, $C(T)$ and $S(T)$ curves 
with positive $q$ values [Figs.~\ref{CTST_TPQ_J3b30p0}(a,c)] 
are almost the same as corresponding curves with negative $q$ [Figs.~\ref{CTST_TPQ_J3b30p0}(b,d)]. 
This fact [$C(T,q)=C(T,-q)$, $S(T,q)=S(T,-q)$] reflects the invariance of the Hamiltonian $\mathcal{H}_0$ under the transformation 
of rotating $\bm{\sigma}_{\bm{r}}$ about the local $\bm{z}_{\bm{r}}$ axis by $\pi/2$ 
and $q \rightarrow -q$. 
For $J_{3\text{s},3} \ne 0$ this invariance does not hold, 
resulting in $C(T,q) \ne C(T,-q)$ [Figs.~\ref{CTST_TPQ_J3b30p1}(a,b)] 
and $S(T,q) \ne S(T,-q)$ [Figs.~\ref{CTST_TPQ_J3b30p1}(c,d)]. 

For $J_{3\text{s},3}/J_{\text{nn}}=0$, 
each curves of $C(T)$ and $S(T)$ (Fig.~\ref{CTST_TPQ_J3b30p0}) 
with $q$ in a range $0 \leq |q| \leq 0.45$ 
have a single broad peak [Figs.~\ref{CTST_TPQ_J3b30p0}(a,b)] 
and an entropy plateau ($S \simeq 0.25$) [Figs.~\ref{CTST_TPQ_J3b30p0}(c,d)], respectively. 
These behaviors are the characteristics of the classical SI [Fig.~\ref{QMC_TPQ}(b)], 
which are expected also for QSI at intermediate temperatures \cite{Kato2015}. 
The discrepancy of the value of the entropy plateau, $S \simeq 0.25$, 
from the the Pauling entropy, $S = \tfrac{1}{2} \ln \tfrac{3}{2}$, may be 
caused by a small size effect. 
Each curves of $C(T)$ and $S(T)$ (Fig.~\ref{CTST_TPQ_J3b30p0}) with $q$ in a range $ |q| \geq 0.5$ 
have a low-$T$ peak and the zero-$T$ limit $S(T \rightarrow 0 ) \simeq 0$, respectively. 
The low-$T$ peak of $C(T)$, which is similar to that of Fig.~\ref{QMC_TPQ}(c), 
implies that a phase transition to a quadrupole ordered state occurs. 
These results shown in Fig.~\ref{CTST_TPQ_J3b30p0} for $J_{3\text{s},3}/J_{\text{nn}}=0$ 
suggest that the quantum phase boundaries on the $q$-axis 
are not very different from the classical phase boundaries, 
$|q_{\text{c}}|=\tfrac{1}{2}$ ($\delta =0$, Fig.~\ref{phase_diagram_Dnn0p0}). 

For $J_{3\text{s},3}/J_{\text{nn}}=0.1$,
each curves of $C(T)$ and $S(T)$ (Fig.~\ref{CTST_TPQ_J3b30p1}) 
with $q$ in a range $-0.4 \leq q \leq 0.45$ 
have a single broad peak [Figs.~\ref{CTST_TPQ_J3b30p1}(a,b)] 
and the entropy plateau ($S \simeq 0.25$) [Figs.~\ref{CTST_TPQ_J3b30p1}(c,d)], respectively. 
These can be understood by the classical SI behaviors 
expected for QSI at intermediate temperatures. 
The upturn of $C(T)$ [Figs.~\ref{CTST_TPQ_J3b30p1}(a,b)] 
and the downturn of $S(T)$ [Figs.~\ref{CTST_TPQ_J3b30p1}(c,d)] 
in a low-$T$ range of $T/J_{\text{nn}} < 0.2$ 
suggest certain QSL behavior \cite{Kato2015} or/and an artifact caused by the small size [Fig.~\ref{QMC_TPQ}(b)]. 
It seems difficult to correctly draw information from low-$T$ data in $T/J_{\text{nn}} < 0.2$. 
This sort of difficulty due to the small system size 
has been observed in studies of 
the frustrated Heisenberg antiferromagnet on the kagome lattice at low temperatures \cite{Sugiura2013,Shimokawa2016,Schnack2018}. 

For $J_{3\text{s},3}/J_{\text{nn}}=0.1$, 
each curves of $C(T)$ and $S(T)$ (Fig.~\ref{CTST_TPQ_J3b30p1}) 
with $q$ in ranges $q \geq 0.6$ and $q \leq -0.55$ 
have a low-$T$ peak and the zero-$T$ limit $S(T \rightarrow 0 ) \simeq 0$, respectively. 
The low-$T$ peak of $C(T)$ [Figs.~\ref{CTST_TPQ_J3b30p1}(a,b)], which is similar to that of Fig.~\ref{QMC_TPQ}(c), 
implies that a phase transition to a quadrupole ordered state occurs. 
The low-$T$ peaks of $C(T)$ with $q = -0.45, -0.5$ [Fig.~\ref{CTST_TPQ_J3b30p1}(b)] 
could also be understood by $T_{\text{c}}$ of the quadrupole LRO, although these are less clear. 
On the other hand, 
the low-$T$ behavior of $C(T)$ with $q = 0.5,0.55$ [Fig.~\ref{CTST_TPQ_J3b30p1}(a)] 
suggests that something different happens at low temperatures. 
These $C(T)$ curves show roughly the behavior of $C(T) \sim \text{const}$ in $T/J_{\text{nn}} < 1$, 
which is reminiscent of $C(T)$ data of the TTO experiments \cite{Taniguchi13}. 
Intriguingly, the Hamiltonian with one of these parameters, $q = 0.55$, 
is one of the candidates for TTO suggested by the CMC simulations [Fig.~\ref{SQcalMCp2}(k0)]. 

\subsubsection{\label{results_TPQ_SQ_q_axisP} $S(\bm{Q})$ on $q$-axis ($q \geq 0$)}
\begin{figure*}[ht]
\centering
\includegraphics[width=18cm,clip]{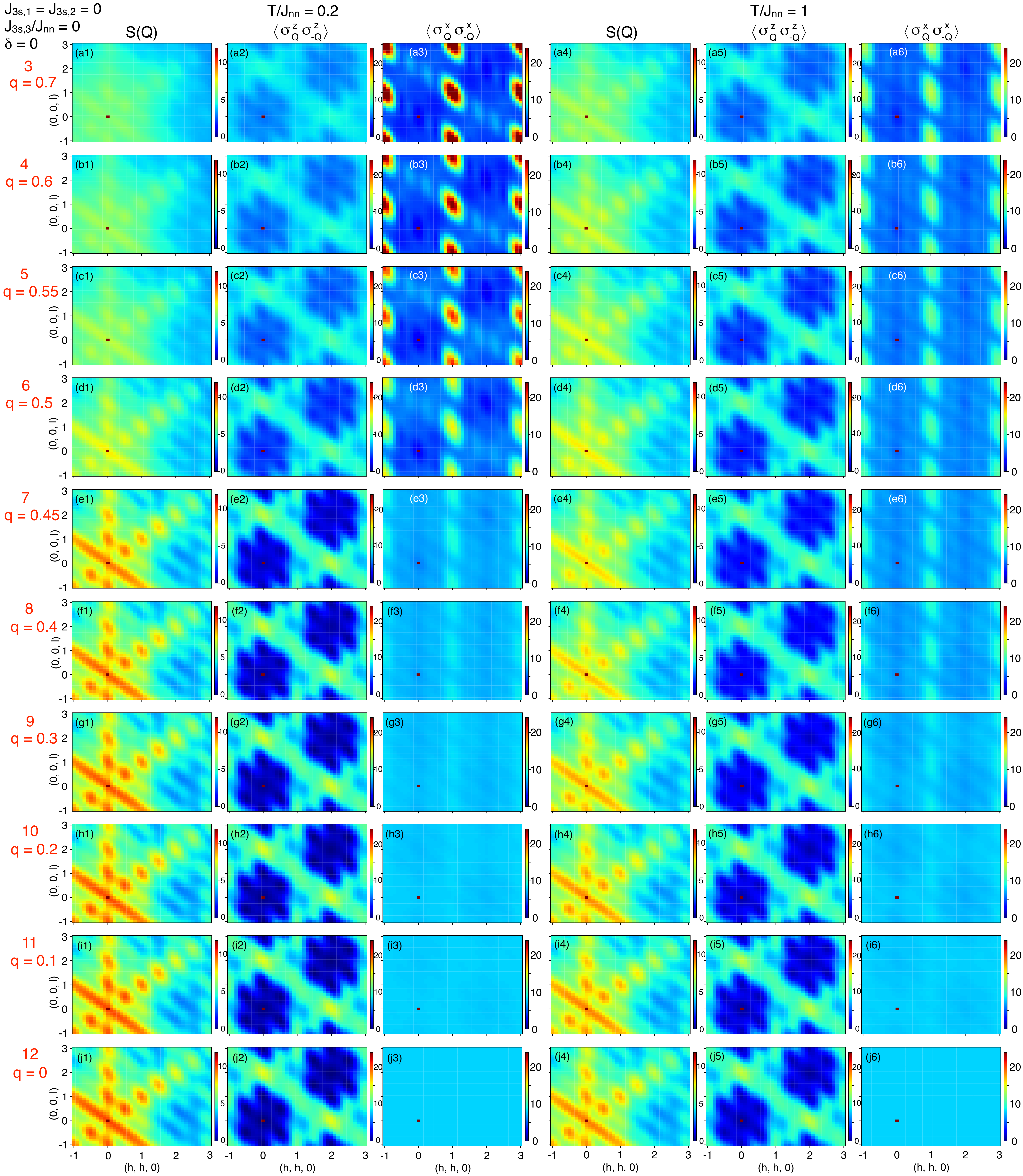}
\caption{ 
Two dimensional slices of 
(w1,w4) $S(\bm{Q})$, 
(w2,w5) $\langle \sigma_{\bm{Q}}^{z} \sigma_{-\bm{Q}}^{z} \rangle$, 
and (w3,w6) $\langle \sigma_{\bm{Q}}^{x} \sigma_{-\bm{Q}}^{x} \rangle$ (w=a--j) 
in the plane $\bm{Q}=(h,h,l)$ 
calculated by the 32-site simulations using the mTPQ state 
for $J_{3\text{s},3}/J_{\text{nn}}=0$ ($J_{3\text{s},1}=J_{3\text{s},2}=0$) 
with parameters $(\delta=0, q \geq 0)$, the points 3--12 in Fig.~\ref{phase_diagram_Dnn0p0}. 
The 2D slice data at $T/J_{\text{nn}}=0.2$ and $1$ are shown in 
(w1--w3) and (w4--w6) (w=a--j), respectively. 
}
\label{SQTPQL2J3b30p0J1qP}
\end{figure*}
\begin{figure*}[ht]
\centering
\includegraphics[width=18cm,clip]{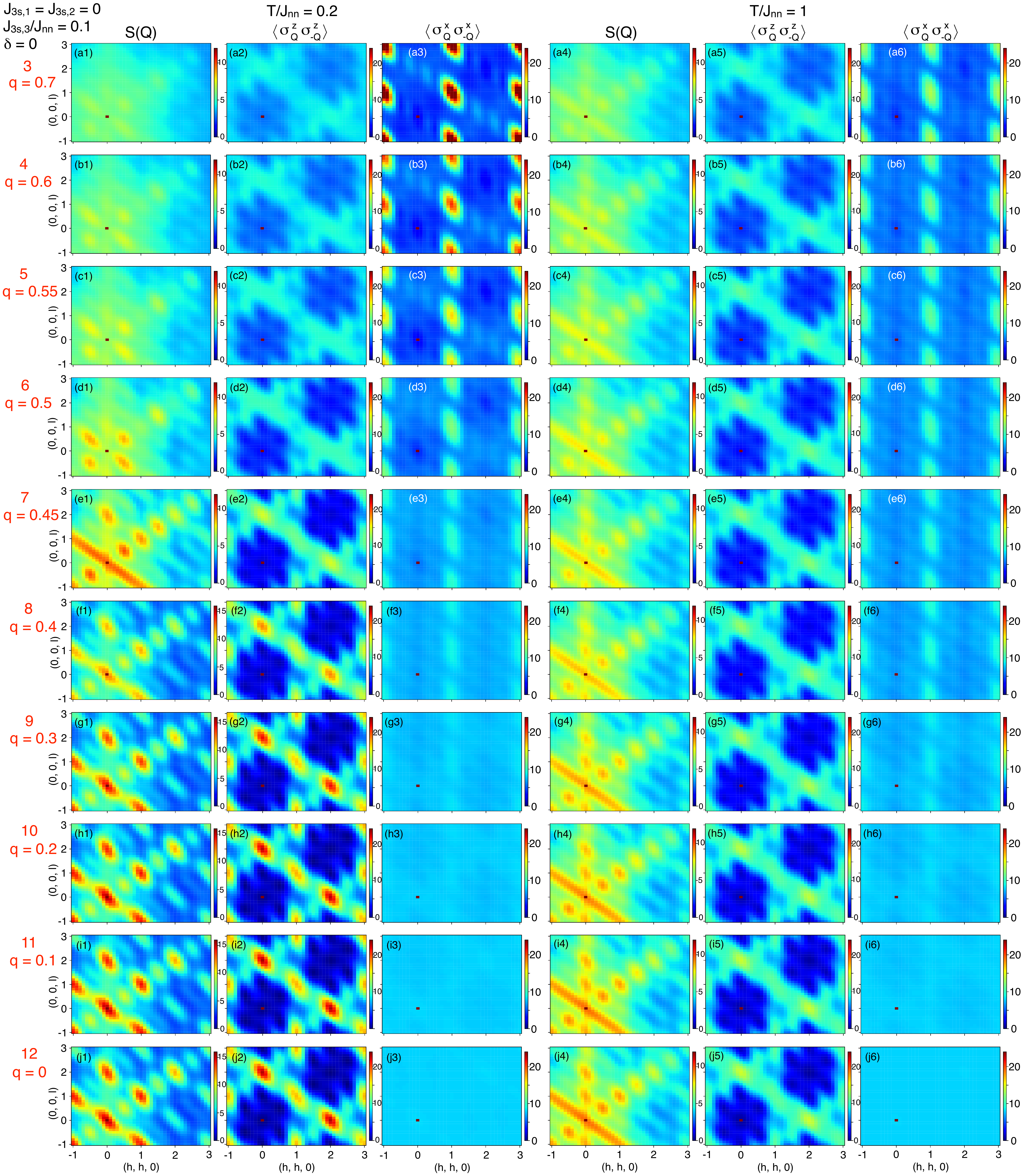}
\caption{ 
Two dimensional slices of 
(w1,w4) $S(\bm{Q})$, 
(w2,w5) $\langle \sigma_{\bm{Q}}^{z} \sigma_{-\bm{Q}}^{z} \rangle$, 
and (w3,w6) $\langle \sigma_{\bm{Q}}^{x} \sigma_{-\bm{Q}}^{x} \rangle$ (w=a--j) 
in the plane $\bm{Q}=(h,h,l)$ 
calculated by the 32-site simulations using the mTPQ state 
for $J_{3\text{s},3}/J_{\text{nn}}=0.1$ ($J_{3\text{s},1}=J_{3\text{s},2}=0$) 
with parameters $(\delta=0, q \geq 0)$, 
the points 3--12 in Fig.~\ref{phase_diagram_Dnn0p0}. 
The 2D slice data at $T/J_{\text{nn}}=0.2$ and $1$ are shown in 
(w1--w3) and (w4--w6) (w=a--j), respectively. 
}
\label{SQTPQL2J3b30p1J1qP}
\end{figure*}
\begin{figure*}[ht]
\centering
\includegraphics[width=18cm,clip]{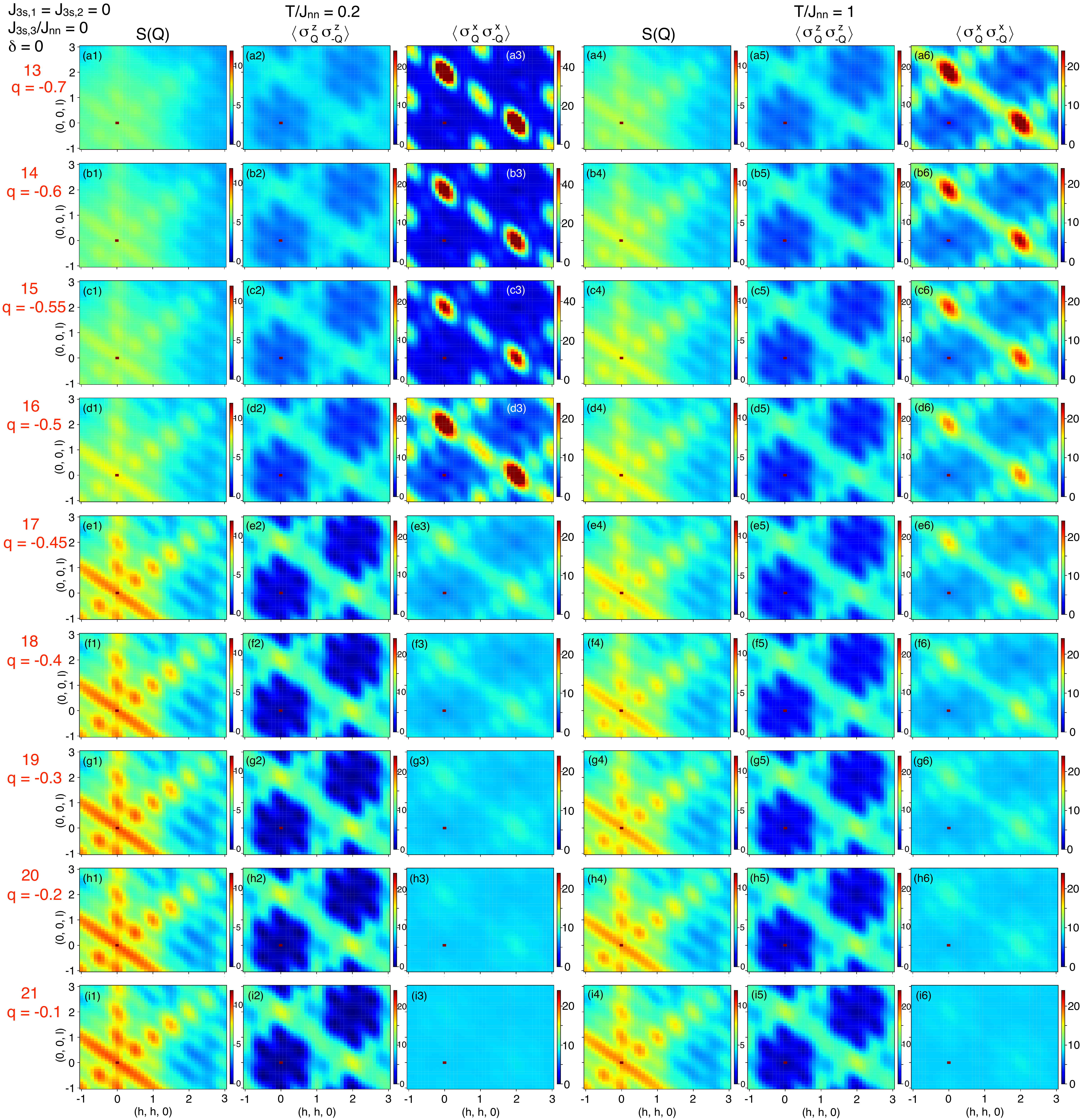}
\caption{ 
Two dimensional slices of 
(w1,w4) $S(\bm{Q})$, 
(w2,w5) $\langle \sigma_{\bm{Q}}^{z} \sigma_{-\bm{Q}}^{z} \rangle$, 
and (w3,w6) $\langle \sigma_{\bm{Q}}^{x} \sigma_{-\bm{Q}}^{x} \rangle$ (w=a--i) 
in the plane $\bm{Q}=(h,h,l)$ 
calculated by the 32-site simulations using the mTPQ state 
for $J_{3\text{s},3}/J_{\text{nn}}=0$ ($J_{3\text{s},1}=J_{3\text{s},2}=0$) 
with parameters $(\delta=0, q < 0)$, the points 13--21 in Fig.~\ref{phase_diagram_Dnn0p0}. 
The 2D slice data at $T/J_{\text{nn}}=0.2$ and $1$ are shown in 
(w1--w3) and (w4--w6) (w=a--i), respectively. 
}
\label{SQTPQL2J3b30p0J1qM}
\end{figure*}
\begin{figure*}[ht]
\centering
\includegraphics[width=18cm,clip]{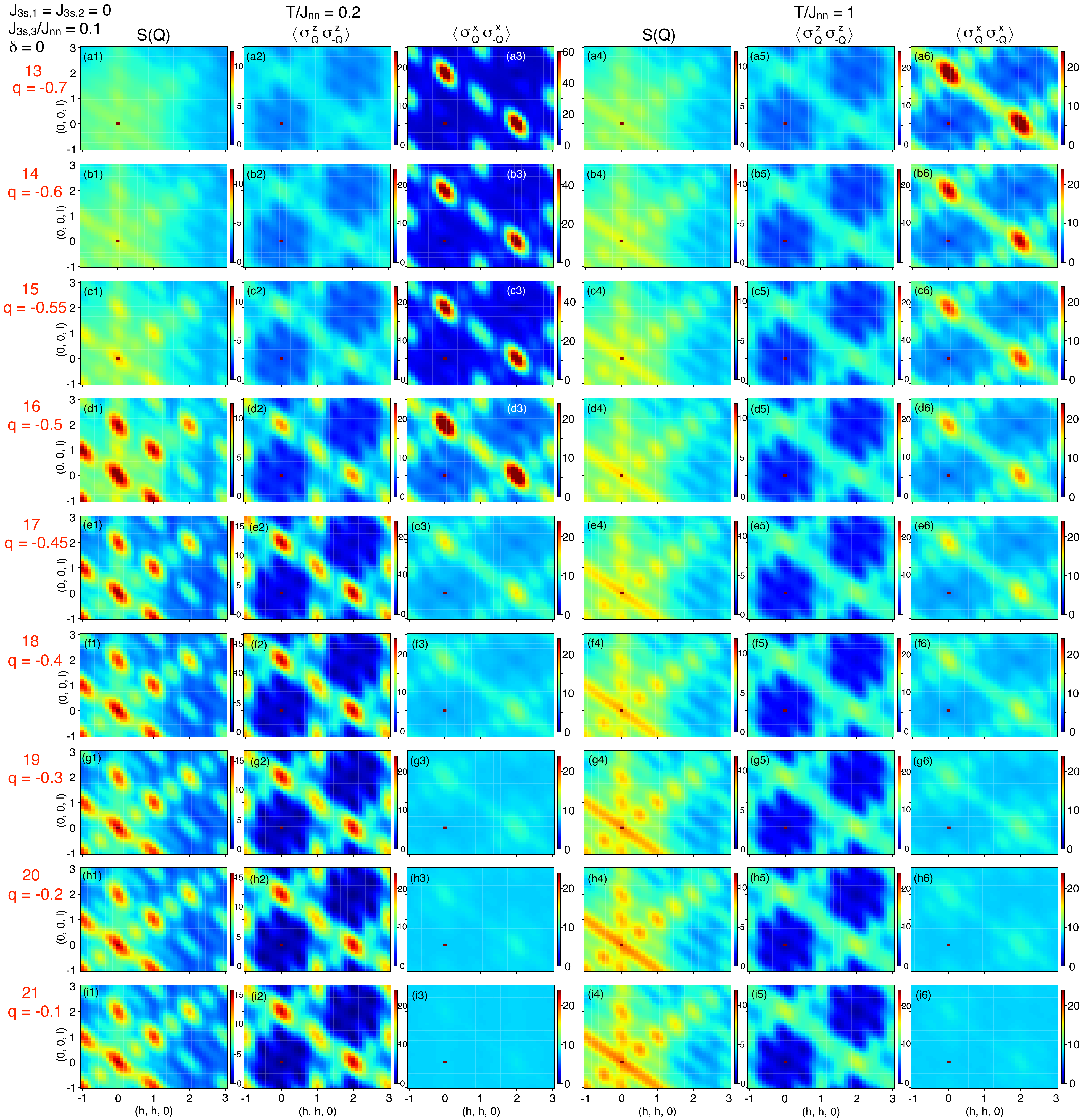}
\caption{ 
Two dimensional slices of 
(w1,w4) $S(\bm{Q})$, 
(w2,w5) $\langle \sigma_{\bm{Q}}^{z} \sigma_{-\bm{Q}}^{z} \rangle$, 
and (w3,w6) $\langle \sigma_{\bm{Q}}^{x} \sigma_{-\bm{Q}}^{x} \rangle$ (w=a--i) 
in the plane $\bm{Q}=(h,h,l)$ 
calculated by the 32-site simulations using the mTPQ state 
for $J_{3\text{s},3}/J_{\text{nn}}=0.1$ ($J_{3\text{s},1}=J_{3\text{s},2}=0$) 
with parameters $(\delta=0, q < 0)$, the points 13--21 in Fig.~\ref{phase_diagram_Dnn0p0}. 
The 2D slice data at $T/J_{\text{nn}}=0.2$ and $1$ are shown in 
(w1--w3) and (w4--w6) (w=a--i), respectively. 
}
\label{SQTPQL2J3b30p1J1qM}
\end{figure*}

We calculated $S(\bm{Q})$ and the expectation values of pseudospin correlations 
using the mTPQ method [Eqs.~(\ref{sigma_xz_Q}) and (\ref{SQ})] 
to take a closer look at temperature and $q$ dependence of the states. 
By choosing two typical temperatures $T/J_{\text{nn}}=1$ and $0.2$, 
which are above and below the specific heat peak (or bump), 
2D slice view data of 
$S(\bm{Q})$ and $\langle \sigma_{\bm{Q}}^{\alpha} \sigma_{-\bm{Q}}^{\alpha} \rangle$ ($\alpha=z,x$) 
in the plane $\bm{Q}=(h,h,l)$ 
were calculated. 
These 2D slices with the parameters $(\delta=0, q \geq 0)$, the points 3--12 in Fig.~\ref{phase_diagram_Dnn0p0}, 
for $J_{3\text{s},3}/J_{\text{nn}}=0$ and $0.1$ ($J_{3\text{s},1}=J_{3\text{s},2}=0$) are shown 
in Figs.~\ref{SQTPQL2J3b30p0J1qP} and \ref{SQTPQL2J3b30p1J1qP}, respectively.

For $J_{3\text{s},3}/J_{\text{nn}}=0$ and $(\delta,q)=(0,0)$, i.e., the classical SI model, 
$S(\bm{Q})$ and $\langle \sigma_{\bm{Q}}^{\alpha} \sigma_{-\bm{Q}}^{\alpha} \rangle$ ($\alpha=z,x$) 
are shown in Figs.~\ref{SQTPQL2J3b30p0J1qP}(j1--j6). 
Since there is no interaction between quadrupole moments, 
$\langle \sigma_{\bm{Q}}^{x} \sigma_{-\bm{Q}}^{x} \rangle$ [Figs.~\ref{SQTPQL2J3b30p0J1qP}(j3,j6)] 
do not depend on $\bm{Q}$. 
Wave vector dependence of $S(\bm{Q})$ [Figs.~\ref{SQTPQL2J3b30p0J1qP}(j1,j4)] 
and $\langle \sigma_{\bm{Q}}^{z} \sigma_{-\bm{Q}}^{z} \rangle$ [Figs.~\ref{SQTPQL2J3b30p0J1qP}(j2,j5)]
show intensity patterns which scarcely appear in simulations of pyrochlore magnets. 
They probably reflect the periodic 32-site cluster, 
because $\langle \sigma_{\bm{Q}}^{z} \sigma_{-\bm{Q}}^{z} \rangle$ [Figs.~\ref{SQTPQL2J3b30p0J1qP}(j2,j5)] 
bear a resemblance to Fig.~6 of Ref.~\cite{Schafer2020}, 
in which a DMRG technique on the 32-site cluster was used 
for the Heisenberg antiferromagnet on the pyrochlore lattice [$(\delta,q)=(1,0)$]. 
Therefore the intensity pattern of $S(\bm{Q})$ [Figs.~\ref{SQTPQL2J3b30p0J1qP}(j1,j4)] 
can be regarded as a 32-site-cluster version of the classical SI. 
We note that for large clusters the pattern should be characterized by the pinch point 
(see Fig.~3 of Ref.~\cite{Kato2015}, Fig.~\ref{SQcalMCp12}(n0), etc.).

For $J_{3\text{s},3}/J_{\text{nn}}=0$ and $(\delta,q)=(0,0.7)$, 
where the system has the ground state with the 3D-PAF ($q>0$) quadrupole order (Fig.~\ref{phase_diagram_Dnn0p0}), 
$S(\bm{Q})$ and $\langle \sigma_{\bm{Q}}^{\alpha} \sigma_{-\bm{Q}}^{\alpha} \rangle$ ($\alpha=z,x$) 
are shown in Figs.~\ref{SQTPQL2J3b30p0J1qP}(a1--a6). 
Bragg-like peaks due to the 3D-PAF ($q>0$) order 
are seen in $\langle \sigma_{\bm{Q}}^{x} \sigma_{-\bm{Q}}^{x} \rangle$ [Fig.~\ref{SQTPQL2J3b30p0J1qP}(a3)] 
at $\bm{Q}=(1,1,1)$ and $(1,1,3)$. 
While $S(\bm{Q})$ [Fig.~\ref{SQTPQL2J3b30p0J1qP}(a1)] 
and $\langle \sigma_{\bm{Q}}^{z} \sigma_{-\bm{Q}}^{z} \rangle$ [Fig.~\ref{SQTPQL2J3b30p0J1qP}(a2)] 
at $T/J_{\text{nn}}=0.2$ 
show magnetic SRO patterns which resemble those of the classical SI [Figs.~\ref{SQTPQL2J3b30p0J1qP}(j1,j2)], 
although intensities become weaker. 

For $J_{3\text{s},3}/J_{\text{nn}}=0$ and $(\delta=0, 0<q<0.7)$, 
$S(\bm{Q})$ and 
$\langle \sigma_{\bm{Q}}^{\alpha} \sigma_{-\bm{Q}}^{\alpha} \rangle$ ($\alpha=z,x$) [Figs.~\ref{SQTPQL2J3b30p0J1qP}(b--i)] 
continuously change in the range $0<q<0.7$, i.e., 
from the classical SI to the 3D-PAF ($q>0$) order. 
At $T/J_{\text{nn}}=0.2$ and around $q=0.5$ [Figs.~\ref{SQTPQL2J3b30p0J1qP}(c1--e1,c2--e2,c3--e3)] 
they change steeply as a function of $q$, 
which is in agreement with the $q$ dependence of $C(T)$ curves [Fig.~\ref{CTST_TPQ_J3b30p0}(a)]. 
This steep change can be ascribed to a first-order phase transition at $T=0$ 
in the thermodynamic limit $N \rightarrow \infty$ \cite{Kadowaki2019}. 

For $J_{3\text{s},3}/J_{\text{nn}}=0.1$, 
at $T/J_{\text{nn}}=1$ 
$S(\bm{Q})$ [Figs.~\ref{SQTPQL2J3b30p1J1qP}(a4--j4)] and 
$\langle \sigma_{\bm{Q}}^{\alpha} \sigma_{-\bm{Q}}^{\alpha} \rangle$ ($\alpha=z,x$) 
[Figs.~\ref{SQTPQL2J3b30p1J1qP}(a5--j5,a6--j6)] 
are almost the same as 
those for $J_{3\text{s},3}/J_{\text{nn}}=0$ [Figs.~\ref{SQTPQL2J3b30p0J1qP}(a4--j4,a5--j5,a6--j6)]. 
This means that 
at this temperature $k_{\text{B}} T$ is much larger than the energy scale of the three-spin interaction. 
On the other hand, 
at $T/J_{\text{nn}}=0.2$  
$S(\bm{Q})$ [Figs.~\ref{SQTPQL2J3b30p1J1qP}(a1--j1)] and 
$\langle \sigma_{\bm{Q}}^{\alpha} \sigma_{-\bm{Q}}^{\alpha} \rangle$ ($\alpha=z,x$) 
[Figs.~\ref{SQTPQL2J3b30p1J1qP}(a2--j2,a3--j3)] 
are very different from 
those for $J_{3\text{s},3}/J_{\text{nn}}=0$ [Figs.~\ref{SQTPQL2J3b30p0J1qP}(a1--j1,a2--j2,a3--j3)]. 
This implies that $k_{\text{B}} T$ becomes comparable to (or lower than) the energy scale of the three-spin interaction. 

For $J_{3\text{s},3}/J_{\text{nn}}=0.1$ and at $T/J_{\text{nn}}=0.2$, 
$S(\bm{Q})$ with $q$ in a range $0 \leq q \leq 0.4$ [Figs.~\ref{SQTPQL2J3b30p1J1qP}(f1--j1)] 
show mutually similar intensity patterns, 
which are very different from 
those for $J_{3\text{s},3}/J_{\text{nn}}=0$ [Figs.~\ref{SQTPQL2J3b30p0J1qP}(f1--j1)]. 
This difference can be brought about 
by lifting the SI degeneracy due to the three-spin interaction, 
which is probably small, 
because $S(T)$ curves ($0 \leq q \leq 0.4$) plotted in Figs.~\ref{CTST_TPQ_J3b30p0}(c) and \ref{CTST_TPQ_J3b30p1}(c) 
show only slight difference between $J_{3\text{s},3}/J_{\text{nn}}=0.1$ and $0$ 
in $T/J_{\text{nn}}>0.2$. 

For $J_{3\text{s},3}/J_{\text{nn}}=0.1$ and at $T/J_{\text{nn}}=0.2$, 
$S(\bm{Q})$ with $q=0.5,0.55,0.6$ [Figs.~\ref{SQTPQL2J3b30p1J1qP}(b1,c1,d1)] 
show a novel intensity pattern 
characterized by magnetic spin correlations with $\bm{k} \sim (\tfrac{1}{2},\tfrac{1}{2},\tfrac{1}{2})$. 
These spin correlations roughly agree with those of the CMC results 
[Figs.~\ref{SQcalMCp2}(k0), \ref{SQcalMCp1}(c0), and \ref{SQcalMCp3}(c0)]. 
Therefore we may conclude that the spin correlations of TTO [Figs.~\ref{SQobs}(a0,c0,e0)] 
can be basically accounted for by these TPQ and CMC results, 
although detailed structures of the intensity patterns are not the same. 
In other words, 
the TPQ results suggest that 
the effective Hamiltonian minimally describing TTO is $\mathcal{H}_0 + \mathcal{H}_{3\text{s}}$ 
with $J_{3\text{s},1}=J_{3\text{s},2}=0$, 
$J_{3\text{s},3}/J_{\text{nn}} \sim 0.1$ (or $-0.1$) 
and the parameters $(\delta,q)$ in the region 
which is enclosed by the red dashed line in Fig.~\ref{phase_diagram_Dnn0p0}, 
where $\delta \ne 0$ will be discussed in Appendix~\ref{appendix_TPQ}. 
Another interesting point one can see from Figs.~\ref{SQTPQL2J3b30p1J1qP}(c1--c3,d1--d3) is that 
$\langle \sigma_{\bm{Q}}^{x} \sigma_{-\bm{Q}}^{x} \rangle$ with $q=0.5,0.55$ [Figs.~\ref{SQTPQL2J3b30p1J1qP}(c3,d3)] 
show broad peaks around $\Gamma$ points, i.e., 
electric quadrupole correlations with $\bm{k} \sim \bm{0}$. 
These results [Figs.~\ref{SQTPQL2J3b30p1J1qP}(c1--c3,d1--d3)] 
imply that the magnetic dipole correlations and electric quadrupole correlations coexist 
at $T/J_{\text{nn}}=0.2$. 
This coexistence may possibly continue down to $T=0$. 

\subsubsection{\label{results_TPQ_SQ_q_axisM} $S(\bm{Q})$ on $q$-axis ($q<0$)}

Two dimensional slices of $S(\bm{Q})$ 
and $\langle \sigma_{\bm{Q}}^{\alpha} \sigma_{-\bm{Q}}^{\alpha} \rangle$ ($\alpha=z,x$) 
calculated with the parameters $(\delta=0, q < 0)$, the points 13--21 in Fig.~\ref{phase_diagram_Dnn0p0}, 
for $J_{3\text{s},3}/J_{\text{nn}}=0$ and $0.1$ ($J_{3\text{s},1}=J_{3\text{s},2}=0$)
are shown in Figs.~\ref{SQTPQL2J3b30p0J1qM} and \ref{SQTPQL2J3b30p1J1qM}, respectively. 

For $J_{3\text{s},3}/J_{\text{nn}}=0$, 
$S(\bm{Q})$ [Figs.~\ref{SQTPQL2J3b30p0J1qM}(a1--i1,a4--i4)] 
and $\langle \sigma_{\bm{Q}}^{z} \sigma_{-\bm{Q}}^{z} \rangle$ [Figs.~\ref{SQTPQL2J3b30p0J1qM}(a2--i2,a5--i5)] 
are the same as 
those with $(\delta=0, q>0)$ [Figs.~\ref{SQTPQL2J3b30p0J1qP}(a1--i1,a4--i4) and Figs.~\ref{SQTPQL2J3b30p0J1qP}(a2--i2,a5--i5)], 
while  
$\langle \sigma_{\bm{Q}}^{x} \sigma_{-\bm{Q}}^{x} \rangle$ [Figs.~\ref{SQTPQL2J3b30p0J1qM}(a3--i3,a6--i6)] 
are different from 
those with $(\delta=0, q>0)$ [Figs.~\ref{SQTPQL2J3b30p0J1qP}(a3--i3,a6--i6)].
These are consequences of the invariance of $\mathcal{H}_0$ under the transformation 
of rotating $\bm{\sigma}_{\bm{r}}$ about the local $\bm{z}_{\bm{r}}$ axis by $\pi/2$ 
and $q \rightarrow -q$, 
confirming the correctness of the simulations using the TPQ states. 

For $J_{3\text{s},3}/J_{\text{nn}} = 0.1$ and at $T/J_{\text{nn}} = 0.2$, 
$S(\bm{Q})$ and $\langle \sigma_{\bm{Q}}^{z} \sigma_{-\bm{Q}}^{z} \rangle$ with $q \leq -0.45$ 
[Figs.~\ref{SQTPQL2J3b30p1J1qM}(a1--e1,a2--e2)] 
are different from 
those with $q \geq 0.45$ [Figs.~\ref{SQTPQL2J3b30p1J1qP}(a1--e1,a2--e2)], 
which results from breaking of the invariance for $\mathcal{H}_0 + \mathcal{H}_{3\text{s}}$. 
In relation to the analysis of TTO, 
none of $S(\bm{Q})$ [Figs.~\ref{SQTPQL2J3b30p1J1qM}(a1--e1)] shows 
spin correlations with $\bm{k} \sim (\tfrac{1}{2},\tfrac{1}{2},\tfrac{1}{2})$, 
which is in agreement with the CMC results with $q<0$ (Figs.~\ref{SQcalMCp5} and \ref{SQcalMCp15}). 
Therefore we conclude again that the $q<0$ side of the phase diagram (Fig.~\ref{phase_diagram_Dnn0p0})
can be excluded from studies of TTO. 

For $J_{3\text{s},3}/J_{\text{nn}}=0$ and $(\delta=0, -0.7<q<0)$, 
$S(\bm{Q})$ and 
$\langle \sigma_{\bm{Q}}^{\alpha} \sigma_{-\bm{Q}}^{\alpha} \rangle$ ($\alpha=z,x$) [Figs.~\ref{SQTPQL2J3b30p0J1qM}(b--i)] 
continuously change in the range $-0.7<q<0$, i.e., 
from the classical SI to the 3D-PAF ($q<0$) order. 
At $T/J_{\text{nn}}=0.2$ and around $q=-0.5$ [Figs.~\ref{SQTPQL2J3b30p0J1qM}(c1--e1,c2--e2,c3--e3)] 
they change very steeply as a function of $q$, 
which is consistent with the $q$ dependence of $C(T)$ [Fig.~\ref{CTST_TPQ_J3b30p0}(b)]. 
This steep change can be ascribed to a first-order phase transition at $T=0$ 
in the thermodynamic limit $N \rightarrow \infty$ \cite{Kadowaki2019}. 

For $J_{3\text{s},3}/J_{\text{nn}}=0.1$ and $(\delta=0, -0.7<q<0)$, 
$S(\bm{Q})$ and 
$\langle \sigma_{\bm{Q}}^{\alpha} \sigma_{-\bm{Q}}^{\alpha} \rangle$ ($\alpha=z,x$) [Figs.~\ref{SQTPQL2J3b30p1J1qM}(b--i)] 
continuously change in the range $-0.7<q<0$. 
At $T/J_{\text{nn}}=0.2$ and around $q=-0.5$ [Figs.~\ref{SQTPQL2J3b30p1J1qM}(c1--e1,c2--e2,c3--e3)] 
they change very steeply as a function of $q$, 
which is consistent with the $q$ dependence of $C(T)$ [Fig.~\ref{CTST_TPQ_J3b30p1}(b)]. 
This steep change suggests a first-order phase transition at $T=0$, 
because the variation of 
$\langle \sigma_{\bm{Q}}^{x} \sigma_{-\bm{Q}}^{x} \rangle$ [Figs.~\ref{SQTPQL2J3b30p1J1qM}(c3--e3)] 
is as steep as 
that for $J_{3\text{s},3}/J_{\text{nn}}=0$ [Figs.~\ref{SQTPQL2J3b30p0J1qM}(c3--e3)]. 
In contrast, 
$S(\bm{Q})$ and $\langle \sigma_{\bm{Q}}^{\alpha} \sigma_{-\bm{Q}}^{\alpha} \rangle$ ($\alpha=z,x$) 
for $J_{3\text{s},3}/J_{\text{nn}}=0.1$ 
vary much more gradually around $q=0.5$ at $T/J_{\text{nn}}=0.2$ [Figs.~\ref{SQTPQL2J3b30p1J1qP}(c1--e1,c2--e2,c3--e3)], 
which is consistent with the corresponding $q$ dependence of $C(T)$ [Fig.~\ref{CTST_TPQ_J3b30p1}(a)].  
These suggest a possibility that at $T=0$ there is another disordered ground state 
in the vicinity of $(\delta,q)=(0,0.5)$, i.e., between the QSI and 3D-PAF ($q>0$) states, 
in the quantum phase diagram with $J_{3\text{s},3}/J_{\text{nn}}=0.1$ (Fig.~\ref{phase_diagram_Dnn0p0}). 

\subsubsection{\label{results_TPQ_SQ_delta} results of $C(T)$, $S(T)$, and $S(\bm{Q})$ for $\delta \ne 0$}

To complement the simulation results on the $q$-axis 
shown in Sections~\ref{results_TPQ_C}, \ref{results_TPQ_SQ_q_axisP}, and \ref{results_TPQ_SQ_q_axisM} 
a few 32-site simulations using the TPQ states 
with the eight sets of the parameters $(\delta = \pm 0.1 ,q)$, 
the points 22--29 in Fig.~\ref{phase_diagram_Dnn0p0}, 
were carried out for $J_{3\text{s},3}/J_{\text{nn}}=0$ and $0.1$ ($J_{3\text{s},1}=J_{3\text{s},2}=0$). 
These results are presented in Appendix~\ref{appendix_TPQ}. 

\section{\label{Discussion_section} Discussion}
An answer to the initial question 
``why does $S(\bm{Q})$ of TTO show the spin correlations with $\bm{k} \sim (\tfrac{1}{2},\tfrac{1}{2},\tfrac{1}{2})$?'' 
has been obtained 
by the results of the CMC simulations and the quantum simulations using the TPQ states 
to a certain extent. 
It is an effect of one of the three-spin interactions, 
the $i=3$ term of $\mathcal{H}_{3\text{s}}$ [Eq.~(\ref{H_3s})]. 
This answer seems to provide basic understanding of TTO 
because of the following affirmative background reasoning or/and narratives. 
The three-spin interaction term is naturally expected 
from a perturbation expansion via virtual CF excitations \cite{Rau_Gingras2019,Molavian2009}. 
The magnitude of $J_{3\text{s},3}$ is consistent with this perturbation theory. 
Since the coupling constant $J_{3\text{s},3}$ is an order smaller than $J_{\text{nn}}$, 
the three-spin interaction affects the spin correlations 
only at low temperatures 
and only if the system is located close to classical phase boundaries of the three states: 
the SI state, 
the 3D-PAF ($q>0$) quadrupole ordered state, 
and 
the state possessing both the quadrupole and magnetic orders. 
This proximity to the phase boundaries may have a profound theoretical meaning \cite{Yan2017,Benton2016,Benton2018}. 
In the CMC simulation, where thermal fluctuations disappear at $T=0$, 
the spin correlations with $\bm{k} \sim (\tfrac{1}{2},\tfrac{1}{2},\tfrac{1}{2})$ appears 
in the intermediate temperature ranges. 
In contrast, 
the quantum simulation suggests that the spin correlations exist down to $T=0$ owing to quantum fluctuations. 

However, 
there remain unresolved problems, mainly because the simulation methods are far from perfect. 
Although there are obviously the peaked structures with $\bm{k} \sim (\tfrac{1}{2},\tfrac{1}{2},\tfrac{1}{2})$ 
in the simulated $S(\bm{Q})$, 
these do not quantitatively reproduce the experimentally observed $S(\bm{Q})$. 
It is likely that the number of model parameters has to be increased for better fitting. 
The small size effect of the 32-site quantum simulation makes its results obscure 
and its interpretation difficult especially at low temperatures.  
As discussed in Ref.~\cite{Rau_Gingras2019} the modeling of TTO is a non-trivial problem. 
In this work, we deal with the excited CF doublet state perturbatively, 
i.e., state vectors in the Hilbert space consisting of one doublet state per site. 
A larger Hilbert space, i.e., two doublet states per site may have to be taken into account \cite{Hallas2020arXiv}. 
Therefore, much work will have to be performed to solve the conundrum of TTO. 

The electric quadrupole (multipole) operators represent the deformation of 
the $f$-electron charge density of Tb$^{3+}$, 
and inevitably couple to displacements of surrounding atoms \cite{Bonville2011,Gritsenko2020,Mirebeau2004,Ruminy2019,Jin2020}. 
An interesting point of the quantum simulation results 
is that the quadrupole correlations coexist with the spin correlations [Figs.~\ref{SQTPQL2J3b30p1J1qP}(c1--c3,d1--d3)]. 
If this is really the case for QSL samples of TTO, 
it will be fascinating to observe these quadrupole correlations 
or/and correlated lattice deformations associated with them, 
which is a challenging experimental task. 

Several neutron scattering experiments were performed 
on TTO samples, of which stoichiometries are mostly unknown. 
They showed that spin correlations are clearly seen in 
energy-resolution-limited (nominally and instrument-dependent) 
elastic scattering at low temperatures. 
Spin correlations were reported to show many features 
including the three main features: 
magnetic SRO with $\bm{k} \sim (\tfrac{1}{2},\tfrac{1}{2},\tfrac{1}{2})$ 
\cite{Yasui2002,Fennell2012,Petit12,Fritsch2013}, 
pinch-point like structures at $\bm{k} \sim \bm{0}$ \cite{Fennell2012,Petit12}, 
and tiny Bragg reflections at $\bm{k} = (\tfrac{1}{2},\tfrac{1}{2},\tfrac{1}{2})$ and $\bm{k} = \bm{0}$ \cite{Taniguchi13,Takatsu2016prl}. 
These may have to be revisited using well controlled TTO samples and under well-tuned instrumental conditions. 

\section{\label{Conclusions_section} Conclusions}
We have studied spin correlations 
characterized by the modulation wave vector $\bm{k} \sim (\tfrac{1}{2},\tfrac{1}{2},\tfrac{1}{2})$ 
observed in the putative QSL pyrochlore magnet Tb$_{2+x}$Ti$_{2-x}$O$_{7+y}$ \cite{Kadowaki2018,Kadowaki2019}. 
Since they could not be accounted for by adding further-neighbor magnetic interactions 
to the NN pseudospin-$\tfrac{1}{2}$ Hamiltonian proposed in Ref.~\cite{Takatsu2016prl}, 
in this work we have explored another possibility of adding a three-spin interaction term 
of a form $\sigma_{\bm{r}}^{\pm} \sigma_{\bm{r}^{\prime}}^z \sigma_{\bm{r}^{\prime \prime}}^z$, 
which is a correction to the Hamiltonian due to the low crystal-field excitation. 

Classical MC simulation and quantum simulation using the TPQ states are applied to analyze 
experimentally observed structure factor $S(\bm{Q})$. 
The simulation results show that 
spin correlations with $\bm{k} \sim (\tfrac{1}{2},\tfrac{1}{2},\tfrac{1}{2})$, 
coexisting with electric quadrupole correlations with $\bm{k} \sim \bm{0}$, 
are induced at low temperatures by the three-spin interaction. 
The results suggest that the QSL state of Tb$_{2+x}$Ti$_{2-x}$O$_{7+y}$ 
is located close to phase boundaries of 
the spin ice, quadrupole-ordered, 
and both quadrupole- and magnetic-ordered states in the classical approximation, 
and that the three-spin interaction brings about a quantum disordered ground state 
with both spin and quadrupole correlations. 

As a by-product, the quantum simulation roughly reproduces 
the puzzling behavior of specific heat $C(T) \sim \text{const}$, 
which was experimentally observed at low temperatures. 
Therefore, we conclude that the classical and quantum simulation results suggest that 
the effective Hamiltonian minimally describing Tb$_{2+x}$Ti$_{2-x}$O$_{7+y}$ is 
$\mathcal{H}_0 + \mathcal{H}_{3\text{s}}$ [Eqs.~(\ref{H_0}) and (\ref{H_3s})] 
with $J_{3\text{s},1}=J_{3\text{s},2}=0$, 
$J_{3\text{s},3}/J_{\text{nn}} \sim 0.1$ (or $-0.1$) 
and the parameters $(\delta,q)$ in the region 
which is enclosed by the red dashed line in Fig.~\ref{phase_diagram_Dnn0p0}. 
A novel viewpoint of the QSL state of Tb$_{2+x}$Ti$_{2-x}$O$_{7+y}$ and/or 
elaborate theories which quantitatively reproduce the spin correlations 
will be hopefully constructed based on this work. 

\begin{acknowledgments}
This work was supported by JSPS KAKENHI grant number 25400345. 
The neutron scattering performed using ILL IN5 (France) 
was transferred from JRR-3M HER (proposal 11567, 15545) 
with the approval of ISSP, Univ. of Tokyo, and JAEA, Tokai, Japan. 
The neutron scattering experiments at J-PARC AMATERAS were carried out 
under a research project number 2016A0327. 
The computation was performed on supercomputers at ISSP University of Tokyo, 
ITC Nagoya University, and Hokkaido University. 
\end{acknowledgments}

\appendix
\section{\label{appendix_def} CF ground state doublet, lattice sites, etc.}
The CF ground state doublet of TTO at each site is written by 
\begin{equation}
| \pm 1 \rangle_{\text{D}} 
= A | \mp 4 \rangle \pm B | \mp 1 \rangle + C | \pm 2 \rangle \mp D | \pm 5 \rangle ,
\label{G_doublet}
\end{equation}
where $| m \rangle$ stands for the $| J=6, m \rangle$ state within 
a $JLS$-multiplet \cite{Jensen91}. 
The coefficients of Eq.~(\ref{G_doublet}) are 
$A=0.9581$, $B=0.1284$, $C=0.1210$, and $D=0.2256$ 
using the CF parameters of Ref.~\cite{Mirebeau07}. 
Magnetic-dipole and electric-quadrupole moment operators \cite{Kusunose08} 
within $| \pm 1 \rangle_{\text{D}} $ 
are proportional to the Pauli matrices $\sigma^{\alpha}$ ($\alpha = x,y,z$)
and the unit matrix \cite{Kadowaki2015,Kadowaki2021Erratum}. 
The magnetic dipole moment operators are given by 
\begin{align}
J_x &= J_y = 0,\nonumber\\
J_z &= - ( 4A^2 + B^2 - 2C^2 - 5D^2) \sigma^z \; .
\label{mag_dipole_moment}
\end{align}
The electric quadrupole moment operators are expressed as 
\begin{align}
\tfrac{1}{2}[ 3 J_z^2 -J(J+1)] &= 
3A^2 - \tfrac{39}{2} B^2 - 15 C^2 + \tfrac{33}{2} D^2 \nonumber\\
\tfrac{\sqrt{3}}{2}[ J_x^2 - J_y^2] &= 
\left( -\tfrac{21 \sqrt{3}}{2} B^2 + 9 \sqrt{10} AC \right) \sigma^x \nonumber\\
\tfrac{\sqrt{3}}{2}[ J_x J_y + J_y J_x ] &= 
- \left( -\tfrac{21 \sqrt{3}}{2} B^2 + 9 \sqrt{10} AC \right) \sigma^y \nonumber\\
\tfrac{\sqrt{3}}{2}[ J_z J_x + J_x J_z ] &= 
- \left( 3 \sqrt{30} BC + 9 \sqrt{\tfrac{33}{2}} AD \right) \sigma^x \nonumber\\
\tfrac{\sqrt{3}}{2}[ J_y J_z + J_z J_y ] &= 
- \left( 3 \sqrt{30} BC + 9 \sqrt{\tfrac{33}{2}} AD \right) \sigma^y \; .
\label{ele_quad_moment}
\end{align}
\begin{table}
\caption{\label{local_axis}
Four crystallographic sites $\bm{d}_{\nu}$ and their 
local symmetry axes $\bm{x}_{\nu}$, $\bm{y}_{\nu}$, and $\bm{z}_{\nu}$. 
These coordinates are defined using (global) cubic XYZ axes shown in Fig.~\ref{phase_diagram_Dnn0p0}(a).
The four sites $\bm{d}_{\nu}$ are illustrated by vertices with light blue numbers ($\nu=1$--$4$) of a tetrahedron in Fig.~\ref{phase_diagram_Dnn0p0}(a). 
}
\begin{ruledtabular}
\begin{tabular}{ccccc}
$\nu$ & $\bm{d}_{\nu}$ & $\bm{x}_{\nu}$ & $\bm{y}_{\nu}$ & $\bm{z}_{\nu}$ \\ \hline
1 & $\tfrac{1}{4}(0,0,0)$ & $\tfrac{1}{\sqrt{6}}(1,1,-2)$ & $\tfrac{1}{\sqrt{2}}(-1,1,0)$ & $\tfrac{1}{\sqrt{3}}(1,1,1)$  \\
2 & $\tfrac{1}{4}(0,1,1)$ & $\tfrac{1}{\sqrt{6}}(1,-1,2)$ & $\tfrac{1}{\sqrt{2}}(-1,-1,0)$ & $\tfrac{1}{\sqrt{3}}(1,-1,-1)$  \\
3 & $\tfrac{1}{4}(1,0,1)$ & $\tfrac{1}{\sqrt{6}}(-1,1,2)$ & $\tfrac{1}{\sqrt{2}}(1,1,0)$ & $\tfrac{1}{\sqrt{3}}(-1,1,-1)$  \\
4 & $\tfrac{1}{4}(1,1,0)$ & $\tfrac{1}{\sqrt{6}}(-1,-1,-2)$ & $\tfrac{1}{\sqrt{2}}(1,-1,0)$ & $\tfrac{1}{\sqrt{3}}(-1,-1,1)$  \\
\end{tabular}
\end{ruledtabular}
\end{table}

The operators $\sigma_{\bm{r}}^{\alpha}$ of Eq.~(\ref{H_0})
act on $| \pm 1 \rangle_{\text{D}}$ 
at each pyrochlore lattice site $\bm{r}=\bm{t}_{n}+\bm{d}_{\nu}$, 
where $\bm{t}_n$ is an FCC translation vector and 
$\bm{d}_{\nu}$ ($\nu=1$, $2$, $3$, and $4$) are four crystallographic sites in the unit cell. 
Coordinates of the sites $\bm{d}_{\nu}$ and their local axes 
$\bm{x}_{\nu}$, $\bm{y}_{\nu}$, and $\bm{z}_{\nu}$ 
are listed in Table~\ref{local_axis}. 
Under these definitions the effective Hamiltonian is described by Eq.~(\ref{H_0}) 
with the phases $\phi_{\bm{r},\bm{r}^{\prime}}$ listed in Table~\ref{phase_Jnnq} \cite{Onoda11}. 
\begin{table}
\caption{\label{phase_Jnnq}
Phases $ \phi_{\bm{r}, \bm{r}^{\prime} }$ of the quadrupole interactions
$ J_{\text{nn}} 2 q \exp[i 2 \phi_{\bm{r}, \bm{r}^{\prime} } ] \sigma_{\bm{r}}^+ \sigma_{\bm{r}^{\prime}}^+ + \text{H.c.}$ [Eq.~(\ref{H_0})], 
where $\bm{r} = \bm{t}_n + \bm{d}_{\nu} $ and $\bm{r}^{\prime} = \bm{t}_{n^{\prime}} + \bm{d}_{\nu^{\prime}} $.
}
\begin{ruledtabular}
\begin{tabular}{cccc}
 $\nu$ & $\nu^\prime$ & $\bm{r}^{\prime} - \bm{r} $ & $ \phi_{\bm{r}, \bm{r}^{\prime}}/(\tfrac{2 \pi}{3})$ \\ \hline
 1 & 2 & $\tfrac{1}{4}(0,1, 1)$ & -1 \\
 1 & 3 & $\tfrac{1}{4}(1,0, 1)$ &  1 \\
 1 & 4 & $\tfrac{1}{4}(1, 1,0)$ &  0 \\
 2 & 3 & $\tfrac{1}{4}(1,-1,0)$ &  0 \\
 2 & 4 & $\tfrac{1}{4}(1,0,-1)$ &  1 \\
 3 & 4 & $\tfrac{1}{4}(0,1,-1)$ & -1 \\
\end{tabular}
\end{ruledtabular}
\end{table}

\section{\label{appendix_three_spin_interaction} three-spin interaction}
Following Refs.~\cite{Rau_Gingras2019,Molavian2009}, 
the three-spin interactions consist of 
terms with a form
$(c \sigma_{\bm{r}}^{+} + c^{*} \sigma_{\bm{r}}^{-}) \sigma_{\bm{r}^{\prime}}^z \sigma_{\bm{r}^{\prime \prime}}^z$, 
where the site triplet $\langle \bm{r}, \bm{r}^{\prime}, \bm{r}^{\prime \prime} \rangle$ 
satisfies geometrical conditions: 
$\langle \bm{r}, \bm{r}^{\prime} \rangle$ and $\langle \bm{r}, \bm{r}^{\prime \prime} \rangle$ 
are NN site pairs, the site $\bm{r}^{\prime}$ is different from $\bm{r}^{\prime \prime}$. 
Under these conditions the three-spin interaction term 
can be expressed by Eq.~(\ref{H_3s}) with unknown phases $\phi^{(i)}_{\bm{r}, \bm{r}^{\prime}, \bm{r}^{\prime \prime} }$. 
By imposing the condition that the three-spin interaction term 
is invariant under the space group symmetry (Fd\={3}m, No. 227), 
it is not difficult to determine the phases using the symmetry method employed 
for the two-spin interaction term of pyrochlore magnets \cite{Onoda11,Onoda2011jpcm,Ross11}. 
The phases $\phi^{(i)}_{\bm{r}, \bm{r}^{\prime}, \bm{r}^{\prime \prime} }$ with $i=1$, $2$, and $3$ are listed 
in Tables~\ref{phase_type1}, \ref{phase_type2}, and \ref{phase_type3}, respectively. 

\begin{table}
\caption{\label{phase_type1}
Phases $\phi^{(1)}_{\bm{r}, \bm{r}^{\prime}, \bm{r}^{\prime \prime} }$ 
of the type $i=1$ three-spin interaction 
$J_{3\text{s},1} 
\exp[i \phi^{(1)}_{\bm{r}, \bm{r}^{\prime}, \bm{r}^{\prime \prime} } ] \sigma_{\bm{r}}^+ \sigma_{\bm{r}^{\prime}}^z \sigma_{\bm{r}^{\prime \prime}}^z + \text{H.c.} $ [Eq.~(\ref{H_3s})], 
where $\bm{r} = \bm{t}_n + \bm{d}_{\nu} $, $\bm{r}^{\prime} = \bm{t}_{n^{\prime}} + \bm{d}_{\nu^{\prime}} $, 
and $\bm{r}^{\prime \prime} = \bm{t}_{n^{\prime \prime}} + \bm{d}_{\nu^{\prime \prime}} $. 
The site triplet $\langle \bm{r}, \bm{r}^{\prime}, \bm{r}^{\prime \prime} \rangle$ 
of the first line is illustrated in Fig.~\ref{J3s_type123}(a).
}
\begin{ruledtabular}
\begin{tabular}{cccccc}
 $\nu$ & $\nu^\prime$ & $\nu^{\prime \prime}$ & $\bm{r}^{\prime} - \bm{r} $ & $\bm{r}^{\prime \prime} - \bm{r} $ & $\phi^{(1)}_{\bm{r}, \bm{r}^{\prime}, \bm{r}^{\prime \prime}}/(\tfrac{2 \pi}{3})$ \\ \hline
 1 & 2 & 2  &   $\tfrac{1}{4}(0,1,1)$    & $\tfrac{1}{4}(0,-1,-1)$ & -1 \\
 1 & 3 & 3  &   $\tfrac{1}{4}(1,0,1)$    & $\tfrac{1}{4}(-1,0,-1)$ &  1 \\
 1 & 4 & 4  &   $\tfrac{1}{4}(1,1,0)$    & $\tfrac{1}{4}(-1,-1,0)$ &  0 \\
 2 & 1 & 1  &   $\tfrac{1}{4}(0,1,1)$    & $\tfrac{1}{4}(0,-1,-1)$ & -1 \\
 2 & 3 & 3  &   $\tfrac{1}{4}(1,-1,0)$   & $\tfrac{1}{4}(-1,1,0)$  &  0 \\
 2 & 4 & 4  &   $\tfrac{1}{4}(1,0,-1)$   & $\tfrac{1}{4}(-1,0,1)$  &  1 \\
 3 & 1 & 1  &   $\tfrac{1}{4}(1,0,1)$    & $\tfrac{1}{4}(-1,0,-1)$ &  1 \\
 3 & 2 & 2  &   $\tfrac{1}{4}(1,-1,0)$   & $\tfrac{1}{4}(-1,1,0)$  &  0 \\
 3 & 4 & 4  &   $\tfrac{1}{4}(0,1,-1)$   & $\tfrac{1}{4}(0,-1,1)$  & -1 \\
 4 & 1 & 1  &   $\tfrac{1}{4}(1,1,0)$    & $\tfrac{1}{4}(-1,-1,0)$ &  0 \\
 4 & 2 & 2  &   $\tfrac{1}{4}(1,0,-1)$   & $\tfrac{1}{4}(-1,0,1)$  &  1 \\
 4 & 3 & 3  &   $\tfrac{1}{4}(0,1,-1)$   & $\tfrac{1}{4}(0,-1,1)$  & -1 \\
\end{tabular}
\end{ruledtabular}
\end{table}
\begin{table}
\caption{\label{phase_type2}
Phases $\phi^{(2)}_{\bm{r}, \bm{r}^{\prime}, \bm{r}^{\prime \prime} }$ 
of the type $i=2$ three-spin interaction 
$J_{3\text{s},2} 
\exp[i \phi^{(2)}_{\bm{r}, \bm{r}^{\prime}, \bm{r}^{\prime \prime} } ] \sigma_{\bm{r}}^+ \sigma_{\bm{r}^{\prime}}^z \sigma_{\bm{r}^{\prime \prime}}^z + \text{H.c.} $ [Eq.~(\ref{H_3s})], 
where $\bm{r} = \bm{t}_n + \bm{d}_{\nu} $, $\bm{r}^{\prime} = \bm{t}_{n^{\prime}} + \bm{d}_{\nu^{\prime}} $, 
and $\bm{r}^{\prime \prime} = \bm{t}_{n^{\prime \prime}} + \bm{d}_{\nu^{\prime \prime}} $. 
The site triplet $\langle \bm{r}, \bm{r}^{\prime}, \bm{r}^{\prime \prime} \rangle$ 
of the first line is illustrated in Fig.~\ref{J3s_type123}(b).
}
\begin{ruledtabular}
\begin{tabular}{cccccc}
 $\nu$ & $\nu^\prime$ & $\nu^{\prime \prime}$ & $\bm{r}^{\prime} - \bm{r} $ & $\bm{r}^{\prime \prime} - \bm{r} $ & $\phi^{(2)}_{\bm{r}, \bm{r}^{\prime}, \bm{r}^{\prime \prime}}/(\tfrac{2 \pi}{3})$ \\ \hline
 1 & 2 & 3  &   $\tfrac{1}{4}(0,1,1)$    & $\tfrac{1}{4}(1,0,1)$   &  0 \\
 1 & 2 & 3  &   $\tfrac{1}{4}(0,-1,-1)$  & $\tfrac{1}{4}(-1,0,-1)$ &  0 \\
 1 & 2 & 4  &   $\tfrac{1}{4}(0,-1,-1)$  & $\tfrac{1}{4}(-1,-1,0)$ &  1 \\
 1 & 2 & 4  &   $\tfrac{1}{4}(0,1,1)$    & $\tfrac{1}{4}(1,1,0)$   &  1 \\
 1 & 3 & 4  &   $\tfrac{1}{4}(-1,0,-1)$  & $\tfrac{1}{4}(-1,-1,0)$ & -1 \\
 1 & 3 & 4  &   $\tfrac{1}{4}(1,0,1)$    & $\tfrac{1}{4}(1,1,0)$   & -1 \\
 2 & 1 & 3  &   $\tfrac{1}{4}(0,-1,-1)$  & $\tfrac{1}{4}(1,-1,0)$  &  1 \\
 2 & 1 & 3  &   $\tfrac{1}{4}(0,1,1)$    & $\tfrac{1}{4}(-1,1,0)$  &  1 \\
 2 & 1 & 4  &   $\tfrac{1}{4}(0,-1,-1)$  & $\tfrac{1}{4}(1,0,-1)$  &  0 \\
 2 & 1 & 4  &   $\tfrac{1}{4}(0,1,1)$    & $\tfrac{1}{4}(-1,0,1)$  &  0 \\
 2 & 3 & 4  &   $\tfrac{1}{4}(-1,1,0)$   & $\tfrac{1}{4}(-1,0,1)$  & -1 \\
 2 & 3 & 4  &   $\tfrac{1}{4}(1,-1,0)$   & $\tfrac{1}{4}(1,0,-1)$  & -1 \\
 3 & 1 & 2  &   $\tfrac{1}{4}(-1,0,-1)$  & $\tfrac{1}{4}(-1,1,0)$  & -1 \\
 3 & 1 & 2  &   $\tfrac{1}{4}(1,0,1)$    & $\tfrac{1}{4}(1,-1,0)$  & -1 \\
 3 & 1 & 4  &   $\tfrac{1}{4}(-1,0,-1)$  & $\tfrac{1}{4}(0,1,-1)$  &  0 \\
 3 & 1 & 4  &   $\tfrac{1}{4}(1,0,1)$    & $\tfrac{1}{4}(0,-1,1)$  &  0 \\
 3 & 2 & 4  &   $\tfrac{1}{4}(-1,1,0)$   & $\tfrac{1}{4}(0,1,-1)$  &  1 \\
 3 & 2 & 4  &   $\tfrac{1}{4}(1,-1,0)$   & $\tfrac{1}{4}(0,-1,1)$  &  1 \\
 4 & 1 & 2  &   $\tfrac{1}{4}(-1,-1,0)$  & $\tfrac{1}{4}(-1,0,1)$  & -1 \\
 4 & 1 & 2  &   $\tfrac{1}{4}(1,1,0)$    & $\tfrac{1}{4}(1,0,-1)$  & -1 \\
 4 & 1 & 3  &   $\tfrac{1}{4}(-1,-1,0)$  & $\tfrac{1}{4}(0,-1,1)$  &  1 \\
 4 & 1 & 3  &   $\tfrac{1}{4}(1,1,0)$    & $\tfrac{1}{4}(0,1,-1)$  &  1 \\
 4 & 2 & 3  &   $\tfrac{1}{4}(-1,0,1)$   & $\tfrac{1}{4}(0,-1,1)$  &  0 \\
 4 & 2 & 3  &   $\tfrac{1}{4}(1,0,-1)$   & $\tfrac{1}{4}(0,1,-1)$  &  0 \\
\end{tabular}
\end{ruledtabular}
\end{table}
\begin{table}
\caption{\label{phase_type3}
Phases $\phi^{(3)}_{\bm{r}, \bm{r}^{\prime}, \bm{r}^{\prime \prime} }$ 
of the type $i=3$ three-spin interaction 
$J_{3\text{s},3} 
\exp[i \phi^{(3)}_{\bm{r}, \bm{r}^{\prime}, \bm{r}^{\prime \prime} } ] \sigma_{\bm{r}}^+ \sigma_{\bm{r}^{\prime}}^z \sigma_{\bm{r}^{\prime \prime}}^z + \text{H.c.} $ [Eq.~(\ref{H_3s})], 
where $\bm{r} = \bm{t}_n + \bm{d}_{\nu} $, $\bm{r}^{\prime} = \bm{t}_{n^{\prime}} + \bm{d}_{\nu^{\prime}} $, 
and $\bm{r}^{\prime \prime} = \bm{t}_{n^{\prime \prime}} + \bm{d}_{\nu^{\prime \prime}} $. 
The site triplet $\langle \bm{r}, \bm{r}^{\prime}, \bm{r}^{\prime \prime} \rangle$ 
of the first line is illustrated in Fig.~\ref{J3s_type123}(c).
}
\begin{ruledtabular}
\begin{tabular}{cccccc}
 $\nu$ & $\nu^\prime$ & $\nu^{\prime \prime}$ & $\bm{r}^{\prime} - \bm{r} $ & $\bm{r}^{\prime \prime} - \bm{r} $ & $\phi^{(3)}_{\bm{r}, \bm{r}^{\prime}, \bm{r}^{\prime \prime}}/(\tfrac{2 \pi}{3})$ \\ \hline
 1 & 2 & 3  &   $\tfrac{1}{4}(0,-1,-1)$  & $\tfrac{1}{4}(1,0,1)$   &  0 \\
 1 & 2 & 3  &   $\tfrac{1}{4}(0,1,1)$    & $\tfrac{1}{4}(-1,0,-1)$ &  0 \\
 1 & 2 & 4  &   $\tfrac{1}{4}(0,-1,-1)$  & $\tfrac{1}{4}(1,1,0)$   &  1 \\
 1 & 2 & 4  &   $\tfrac{1}{4}(0,1,1)$    & $\tfrac{1}{4}(-1,-1,0)$ &  1 \\
 1 & 3 & 4  &   $\tfrac{1}{4}(-1,0,-1)$  & $\tfrac{1}{4}(1,1,0)$   & -1 \\
 1 & 3 & 4  &   $\tfrac{1}{4}(1,0,1)$    & $\tfrac{1}{4}(-1,-1,0)$ & -1 \\
 2 & 1 & 3  &   $\tfrac{1}{4}(0,-1,-1)$  & $\tfrac{1}{4}(-1,1,0)$  &  1 \\
 2 & 1 & 3  &   $\tfrac{1}{4}(0,1,1)$    & $\tfrac{1}{4}(1,-1,0)$  &  1 \\
 2 & 1 & 4  &   $\tfrac{1}{4}(0,-1,-1)$  & $\tfrac{1}{4}(-1,0,1)$  &  0 \\
 2 & 1 & 4  &   $\tfrac{1}{4}(0,1,1)$    & $\tfrac{1}{4}(1,0,-1)$  &  0 \\
 2 & 3 & 4  &   $\tfrac{1}{4}(-1,1,0)$   & $\tfrac{1}{4}(1,0,-1)$  & -1 \\
 2 & 3 & 4  &   $\tfrac{1}{4}(1,-1,0)$   & $\tfrac{1}{4}(-1,0,1)$  & -1 \\
 3 & 1 & 2  &   $\tfrac{1}{4}(-1,0,-1)$  & $\tfrac{1}{4}(1,-1,0)$  & -1 \\
 3 & 1 & 2  &   $\tfrac{1}{4}(1,0,1)$    & $\tfrac{1}{4}(-1,1,0)$  & -1 \\
 3 & 1 & 4  &   $\tfrac{1}{4}(-1,0,-1)$  & $\tfrac{1}{4}(0,-1,1)$  &  0 \\
 3 & 1 & 4  &   $\tfrac{1}{4}(1,0,1)$    & $\tfrac{1}{4}(0,1,-1)$  &  0 \\
 3 & 2 & 4  &   $\tfrac{1}{4}(-1,1,0)$   & $\tfrac{1}{4}(0,-1,1)$  &  1 \\
 3 & 2 & 4  &   $\tfrac{1}{4}(1,-1,0)$   & $\tfrac{1}{4}(0,1,-1)$  &  1 \\
 4 & 1 & 2  &   $\tfrac{1}{4}(-1,-1,0)$  & $\tfrac{1}{4}(1,0,-1)$  & -1 \\
 4 & 1 & 2  &   $\tfrac{1}{4}(1,1,0)$    & $\tfrac{1}{4}(-1,0,1)$  & -1 \\
 4 & 1 & 3  &   $\tfrac{1}{4}(-1,-1,0)$  & $\tfrac{1}{4}(0,1,-1)$  &  1 \\
 4 & 1 & 3  &   $\tfrac{1}{4}(1,1,0)$    & $\tfrac{1}{4}(0,-1,1)$  &  1 \\
 4 & 2 & 3  &   $\tfrac{1}{4}(-1,0,1)$   & $\tfrac{1}{4}(0,1,-1)$  &  0 \\
 4 & 2 & 3  &   $\tfrac{1}{4}(1,0,-1)$   & $\tfrac{1}{4}(0,-1,1)$  &  0 \\
\end{tabular}
\end{ruledtabular}
\end{table}

\section{\label{appendix_CMC_3DPAF_delta} CMC simulation results: $S(\bm{Q})$ in 3D-PAF ($q>0$) phase $\delta \neq 0$}
\begin{figure*}[ht]
\centering
\includegraphics[width=18cm,clip]{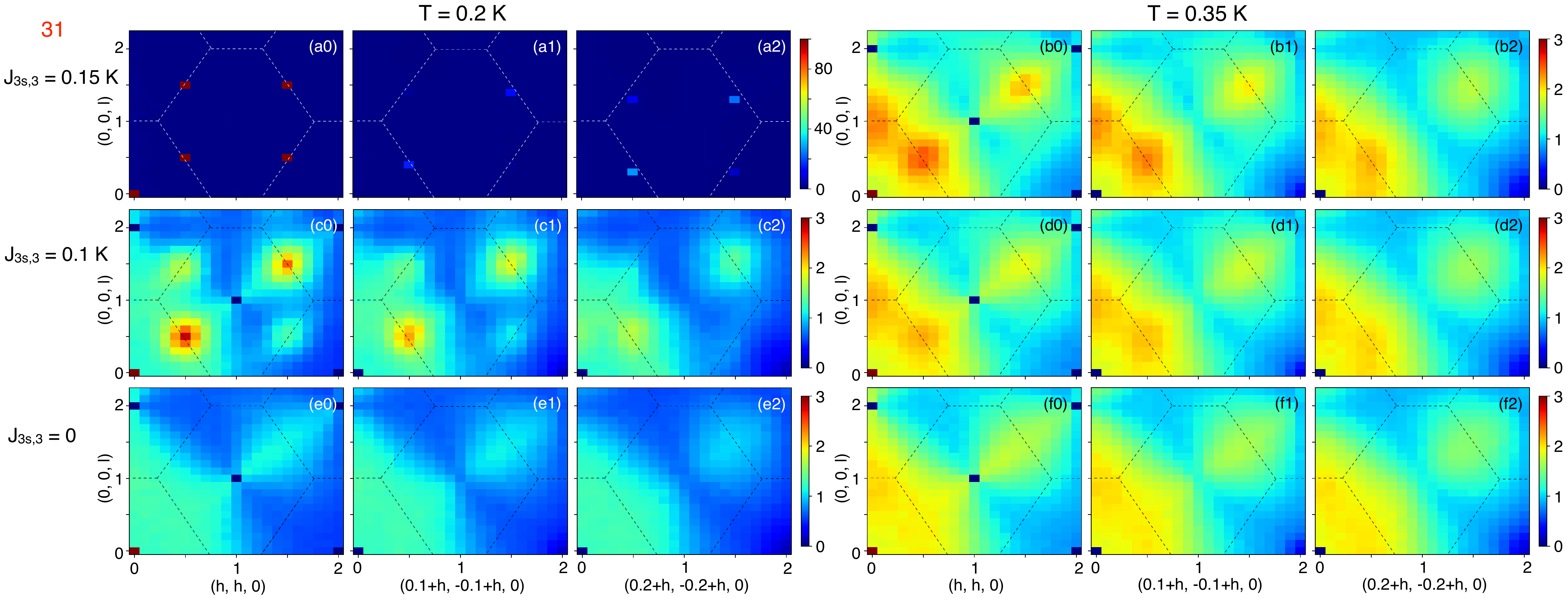}
\caption{ 
Intensity maps of $S(\bm{Q})$ calculated by the 16000-site CMC simulations 
using parameters indicated by red circles shown in Fig.~\ref{CT_MC123456}(a3) 
($J_{3\text{s},3}$; $J_{3\text{s},1}=J_{3\text{s},2}=0$) and 
by the point 31 in Fig.~\ref{phase_diagram_Dnn0p478} [$\frac{J_{\text{nn}}}{J_{\text{nn}} + D_{\text{nn}}}(\delta,q)=(-0.1,0.6)$]. 
They are viewed by 2D slices of $\bm{Q}=(k+h,-k+h,l)$ with fixed $k=0,0.1$, and $0.2$, 
which are shown in (w0), (w1), and (w2) (w=a--f), respectively. 
They are calculated at two temperatures 0.2 K (a,c,e) and 0.35 K (b,d,f), 
below and above the phase transition temperature of the 3D-PAF ($q>0$) LRO.
Intensity maps for $J_{3\text{s},3} = 0.15$, $0.1$, and 0 K [Fig.~\ref{CT_MC123456}(a3)] are shown in (a,b), (c,d), and (e,f), respectively. 
}
\label{SQcalMCp1}
\end{figure*}
\begin{figure*}[ht]
\centering
\includegraphics[width=18cm,clip]{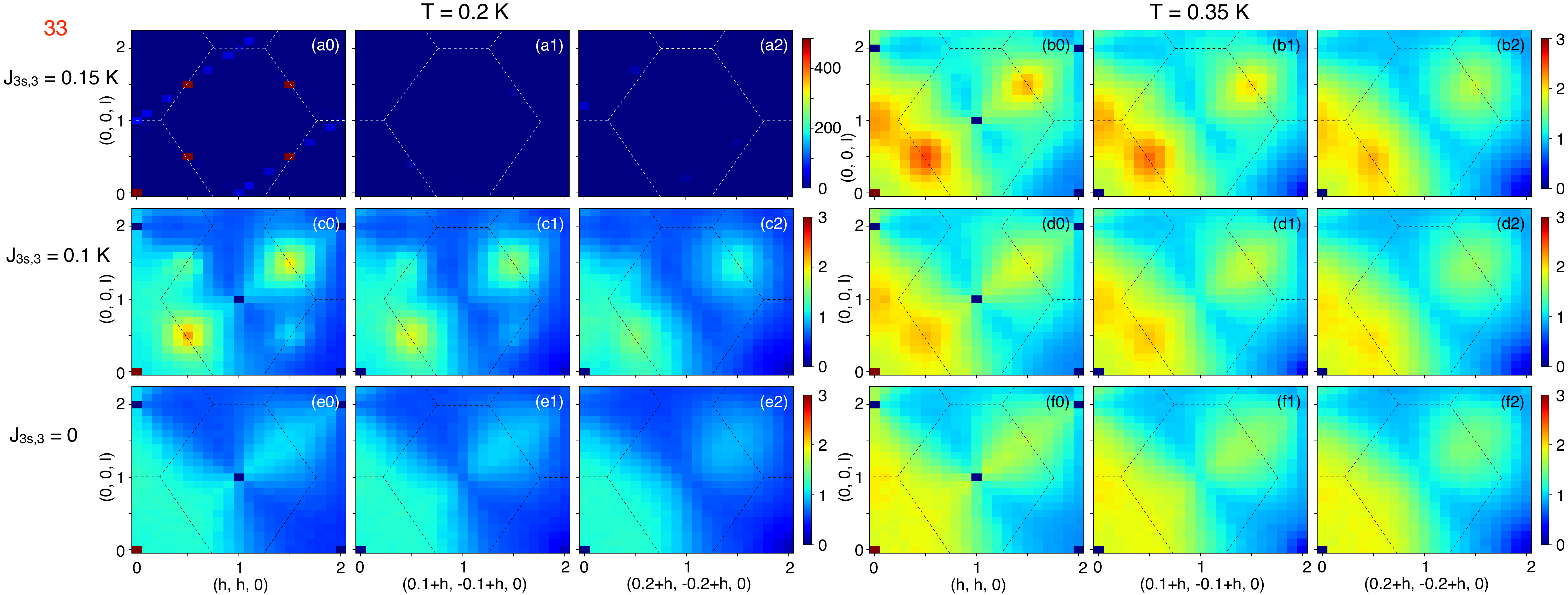}
\caption{ 
Intensity maps of $S(\bm{Q})$ calculated by the 16000-site CMC simulations 
using parameters indicated by red circles shown in Fig.~\ref{CT_MC123456}(c3) 
($J_{3\text{s},3}$; $J_{3\text{s},1}=J_{3\text{s},2}=0$) and 
by the point 33 in Fig.~\ref{phase_diagram_Dnn0p478} [$\frac{J_{\text{nn}}}{J_{\text{nn}} + D_{\text{nn}}}(\delta,q)=(0.1,0.5)$]. 
They are viewed by 2D slices of $\bm{Q}=(k+h,-k+h,l)$ with fixed $k=0,0.1$, and $0.2$, 
which are shown in (w0), (w1), and (w2) (w=a--f), respectively. 
They are calculated at two temperatures 0.2 K (a,c,e) and 0.35 K (b,d,f), 
below and above the phase transition temperature of the 3D-PAF ($q>0$) LRO.
Intensity maps for $J_{3\text{s},3} = 0.15$, $0.1$, and 0 K [Fig.~\ref{CT_MC123456}(c3)] are shown in (a,b), (c,d), and (e,f), respectively. 
}
\label{SQcalMCp3}
\end{figure*}

To complement the simulation results of $S(\bm{Q})$ shown in Figs.~\ref{SQcalMCp2}(i--n), 
we performed a few 16000-site CMC simulations 
with slightly different parameters: 
$\frac{J_{\text{nn}}}{J_{\text{nn}} + D_{\text{nn}}} (\delta,q)=(-0.1,0.6)$ and $(0.1,0.5)$,
corresponding to the points 31 and 33 in Fig.~\ref{phase_diagram_Dnn0p478}, respectively. 
The three-spin interaction constants were fixed to $J_{3\text{s},1}=J_{3\text{s},2}=0$ 
and $J_{3\text{s},3}=0, 0.1, 0.15$ K. 
Figures~\ref{SQcalMCp1} and \ref{SQcalMCp3} show the resulting intensity maps of $S(\bm{Q})$ 
which were calculated with the parameters corresponding to the red circles 
in Figs.~\ref{CT_MC123456}(a3) and \ref{CT_MC123456}(c3), respectively, 
and at two temperatures $0.2$ and $0.35$ K,
below and above the phase transition temperature of the 3D-PAF ($q>0$) LRO. 

The calculated $S(\bm{Q})$ of Figs.~\ref{SQcalMCp1}(a--f) and \ref{SQcalMCp3}(a--f)
bear close resemblances to $S(\bm{Q})$ of Figs.~\ref{SQcalMCp2}(i--n). 
This is in parallel with the analyses of Ref.~\cite{Takatsu2016prl}, 
in which the acceptable parameter range we proposed has the elongated shape (Fig.~\ref{phase_diagram_Dnn0p478}). 
Thus we can conclude that the parameter sets used for Figs.~\ref{SQcalMCp1}(c,d) and \ref{SQcalMCp3}(c,d) 
are also candidates for the further investigation. 
We note that $S(\bm{Q})$ maps of Figs.~\ref{SQcalMCp1}(a,c) and \ref{SQcalMCp3}(a,c), 
which are obtained from simulations at 0.2 K with $J_{3\text{s},3} = 0.15$ and $0.1$,
show certain difference from those of Figs.~\ref{SQcalMCp2}(i,k). 
This may be caused by high degeneracy due to proximity to the SI phase boundary, 
where the $S(\bm{Q})$ map is easily changed by small perturbations. 
To obtain better fit of the calculated $S(\bm{Q})$ to the observed $S(\bm{Q})$ of TTO, 
we have tried several parameter adjustments by adding small parameters: $J_{3\text{s},1}$, $J_{3\text{s},2}$, 
second- and third-neighbor magnetic exchange couplings ($J_2$, $J_3$, $J_4$ \cite{Kadowaki2019}). 
The fit, however, could not be improved. 

\section{\label{appendix_CMC_SI} CMC simulation results: $S(\bm{Q})$ in SI phase}
\begin{figure*}[ht]
\centering
\includegraphics[width=18cm,clip]{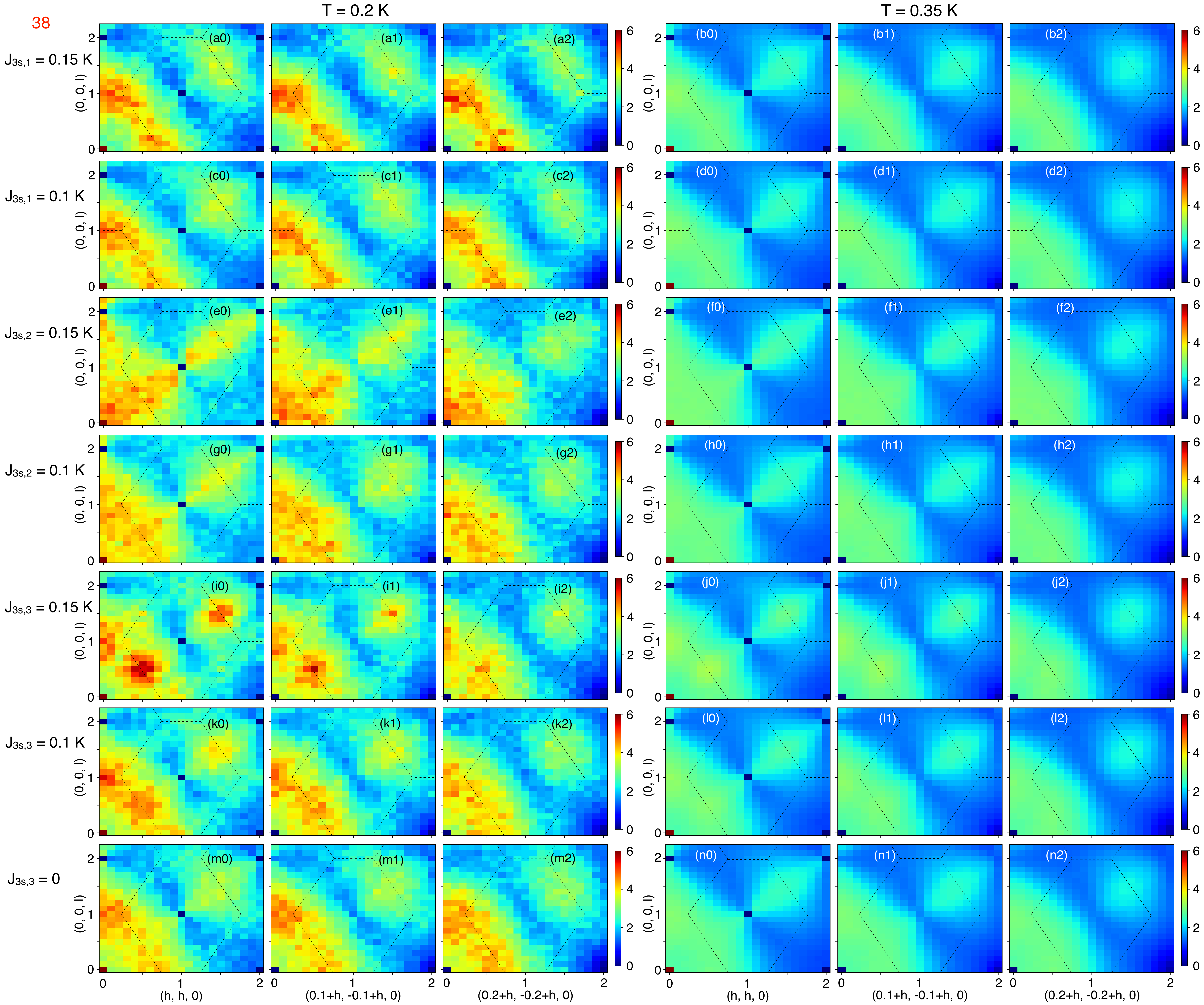}
\caption{ 
Intensity maps of $S(\bm{Q})$ calculated by the 16000-site CMC simulations 
using parameters indicated by red circles shown in Figs.~\ref{CT_MC111213141516}(b1--b3) 
($J_{3\text{s},i}$; $i=1,2,3$; $J_{3\text{s},j \neq i}=0$) and 
by the point 38 in Fig.~\ref{phase_diagram_Dnn0p478} [$\frac{J_{\text{nn}}}{J_{\text{nn}} + D_{\text{nn}}}(\delta,q)=(0.0,0.45)$]. 
They are viewed by 2D slices of $\bm{Q}=(k+h,-k+h,l)$ with fixed $k=0,0.1$, and $0.2$, 
which are shown in (w0), (w1), and (w2) (w=a--n), respectively. 
They are calculated at two temperatures 0.2 K (a,c,e,g,i,k,m) and 0.35 K (b,d,f,h,j,l,n), 
below and above the specific heat peak.
Intensity maps for $J_{3\text{s},1} = 0.15$ and $0.1$ K [Fig.~\ref{CT_MC111213141516}(b1)] are shown in (a,b) and (c,d), respectively. 
Intensity maps for $J_{3\text{s},2} = 0.15$ and $0.1$ K [Fig.~\ref{CT_MC111213141516}(b2)] are shown in (e,f) and (g,h), respectively. 
Intensity maps for $J_{3\text{s},3} = 0.15$, $0.1$, and 0 K [Fig.~\ref{CT_MC111213141516}(b3)] are shown in (i,j), (k,l), and (m,n), respectively. 
}
\label{SQcalMCp12}
\end{figure*}
\begin{figure*}[ht]
\centering
\includegraphics[width=18cm,clip]{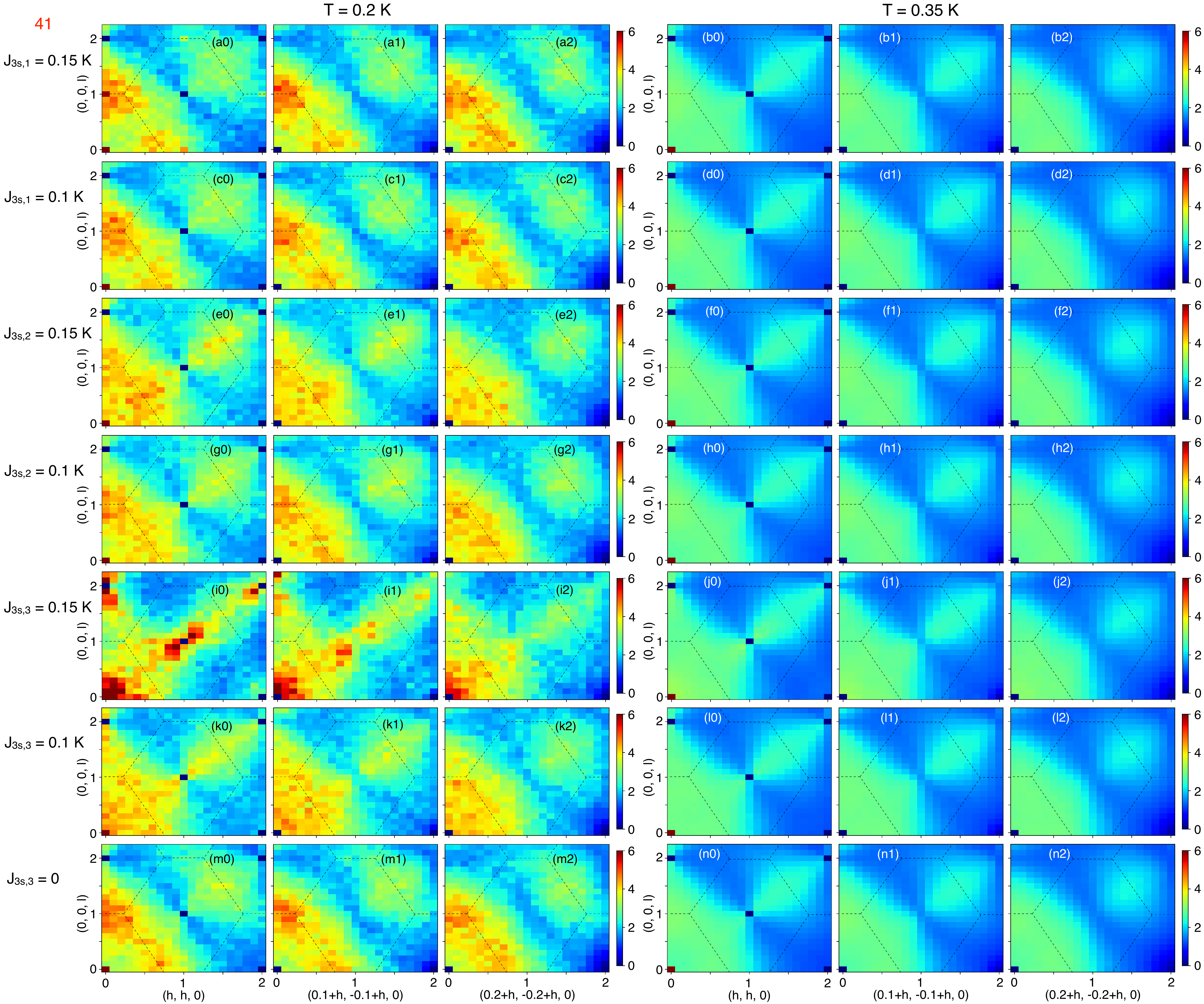}
\caption{ 
Intensity maps of $S(\bm{Q})$ calculated by the 16000-site CMC simulations 
using parameters indicated by red circles shown in Figs.~\ref{CT_MC111213141516}(e1--e3) 
($J_{3\text{s},i}$; $i=1,2,3$; $J_{3\text{s},j \neq i}=0$) and 
by the point 41 in Fig.~\ref{phase_diagram_Dnn0p478} [$\frac{J_{\text{nn}}}{J_{\text{nn}} + D_{\text{nn}}}(\delta,q)=(0.0,-0.45)$]. 
They are viewed by 2D slices of $\bm{Q}=(k+h,-k+h,l)$ with fixed $k=0,0.1$, and $0.2$, 
which are shown in (w0), (w1), and (w2) (w=a--n), respectively. 
They are calculated at two temperatures 0.2 K (a,c,e,g,i,k,m) and 0.35 K (b,d,f,h,j,l,n), 
below and above the specific heat peak.
Intensity maps for $J_{3\text{s},1} = 0.15$ and $0.1$ K [Fig.~\ref{CT_MC111213141516}(b1)] are shown in (a,b) and (c,d), respectively. 
Intensity maps for $J_{3\text{s},2} = 0.15$ and $0.1$ K [Fig.~\ref{CT_MC111213141516}(b2)] are shown in (e,f) and (g,h), respectively. 
Intensity maps for $J_{3\text{s},3} = 0.15$, $0.1$, and 0 K [Fig.~\ref{CT_MC111213141516}(b3)] are shown in (i,j), (k,l), and (m,n), respectively. 
}
\label{SQcalMCp15}
\end{figure*}

Several 16000-site CMC simulations were performed 
to study effects of each three-spin interaction on $S(\bm{Q})$ 
on the SI phase sides of neighborhoods of the SI and 3D-PAF phase boundaries. 
Considering the results of Sec.~\ref{results_CMC_C_SI}, 
the parameters $(\delta,q)$ were fixed to the two sets: 
$\frac{J_{\text{nn}}}{J_{\text{nn}} + D_{\text{nn}}} (\delta,q)=(0.0,0.45)$ and $(0.0,-0.45)$, 
the points 38 and 41 in Fig.~\ref{phase_diagram_Dnn0p478}. 
The three-spin interaction constant was fixed to two typical values: 
$J_{3\text{s},i}=0.1$ and $0.15$ K ($J_{3\text{s},j \neq i}=0$). 
Figure~\ref{SQcalMCp12} shows the resulting intensity maps of $S(\bm{Q})$ 
which are calculated with the parameters corresponding to the red circles in Figs.~\ref{CT_MC111213141516}(b1--b3), 
and at $0.2$ and $0.35$ K,
below and above the specific heat peak. 
Figure~\ref{SQcalMCp15} shows the resulting intensity maps of $S(\bm{Q})$ 
which are calculated with the parameters corresponding to the red circles in Figs.~\ref{CT_MC111213141516}(e1--e3), 
and at $0.2$ and $0.35$ K,
below and above the specific heat peak. 
One can notice that statistical errors of $S(\bm{Q})$ at 0.2K 
shown in Figs.~\ref{SQcalMCp12} and \ref{SQcalMCp15} 
are much larger than those in Figs.~\ref{SQcalMCp2} and \ref{SQcalMCp5}. 
This indicates that pseudospin fluctuations 
are considerably slowed down in the SI phase within the CMC simulation. 

When the three-spin interactions are set to zero, 
the calculated intensity maps with $\frac{J_{\text{nn}}}{J_{\text{nn}} + D_{\text{nn}}} q=0.45$ and $-0.45$, 
which are Figs.~\ref{SQcalMCp12}(m,n) and \ref{SQcalMCp15}(m,n), respectively, 
show almost the same characteristics: 
the pinch-point structure of SI is seen around the $\Gamma$ points $(0,0,2)$ and $(1,1,1)$, 
the intensity pattern becomes strengthened as temperature is lowered below the specific heat peak. 
The intensity pattern is scarcely affected by the sign of the parameter $q$. 

When the three-spin interactions are switched on, 
$S(\bm{Q})$ at 0.35 K [Figs.~\ref{SQcalMCp12}(b,d,f,h,j,l) and \ref{SQcalMCp15}(b,d,f,h,j,l)] 
depend little on $J_{3\text{s},i}$. 
On the other hand, 
$S(\bm{Q})$ at 0.2 K [Figs.~\ref{SQcalMCp12}(a,c,e,g,i,k) and \ref{SQcalMCp15}(a,c,e,g,i,k)] 
show various intensity patterns depending on $J_{3\text{s},i}$, 
which are attributable to the lifting of the degeneracy of the SI manifold. 
In relation to the analysis of TTO, 
there is only one somewhat interesting $S(\bm{Q})$ shown in Fig.~\ref{SQcalMCp12}(i), 
of which the parameters are $J_{3\text{s},3}$ = 0.15 K and $q>0$. 
These parameters are very close to the suggested range for the further investigation 
discussed in Sec.~\ref{results_CMC_SQ_3DPAF}. 

\section{\label{appendix_TPQ} other results of quantum simulation using TPQ states: specific heat, entropy, and $S(\bm{Q})$ for $\delta \ne 0$}
\begin{figure}[hbt]
\centering
\includegraphics[width=8cm,clip]{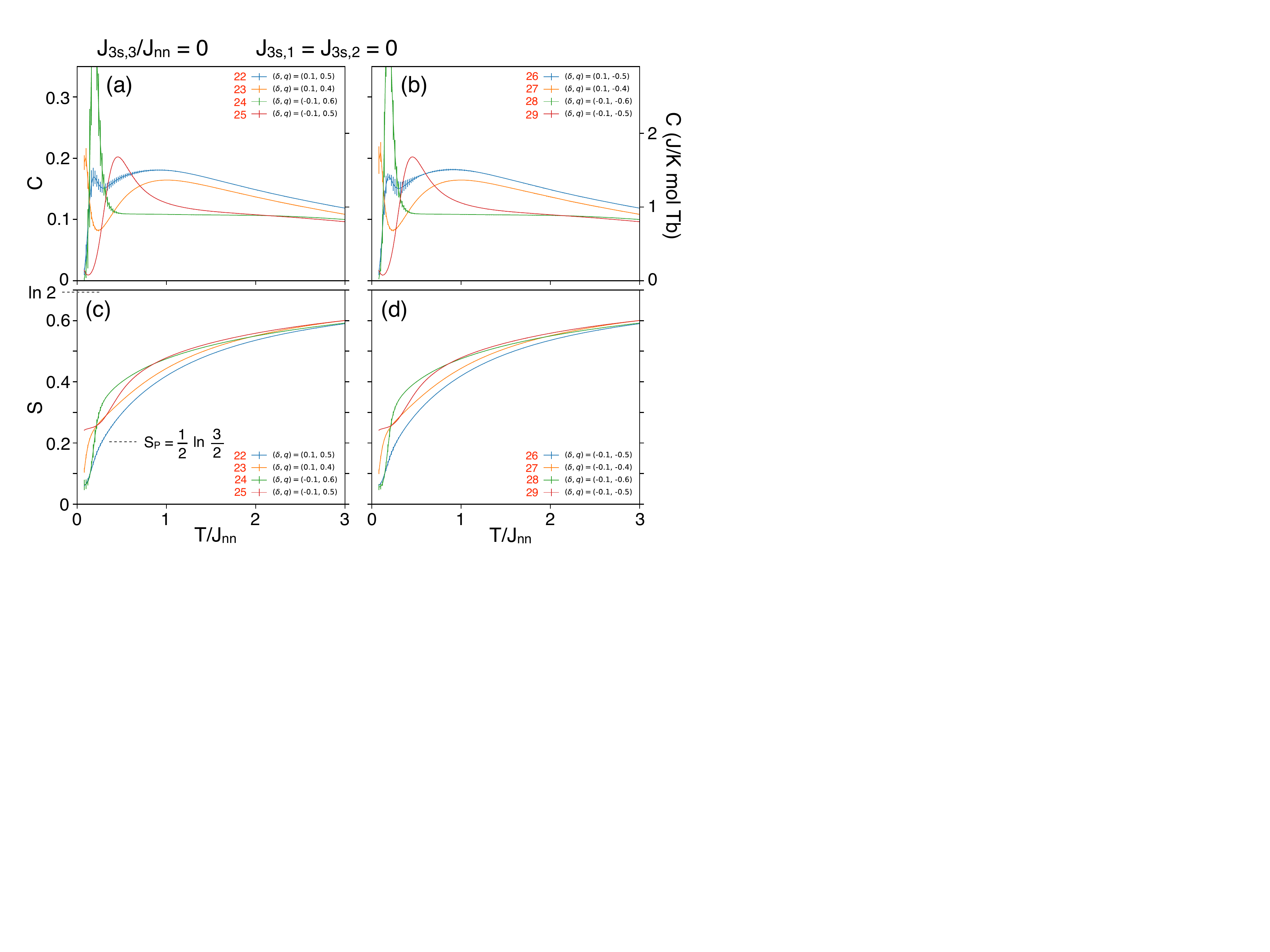}
\caption{
Temperature dependence of specific heat $C(T)$ and entropy $S(T)$ 
obtained by the 32-site simulations using the cTPQ state 
for $J_{3\text{s},3}/J_{\text{nn}}=0$ ($J_{3\text{s},1}=J_{3\text{s},2}=0$) 
with parameters $(\delta \ne 0,q)$, the points 22--29 in Fig.~\ref{phase_diagram_Dnn0p0}. 
In (a) and (b) $C(T)$ for $q \geq 0$ and $q \leq 0$ are shown, respectively. 
In (c) and (d) $S(T)$ for $q \geq 0$ and $q \leq 0$ are shown, respectively.
}
\label{CTST_TPQ_J3b30p0_delta}
\end{figure}
\begin{figure}[hbt]
\centering
\includegraphics[width=8cm,clip]{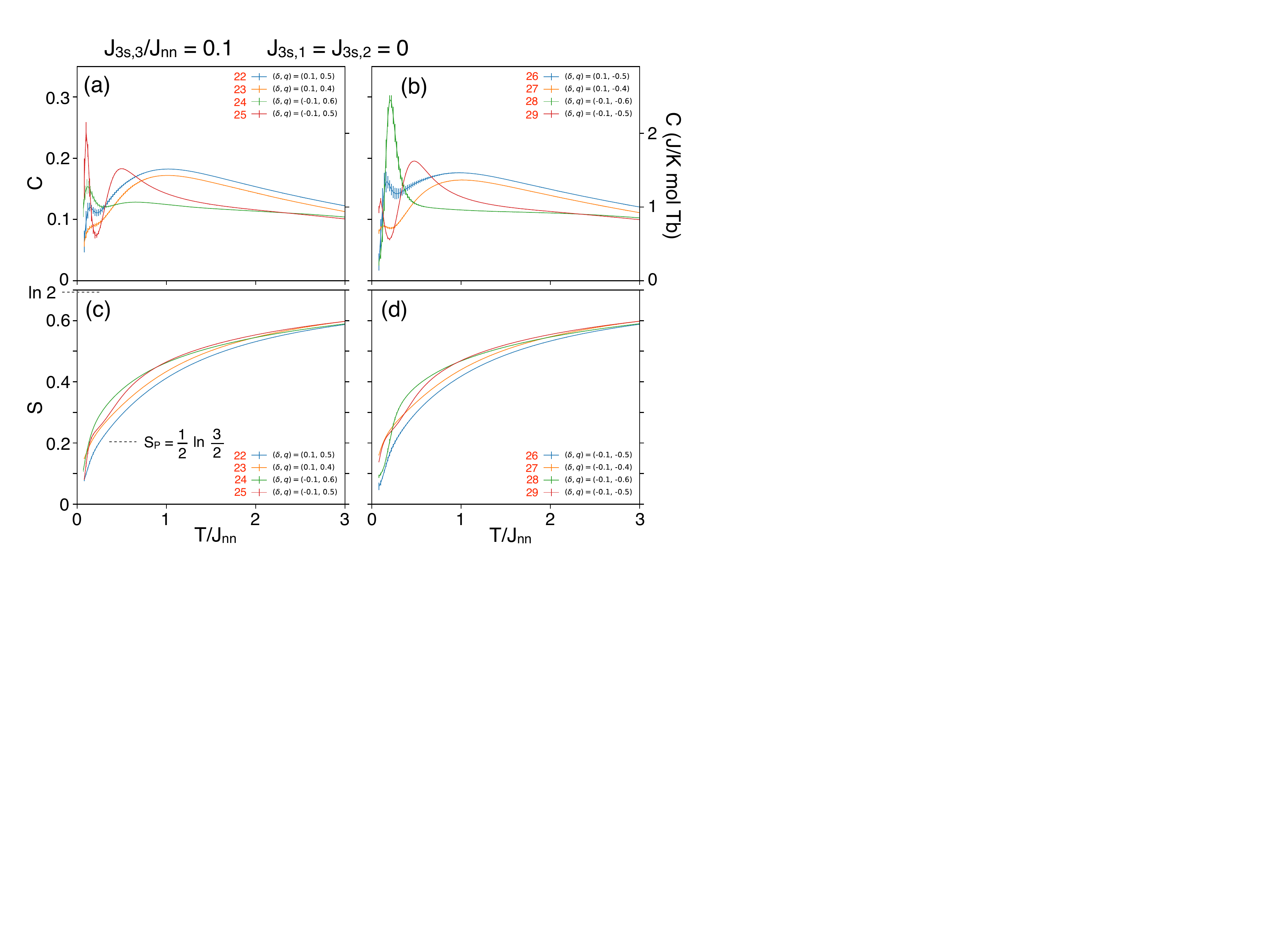}
\caption{
Temperature dependence of specific heat $C(T)$ and entropy $S(T)$ 
obtained by the 32-site simulations using the cTPQ state 
for $J_{3\text{s},3}/J_{\text{nn}}=0.1$ ($J_{3\text{s},1}=J_{3\text{s},2}=0$) 
with parameters $(\delta \ne 0,q)$, the points 22--29 in Fig.~\ref{phase_diagram_Dnn0p0}. 
In (a) and (b) $C(T)$ for $q \geq 0$ and $q \leq 0$ are shown, respectively. 
In (c) and (d) $S(T)$ for $q \geq 0$ and $q \leq 0$ are shown, respectively.
}
\label{CTST_TPQ_J3b30p1_delta}
\end{figure}
\begin{figure*}[ht]
\centering
\includegraphics[width=18cm,clip]{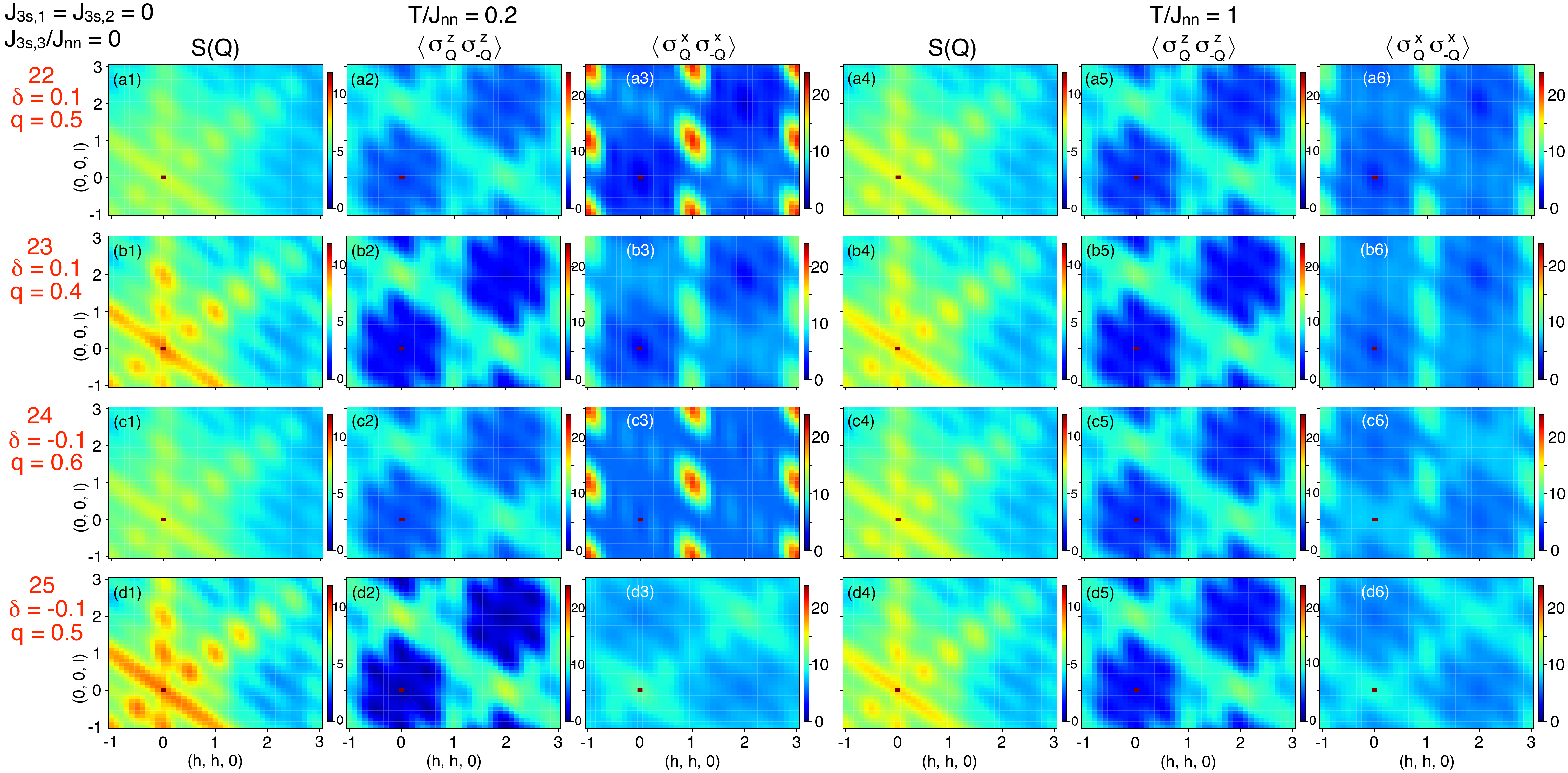}
\caption{ 
Two dimensional slices of 
(w1,w4) $S(\bm{Q})$, 
(w2,w5) $\langle \sigma_{\bm{Q}}^{z} \sigma_{-\bm{Q}}^{z} \rangle$, 
and (w3,w6) $\langle \sigma_{\bm{Q}}^{x} \sigma_{-\bm{Q}}^{x} \rangle$ (w=a--d) 
in the plane $\bm{Q}=(h,h,l)$ 
calculated by the 32-site simulations using the mTPQ state 
for $J_{3\text{s},3}/J_{\text{nn}}=0$ ($J_{3\text{s},1}=J_{3\text{s},2}=0$) 
with parameters $(\delta= \pm 0.1 , q > 0)$, the points 22--25 in Fig.~\ref{phase_diagram_Dnn0p0}. 
The 2D slice data at $T/J_{\text{nn}}=0.2$ and $1$ are shown in 
(w1--w3) and (w4--w6) (w=a--d), respectively. 
}
\label{SQTPQp2_p3p13p1p11J3b30p0J1qP}
\end{figure*}
\begin{figure*}[ht]
\centering
\includegraphics[width=18cm,clip]{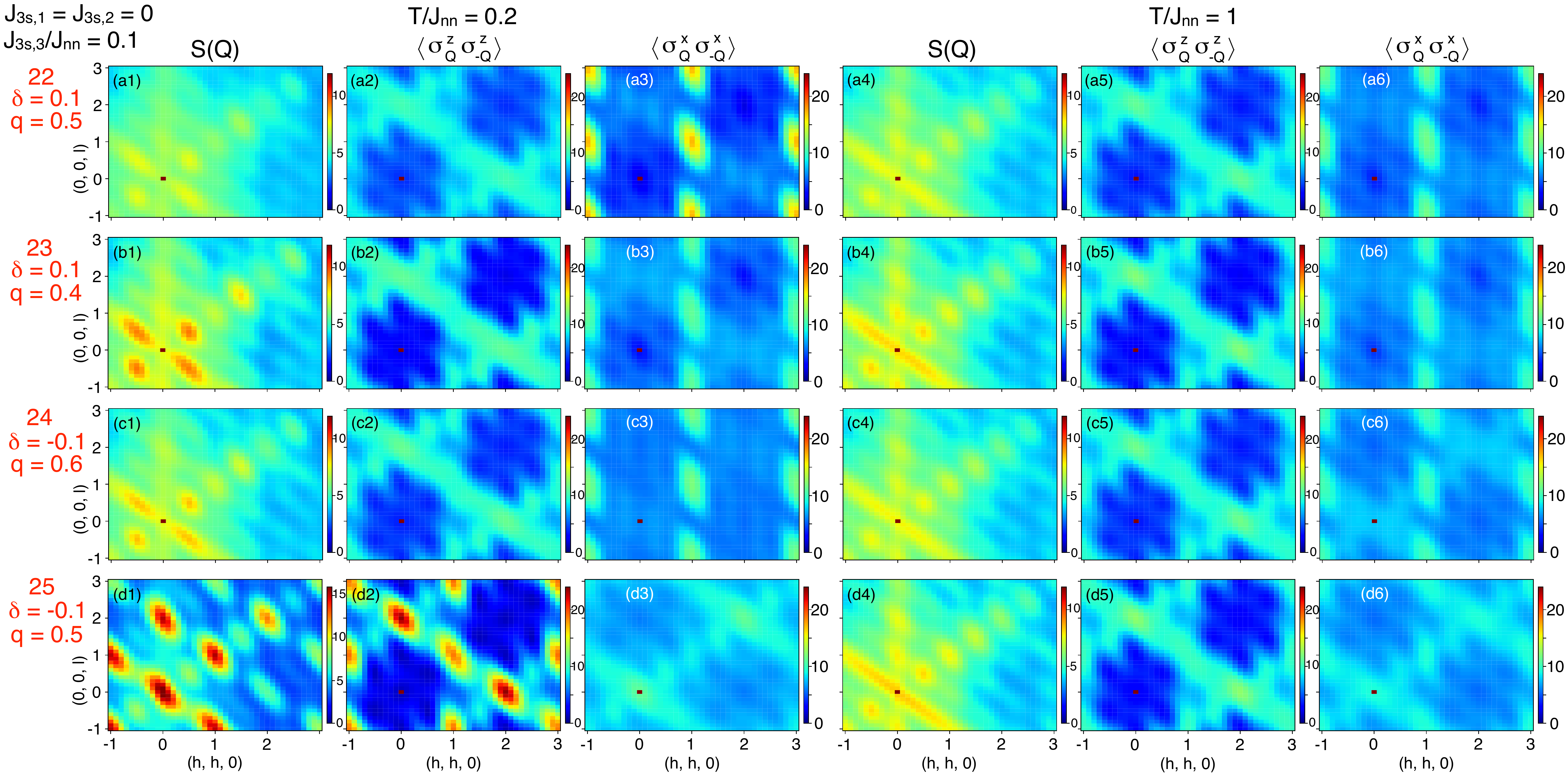}
\caption{ 
Two dimensional slices of 
(w1,w4) $S(\bm{Q})$, 
(w2,w5) $\langle \sigma_{\bm{Q}}^{z} \sigma_{-\bm{Q}}^{z} \rangle$, 
and (w3,w6) $\langle \sigma_{\bm{Q}}^{x} \sigma_{-\bm{Q}}^{x} \rangle$ (w=a--d) 
in the plane $\bm{Q}=(h,h,l)$ 
calculated by the 32-site simulations using the mTPQ state 
for $J_{3\text{s},3}/J_{\text{nn}}=0.1$ ($J_{3\text{s},1}=J_{3\text{s},2}=0$) 
with parameters $(\delta= \pm 0.1 , q > 0)$, the points 22--25 in Fig.~\ref{phase_diagram_Dnn0p0}. 
The 2D slice data at $T/J_{\text{nn}}=0.2$ and $1$ are shown in 
(w1--w3) and (w4--w6) (w=a--d), respectively. 
}
\label{SQTPQp2_p3p13p1p11J3b30p1J1qP}
\end{figure*}
\begin{figure*}[ht]
\centering
\includegraphics[width=18cm,clip]{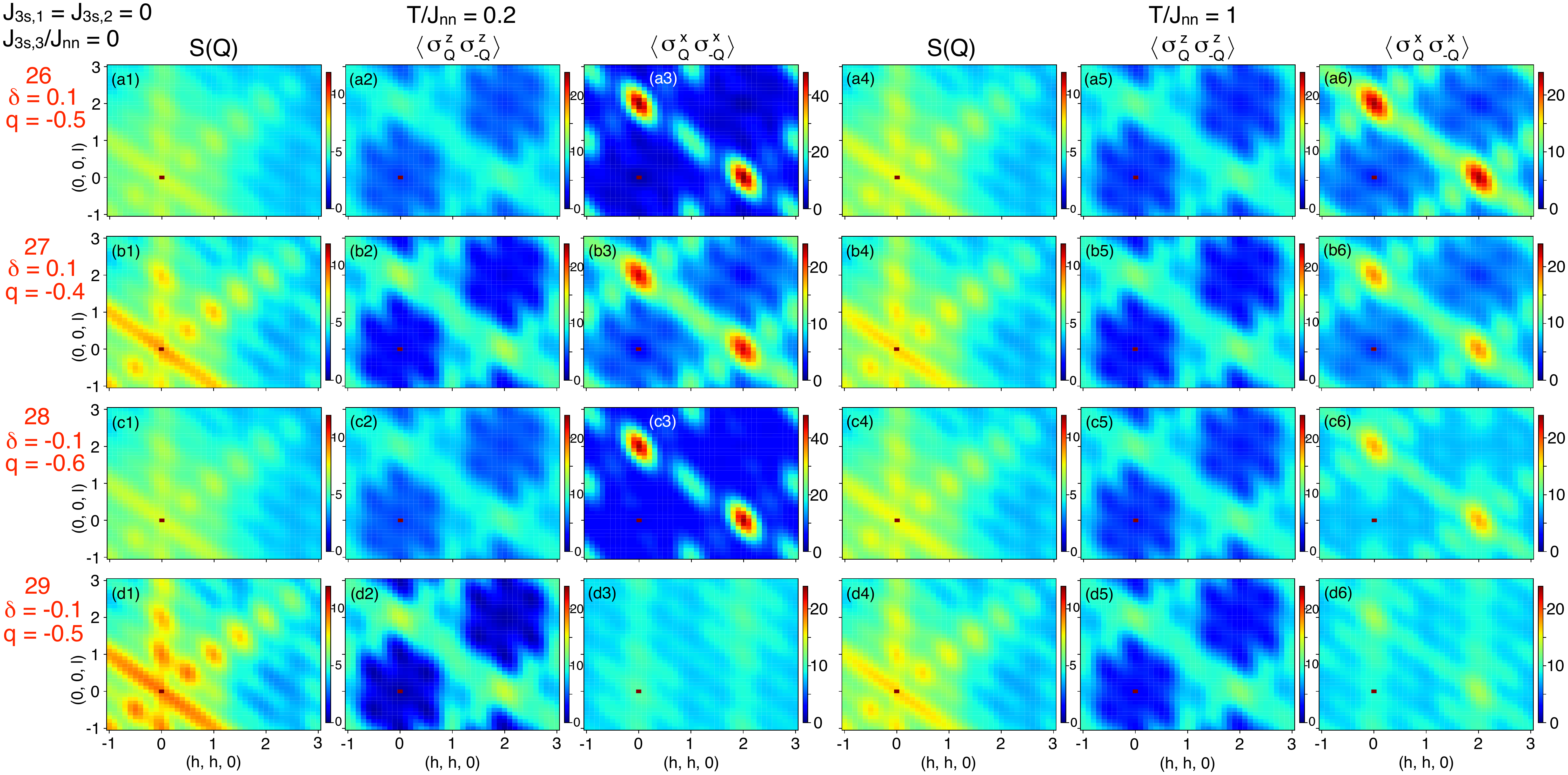}
\caption{ 
Two dimensional slices of 
(w1,w4) $S(\bm{Q})$, 
(w2,w5) $\langle \sigma_{\bm{Q}}^{z} \sigma_{-\bm{Q}}^{z} \rangle$, 
and (w3,w6) $\langle \sigma_{\bm{Q}}^{x} \sigma_{-\bm{Q}}^{x} \rangle$ (w=a--d) 
in the plane $\bm{Q}=(h,h,l)$ 
calculated by the 32-site simulations using the mTPQ state 
for $J_{3\text{s},3}/J_{\text{nn}}=0$ ($J_{3\text{s},1}=J_{3\text{s},2}=0$) 
with parameters $(\delta= \pm 0.1 , q < 0)$, the points 26--29 in Fig.~\ref{phase_diagram_Dnn0p0}. 
The 2D slice data at $T/J_{\text{nn}}=0.2$ and $1$ are shown in 
(w1--w3) and (w4--w6) (w=a--d), respectively. 
}
\label{SQTPQp6p16p4p14J3b30p0J1qM}
\end{figure*}
\begin{figure*}[ht]
\centering
\includegraphics[width=18cm,clip]{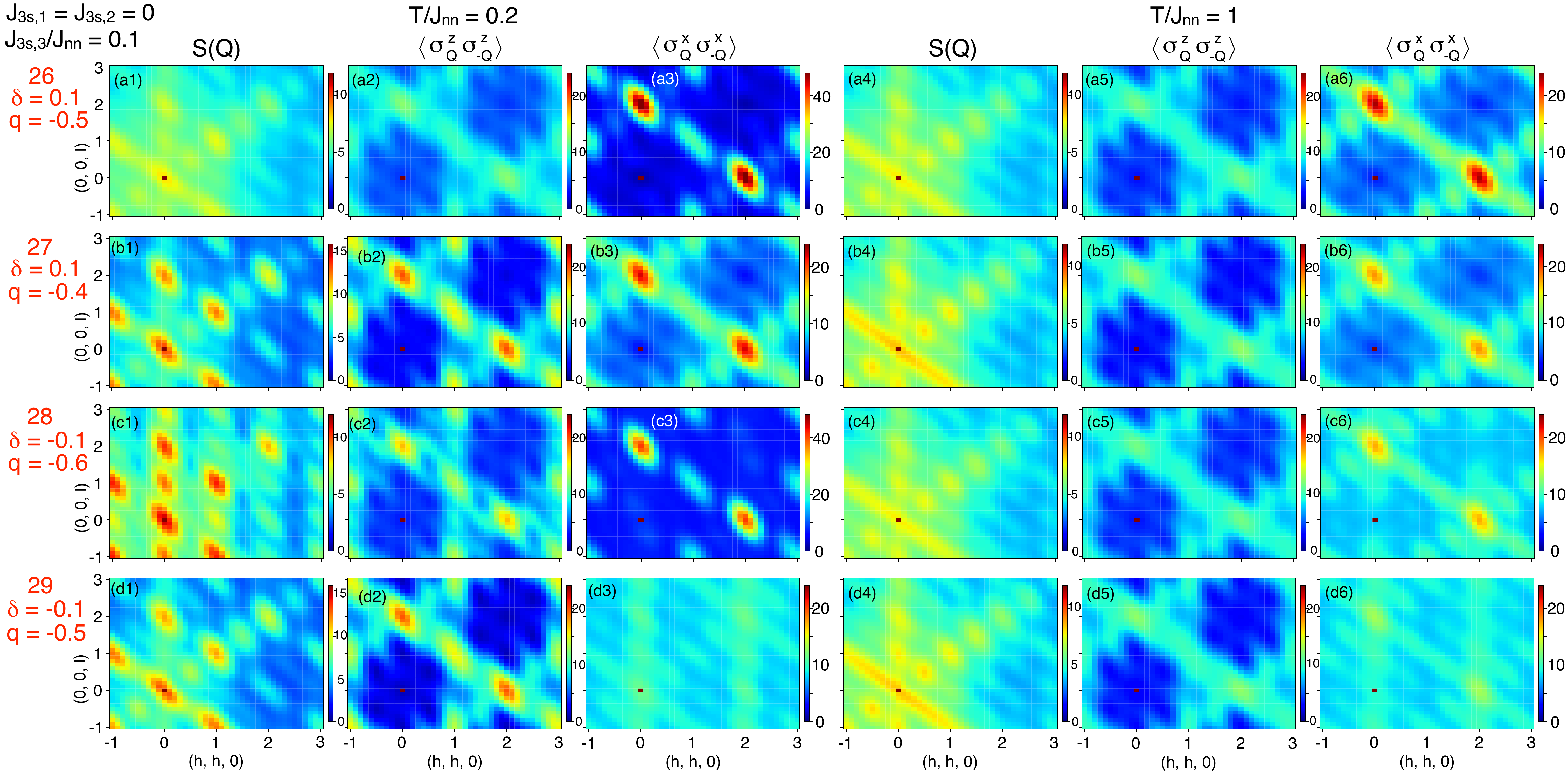}
\caption{ 
Two dimensional slices of 
(w1,w4) $S(\bm{Q})$, 
(w2,w5) $\langle \sigma_{\bm{Q}}^{z} \sigma_{-\bm{Q}}^{z} \rangle$, 
and (w3,w6) $\langle \sigma_{\bm{Q}}^{x} \sigma_{-\bm{Q}}^{x} \rangle$ (w=a--d) 
in the plane $\bm{Q}=(h,h,l)$ 
calculated by the the 32-site simulations using the mTPQ state 
for $J_{3\text{s},3}/J_{\text{nn}}=0.1$ ($J_{3\text{s},1}=J_{3\text{s},2}=0$) 
with parameters $(\delta= \pm 0.1 , q < 0)$, the points 26--29 in Fig.~\ref{phase_diagram_Dnn0p0}. 
The 2D slice data at $T/J_{\text{nn}}=0.2$ and $1$ are shown in 
(w1--w3) and (w4--w6) (w=a--d), respectively. 
}
\label{SQTPQp6p16p4p14J3b30p1J1qM}
\end{figure*}

To complement the simulation results on the $q$-axis, 
a few 32-site simulations using the TPQ states 
with the eight sets of the parameters $(\delta = \pm 0.1 ,q)$, 
the points 22--29 in Fig.~\ref{phase_diagram_Dnn0p0}, 
were carried out for $J_{3\text{s},3}/J_{\text{nn}}=0$ and $0.1$ ($J_{3\text{s},1}=J_{3\text{s},2}=0$). 
Temperature dependence of specific heat $C(T)$ and entropy $S(T)$ 
are plotted in Figs.~\ref{CTST_TPQ_J3b30p0_delta} and \ref{CTST_TPQ_J3b30p1_delta}. 
Two dimensional slices of $S(\bm{Q})$ 
and $\langle \sigma_{\bm{Q}}^{\alpha} \sigma_{-\bm{Q}}^{\alpha} \rangle$ ($\alpha=z,x$) 
calculated with $q > 0$, the points 22--25 in Fig.~\ref{phase_diagram_Dnn0p0}, 
for $J_{3\text{s},3}/J_{\text{nn}}=0$ and $0.1$ 
are shown in Figs.~\ref{SQTPQp2_p3p13p1p11J3b30p0J1qP} and \ref{SQTPQp2_p3p13p1p11J3b30p1J1qP}, respectively. 
Two dimensional slices of $S(\bm{Q})$ 
and $\langle \sigma_{\bm{Q}}^{\alpha} \sigma_{-\bm{Q}}^{\alpha} \rangle$ ($\alpha=z,x$) 
calculated with $q < 0$, the points 26--29 in Fig.~\ref{phase_diagram_Dnn0p0}, 
for $J_{3\text{s},3}/J_{\text{nn}}=0$ and $0.1$ 
are shown in Figs.~\ref{SQTPQp6p16p4p14J3b30p0J1qM} and \ref{SQTPQp6p16p4p14J3b30p1J1qM}, respectively. 

For $J_{3\text{s},3}/J_{\text{nn}}=0$, 
since $\mathcal{H}_0$ is invariant under the transformation 
of rotating $\bm{\sigma}_{\bm{r}}$ about the local $\bm{z}_{\bm{r}}$ axis by $\pi/2$ 
and $q \rightarrow -q$, 
$C(T)$ and $S(T)$ curves with $q>0$ [Figs.~\ref{CTST_TPQ_J3b30p0_delta}(a,c)] 
are almost the same as corresponding curves with $q<0$ [Figs.~\ref{CTST_TPQ_J3b30p0_delta}(b,d)]. 
For $J_{3\text{s},3} \ne 0$ the invariance does not hold, 
resulting in $C(T,\delta,q) \ne C(T,\delta,-q)$ [Figs.~\ref{CTST_TPQ_J3b30p1_delta}(a,b)] 
and $S(T,\delta,q) \ne S(T,\delta,-q)$ [Figs.~\ref{CTST_TPQ_J3b30p1_delta}(c,d)]. 

For $J_{3\text{s},3}/J_{\text{nn}}=0$, 
$S(\bm{Q})$ and $\langle \sigma_{\bm{Q}}^{z} \sigma_{-\bm{Q}}^{z} \rangle$ with $q>0$ 
[Figs.~\ref{SQTPQp2_p3p13p1p11J3b30p0J1qP}(a1--d1,a4--d4) and \ref{SQTPQp2_p3p13p1p11J3b30p0J1qP}(a2--d2,a5--d5)] 
are the same as 
those with $q<0$ 
[Figs.~\ref{SQTPQp6p16p4p14J3b30p0J1qM}(a1--d1,a4--d4) and \ref{SQTPQp6p16p4p14J3b30p0J1qM}(a2--d2,a5--d5)], 
while 
$\langle \sigma_{\bm{Q}}^{x} \sigma_{-\bm{Q}}^{x} \rangle$ with $q>0$ [Figs.~\ref{SQTPQp2_p3p13p1p11J3b30p0J1qP}(a3--d3,a6--d6)] 
are different from 
those with $q<0$ [Figs,~\ref{SQTPQp6p16p4p14J3b30p0J1qM}(a3--d3,a6--d6)]. 
These are consequences of the invariance of $\mathcal{H}_0$. 
For $J_{3\text{s},3}/J_{\text{nn}} = 0.1$ and at $T/J_{\text{nn}} = 0.2$, 
since the invariance does not hold for $J_{3\text{s},3} \ne 0$, 
$S(\bm{Q})$ and $\langle \sigma_{\bm{Q}}^{z} \sigma_{-\bm{Q}}^{z} \rangle$ with $q>0$ 
[Figs.~\ref{SQTPQp2_p3p13p1p11J3b30p1J1qP}(a1--d1,a2--d2)] 
are different from 
those with $q<0$ [Figs.~\ref{SQTPQp6p16p4p14J3b30p1J1qM}(a1--d1,a2--d2)]. 

In relation to the analysis of TTO, 
from experience in Sections~\ref{results_TPQ_C}, \ref{results_TPQ_SQ_q_axisP}, and \ref{results_TPQ_SQ_q_axisM} 
we think that the important parameters can be found 
by inspection of $C(T)$ curves, i.e., 
by selecting $C(T)$ satisfying three conditions: 
$J_{3\text{s},3}/J_{\text{nn}} = 0.1$, 
$q>0$, 
temperature dependence of $C(T)$ is similar to that with $q = 0.5$ or $0.55$ shown in Fig.~\ref{CTST_TPQ_J3b30p1}(a). 
By inspecting Fig.~\ref{CTST_TPQ_J3b30p1_delta}(a), 
it is obvious that these conditions are met by three $C(T)$ curves 
with $J_{3\text{s},3}/J_{\text{nn}} = 0.1$ and 
with $(\delta,q)=(0.1,0.5)$, $(0.1,0.4)$, and $(-0.1,0.6)$, 
the points 22--24 in Fig.~\ref{phase_diagram_Dnn0p0}. 
The corresponding three 2D slices of $S(\bm{Q})$ at $T/J_{\text{nn}} = 0.2$ [Figs.~\ref{SQTPQp2_p3p13p1p11J3b30p1J1qP}(a1,b1,c1)] 
show spin correlations with $\bm{k} \sim (\tfrac{1}{2},\tfrac{1}{2},\tfrac{1}{2})$, 
which resemble those shown in Figs.~\ref{SQTPQL2J3b30p1J1qP}(b1,c1,d1). 
Therefore, we conclude that 
the TPQ results suggest that 
the effective Hamiltonian minimally describing TTO is $\mathcal{H}_0 + \mathcal{H}_{3\text{s}}$ 
with $J_{3\text{s},1}=J_{3\text{s},2}=0$, 
$J_{3\text{s},3}/J_{\text{nn}} \sim 0.1$ (or $-0.1$) 
and the parameters $(\delta,q)$ in the region 
which is enclosed by the red dashed line in Fig.~\ref{phase_diagram_Dnn0p0}. 

We chose the eight parameter sets $(\delta = \pm 0.1 ,q)$,
the points 22--29 in Fig.~\ref{phase_diagram_Dnn0p0}: 
the two points 22 and 24 are in the 3D PAF ($q>0$) phase, 
the two points 26 and 28 are in the 3D PAF ($q<0$) phase, 
the four points 23, 25, 27, and 29 are in the SI phase, 
because simulation results were expected to be 
similar to those of the four points 5, 7, 15, and 17 on the $q$-axis ($\delta =0$). 
But this was not the case. 
For example, 
the intensity patterns of $\langle \sigma_{\bm{Q}}^{x} \sigma_{-\bm{Q}}^{x} \rangle$ 
with $(\delta,q)=(-0.1,0.5)$ [Figs.~\ref{SQTPQp2_p3p13p1p11J3b30p0J1qP}(d3,d6) and \ref{SQTPQp2_p3p13p1p11J3b30p1J1qP}(d3,d6)] 
and 
with $(\delta,q)=(-0.1,-0.5)$ [Figs.~\ref{SQTPQp6p16p4p14J3b30p0J1qM}(d3,d6) and \ref{SQTPQp6p16p4p14J3b30p1J1qM}(d3,d6)] 
are very different from those with $\delta=0$. 
Since these seem to be caused by certain quantum corrections, 
simulations with systematic $(\delta,q)$-variation have to be performed 
to gain detailed information. 
Thus instead of making further comments, we make two plausible remarks. 
A quantum correction would explain the reason 
why the region enclosed by red dashed line in Fig.~\ref{phase_diagram_Dnn0p0} 
is not parallel to the classical phase boundary. 
All the results using the TPQ methods support that 
the $q<0$ side of the phase diagram (Fig.~\ref{phase_diagram_Dnn0p0})
can be excluded from studies of TTO. 

\bibliography{TTO_HK_p2}

\end{document}